\documentclass[namedreferences]{SolarPhysics}
\usepackage[optionalrh]{spr-sola-addons} 
\usepackage{graphicx}                    
\usepackage{color}                       
\usepackage{url}                         
\newcommand{\aap}{ {\it Astron. Astrophys.}}

\newcommand{\apj}{ {\it Astrophys. J.}}
\newcommand{\apjl}{ {\it Astrophys. J. Lett.}}
\newcommand{\apjs}{ {\it Astrophys. J. Suppl. Ser.}}

\newcommand{\solphys}{{\it Solar Phys.}}

\newcommand{\sovmat}{ {\it Sov. Math. Dokl.}}
\newcommand{\ssr}{ {\it Space Sci. Rev.}}

\begin{document}

\begin{article}

\begin{opening}

\title{Measurements of electron anisotropy in solar flares using albedo
with RHESSI X-ray data}
%
\author{E.~C.~M.~\surname{Dickson}\sep
E.~P.~\surname{Kontar}}
\runningauthor{Dickson and Kontar}
\runningtitle{Measurements of Solar Flare Anisotropy Using Albedo with
RHESSI}

\institute{SUPA School of Physics and Astronomy, University of
Glasgow, Glasgow G12 8QQ, Scotland, UK
email: \url{edickson.glasgow@gmail.com}\\ }

\begin{abstract}
The angular distribution of electrons accelerated in solar flares is a key parameter in the understanding 
of the acceleration and propagation mechanisms that occur there. 
However, the anisotropy of energetic electrons is still a poorly known quantity, with observational studies producing evidence for an isotropic distribution
and theoretical models mainly considering the strongly beamed case. We use the effect of photospheric albedo to infer the pitch angle distribution of X-ray emitting electrons using Hard X-ray data from {\it RHESSI}. A bi-directional approximation is applied and a regularized inversion is performed for eight large flare events to deduce the electron spectra in both downward (towards the photosphere) and upward (away from the photosphere) directions. The electron spectra and the electron anisotropy ratios are calculated for broad energy range from about $10$ and up to $\sim 300$ keV near the peak of the flares. The variation of electron anisotropy over short periods of time intervals lasting 4, 8 and 16 seconds near the impulsive peak has been examined. The results show little evidence for strong anisotropy and the mean electron flux spectra are consistent with the isotropic electron distribution.  The $3\sigma $-level uncertainties, although energy and event dependent, are found to suggest that anisotropic distribution with anisotropy larger than $\sim 3$ are not consistent with the hard X-ray data. At energies above $150-200$ keV, the uncertainties are larger and
thus the possible electron anisotropies could be larger.
\end{abstract}

%

\end{opening}

%

\section{Introduction}

Solar flares are one of the most energetic processes which occur in the
solar system. However the details of the acceleration processes responsible are still
poorly known. The X-rays which are observed at the Earth are commonly produced by the
accelerated electrons interacting with the solar plasma and producing bremsstrahlung
radiation. To understand solar flares it is therefore important to understand the
distribution of these accelerated electrons.  In general, this distribution will vary in
space, energy and in pitch angle (see \opencite{2011SSRv..159..107H}, \opencite{2011SSRv..159..301K},
as the recent reviews of electron properties in solar flares).

Several techniques have been used to estimate the anisotropy in the
pitch-angle distribution of X-ray emitting electrons in solar flares
\cite{2011SSRv..159..107H,2011SSRv..159..301K}.
The most commonly used method is to look at the centre-to-limb variation
of solar X-ray properties \cite{1974SoPh...39..155D,1987ApJ...322.1010V}.
That is comparing the characteristics, most commonly total X-ray flux
\cite{1974SoPh...35..431P}, or the spectral index of solar flares at the limb to disk centre events.
Studies concentrating on lower energy emission (below 300 keV) tended to find no significant
evidence of directivity. These studies have also been performed using SMM data
\cite{1988SoPh..118...49D} studying flares with energies
above 300 keV
\cite{1994ApJS...90..611V,1988ApJ...334.1049B,1985SvAL...11..322B}.
Some evidence for directivity at high energies has been reported
\cite{1991ApJ...379..381M,1991ICRC....3...69V}.
More recently RHESSI (Ramaty High Energy Solar Spectroscopic Imager) data \cite{2002SoPh..210....3L} has been used
to determine the X-ray anisotropy \cite{2007A&A...466..705K}. An obvious
disadvantage of the statistical method
is that the variation can only be seen as an average over a large number
of solar flares, so little
can be said about X-ray or electron anisotropy in a given flare.

An approach which allows individual flares to be studied is the
stereoscopic method \cite{1973ApJ...185..335C}.
Here each individual flare is measured directly by two spacecraft at two
different locations,
ideally well separated in space. Studies that have been performed using
this method do not show any clear evidence
of directivity \cite{1998ApJ...500.1003K,1994ApJ...426..758L}. A
disadvantage of this
approach is the difficulty in cross calibrating, often leading to large
errors. Another drawback is,
as in the centre-to-limb method, this technique does not give direct
information about the downward electron
distribution.

As an anisotropic electron distribution will produce polarised X-rays,
measuring the polarisation can therefore give a measure of the
anisotropy of the electron distribution \cite{1983ApJ...269..715L}.
An isotropic source should show low polarisation, whereas a beam should
produce significant
polarisation
\cite{1978ApJ...219..705B,1983ApJ...269..715L,2008ApJ...674..570E}.
Studies of polarisation have been performed
using various X-ray satellites, recently using the \emph{Coronas-F} satellite
\cite{2006SoSyR..40...93Z}.
 The reported measurements vary substantially from observation to
observation,
adding to the scepticism of these measurements. A major drawback of this
approach is the observational difficulty in measuring polarisation at
HXR energies for transient events like solar flares.
Several attempts have been made
using RHESSI \cite{2002SoPh..210..125M,2006SoPh..239..149S},
but so far there have been no conclusive measurements made.
HXR polarisation has not yet been used to its full potential and future
observations could provide a more definitive answer.

Another important process, which can be used to diagnose the angular
distribution,
is photospheric albedo \cite{2006ApJ...653L.149K}. Photospheric albedo
of X-rays results from
initially downward directed X-rays, which are Compton scattered off
electrons in the solar photosphere, being observed at Earth. This
effect distorts the HXR
spectrum \cite{1972ApJ...171..377T,1978ApJ...219..705B} and leads
to an appearance of apparent low energy cutoff in the deduced electron
spectrum \cite{2008SoPh..252..139K},
when the observed X-ray spectrum is not corrected for an albedo
\cite{2006A&A...446.1157K}.
As the spectral shapes of reflected and primary hard X-ray spectra are
sufficiently
distinct, these two components can be distinguished and the albedo
reflected flux
could be used as a measure of the downward going electrons (Figure 1).
RHESSI provides sufficient energy resolution, broad energy coverage,
and sensitivity to better constrain directivity
of energetic electrons in individual solar flare events.

Here we use this albedo method to examine the directivity of energetic
electrons in solar flares.
We first perform some forward modelling to examine the effect
directivity of the electron spectrum and the role of the albedo
contribution has on solar flare spectra. The RHESSI flare catalogue has
been searched for suitable flares
between 2002 and 2008 and we use the spectral data from the impulsive
phases of several well observed flares
to perform a bi-directional inversion, estimating the fluxes of
electrons travelling towards and away from the photosphere.

\section{Solar flare spectrum and electron anisotropy}

The spatially integrated X-ray photon flux spectrum at the Earth
$I(\epsilon,\theta_0)$,
[phot\-ons\-~cm$^{-2}$\-~s$^{-1}$~\-keV$^{-1}$] is
straightforwardly related to the mean electron flux spectrum of
energetic electron
${\overline F}(E,\beta)$ [electrons
~cm$^{-2}$\-~s$^{-1}$~\-keV$^{-1}$~\-sr$^{-1}$]
via the linear integral relation (e.g.
\inlinecite{2004ApJ...613.1233M}, \inlinecite{2011arXiv1110.4993J}).
For a given electron distribution ${\overline F}(E,\beta)$, where $\beta
$ is the electron pitch angle,
the resulting emitted X-ray flux at distance $R$ is
\begin{equation}
\label{eq:Idef}
I(\epsilon,\theta_0) =\frac{{\bar n} V } {4\pi R^2}
\int_0^{2\pi} \int_0^{\pi}\int_\epsilon^\infty
{\overline F}(E,\beta) Q(\epsilon,E,\theta(\beta,\phi))sin(\beta)dE d\beta d\phi \;~,
\end{equation}
where $V$ is the source volume, $\bar
n$ is the mean plasma density,
$Q(\epsilon,E,\theta)$ is the angular dependant bremsstrahlung
cross-section, $\epsilon$ is photon energy, $E$ is electron energy
and $\theta$ is the angle between the initial electron velocity vector
and the direction of the
emitted photon. The cross-section used here is the electron-ion cross
section, formula 2BN from \inlinecite{1959RvMP...31..920K}
with Coulomb correction by \inlinecite{1939AnP...426..178E} added.
The relation between the angles $\beta$, $\theta_0$, $\phi$ and $\theta$
is given
by
\begin{equation}
cos\theta= cos\beta \, cos\theta_0 + sin\beta \, sin\theta_0 \, cos\phi \;~,
\end{equation}
where $\theta_0$ is the angle between the emitted photon and the direction of the photosphere.
The exact plasma density distribution $n$ and flaring volume $V$ are
often poorly known
and therefore it is preferable if the value ${\bar n}V{\bar
F}(E,\Omega)$ is inferred,
this is the mean electron flux spectrum \cite{2003ApJ...595L.115B},
a density weighted electron flux, multiplied by ${\bar n}V$ the number of electrons
in the emitting volume. This value is model independent, and its detailed
structure can be related to the electron acceleration and
propagation physics which is central to the understanding
of solar flares.

\begin{figure}[htbp!]
\centering
\includegraphics[scale=0.3]{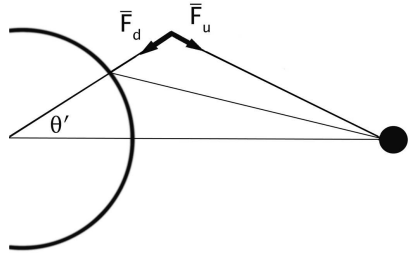}
\caption{The geometry of the X-ray emitting source above the photosphere
and bi-directional approximation.
X-rays are emitted in all directions and observed directly at Earth or
Compton back-scattered in the solar photosphere and then observed at
Earth. The true angular distribution of electrons $ {\overline
F}(E,\beta)$ is approximated
by downward $F_{d}$ and upward $ F_{u}$ going electrons.}
\label{fig:geo}
\end{figure}

The electron distribution of emitting electrons is assumed to be
separable in energy
and angle. The energy dependence is taken to be a single power-law as
solar flare x-ray observations are often well fit by power laws
implying close to power-law electron spectra. For simplicity the pitch
angle distribution is taken
to be a gaussian beam centred downwards. This pitch angle distribution
is assumed as
electrons tend to stream along magnetic field lines resulting in a
downwards beam, and the gaussian parametrization of this is popular as it
has useful analytic properties. The total distribution has the form
similar to \inlinecite{1983ApJ...269..715L}
\begin{equation}
\label{eq:mod}
{\overline F}(E,\beta) \propto E^{-\delta} \exp \left( \frac{-(1 - \mu)^2}{\Delta
\mu^2}
\right) \;~,
\end{equation}
where $ \mu = cos(\beta)$ and $ \Delta\mu$ characterises the width of
the Gaussian. It should be noted that Equation~(3) refers to the mean electron flux spectrum
not the injected electron spectrum considered in \opencite{1983ApJ...269..715L}.
Three different cases were assumed for the angular
variation of the electron spectrum  (Figure 2): the isotropic case ($\Delta \mu
=10$), an intermediate anisotropic case ($\Delta \mu =0.4$) and a
highly beamed case ($\Delta \mu =0.1$). The electron spectral index
$\delta$ was assumed to be $\delta =2$ as the typical flare spectral
index of X-rays in large flares is around $\gamma=3$ (see e.g. \opencite{1985SoPh..100..465D}) and for the mean electron flux spectrum $\gamma \approx \delta + 1$.

As the assumed electron distribution has azimuthal symmetry an average
cross-section integrated over $\phi$ can immediately be defined leaving
only the angles $\theta_0$ and $\beta$ needed to characterise the
angular distribution (c.f. \opencite{2004ApJ...613.1233M})

\begin{equation}
\label{cs} Q'(\epsilon,E,\theta_0,\beta) =
\int_{\phi=0}^{2\pi}Q(\epsilon,E,\theta(\phi)) d\phi\;~.
\end{equation}

The angular distribution of the emitted X-rays with respect to the
downward direction can now be found by applying this cross section to
the assumed electron spectrum and integrating over electron energy and
pitch angle (Figure 3)
\begin{equation}
\label{Idef3} I(\epsilon,\theta_0) =\frac{{\bar n} V } {4\pi R^2}
\int_0^\pi\int_\epsilon^\infty
{\overline F}(E,\beta) Q'(\epsilon,E,\beta,\theta_0)sin(\beta)dEd\beta \;~.
\end{equation}

The primary emission observed for a flare at heliocentric angle
$\theta^\prime$ is expected to come from a small range in angle in the
direction of the observer. Considering the geometry, it is clear that the
upward directed photon distribution can be approximated as
\begin{equation}
I_U (\epsilon)= I(\epsilon,\theta_0=180^{\circ}-\theta^\prime)\;~.
\end{equation}
The albedo reflected component, on the other hand, results from the
photons directed down towards the photosphere. This is likely to be a
broader distribution so an average is taken over a downwards directed
cone concentric with the mean direction of the electron distribution and
with half angle $\alpha$, in this study a value of $90^\circ$ is used.
The downward directed flux can then be defined as
\begin{equation}
I_D (\epsilon)= \frac{\displaystyle \int_{\theta_0 = 0}^\alpha
I(\epsilon,\theta_0) \sin \theta_0 d \theta_0} {\displaystyle
\int_{\theta_0 =
0}^\alpha\sin\theta_0 d\theta_0} \;~.
\end{equation}

The reflected component, due to albedo, of a given X-ray spectrum
incident on the photosphere can be characterised by using a Green's
function $A$ dependant on observation angle $\theta^{\prime}$
\cite{2006ApJ...653L.149K}. Thus, the total observed X-ray spectrum will
be given by the sum of the directly observed and reflected components:
\begin{equation}
I_O = I_U+ A\; I_D \;~,
\end{equation}
where $I_O$ is the total and $I_U$, $I_D$ are upward and downward
directed components (Figure 6).

In practice the photon spectrum is calculated numerically with a finite
resolution. For energy a pseudo-logarithmic binning scheme with 100 bins
starting at $10$ keV and going up to 5~MeV is used for both electrons and
photons. Due to the highly anisotropic nature of the cross-section at
high energies fine resolution in angle was needed. Angles $\beta$ and
$\theta_0$ were binned in 90 evenly spaced bins between 0 and $\pi$
whereas $\phi$ was binned in 180 evenly spaced bins between 0 and $2\pi$.

The upward and downward components of the photon flux are now given by
vectors. The observer directed component is defined as
\begin{equation}
{\bf I_{U}} = \{I_U(\epsilon_i),...,I_U(\epsilon_n)\}, \qquad i=1,..,n
\end{equation}
where $\epsilon_i$ corresponds to the centre energy of the photon bin
and $n$ is the number of bins in photons space. The photosphere directed
component is given by ${\bf I_{D}}$ defined in the same way.
Similarly the angle dependant Green's function is here calculated in the
form of a $ n \times n$ Green's matrix $\bf A$.
The vector representing the total observed flux is simply given by
\begin{equation}
\bf{I_O} = {\bf I _U}+ A{\bf I _D}\;~,
\end{equation}
where $A$ is the albedo matrix constructed from Green's functions.

\subsection{Photon Spectral Index}

Due to the effect of albedo one of the most notable variations with
changing anisotropy is in the photon spectral index, the derivative with
respect to energy of the photon flux. It is commonly defined as
\cite{1988ApJ...331..554B,2003A&A...407..725C}
\begin{equation}
\gamma (\epsilon) = - \frac{\epsilon}{I} \, \frac{dI}{d\epsilon} =
-\frac{d\,log\,I}{d\,log\,\epsilon} \;~,
\end{equation}
which is calculated for both the primary photon spectrum for a range of
viewing angles (Figure 5) and for the total observed spectrum including albedo
component for a range of positions on the solar disk (Figure 7).

\begin{figure}[htbp]
\centering
\includegraphics[width=39mm,height=39mm]{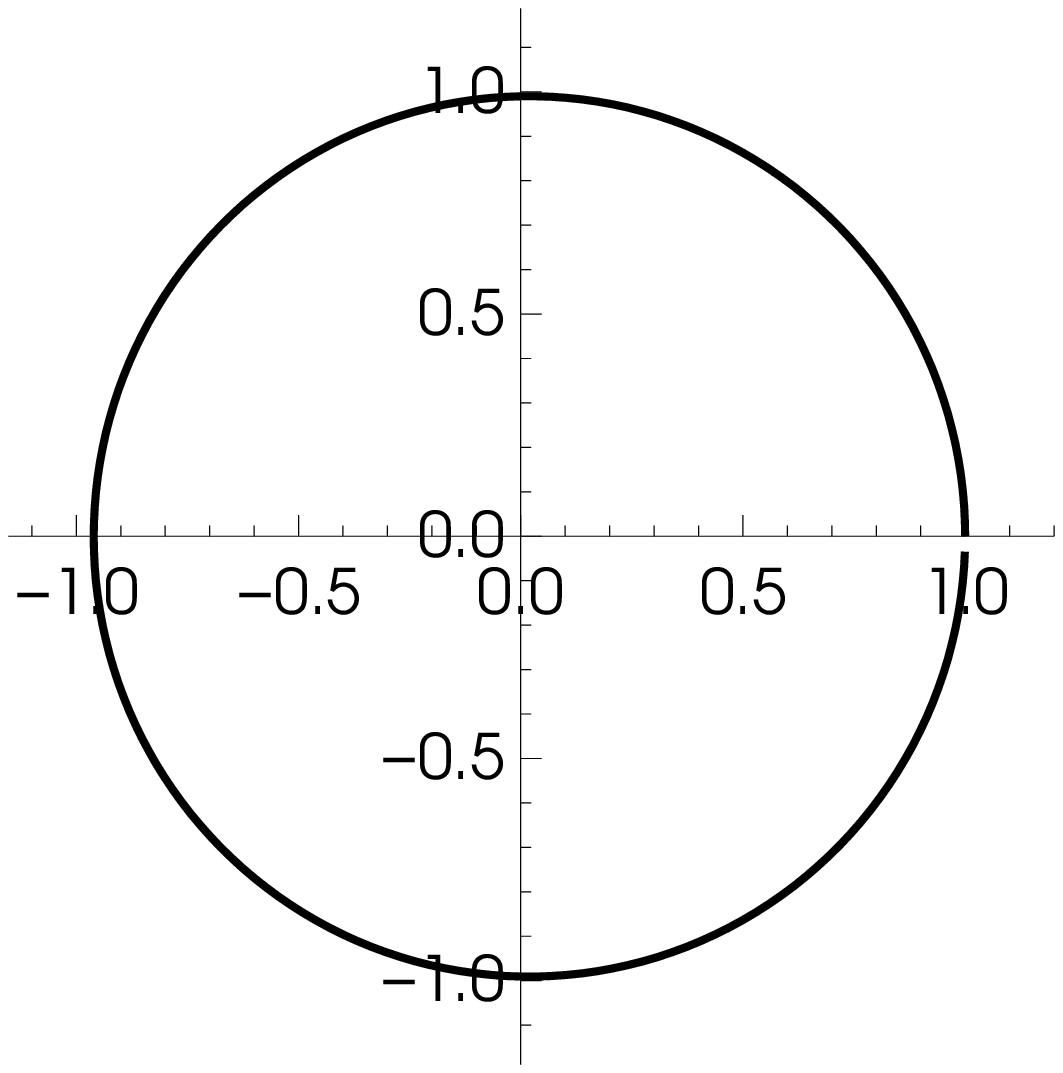}
\includegraphics[width=39mm,height=39mm]{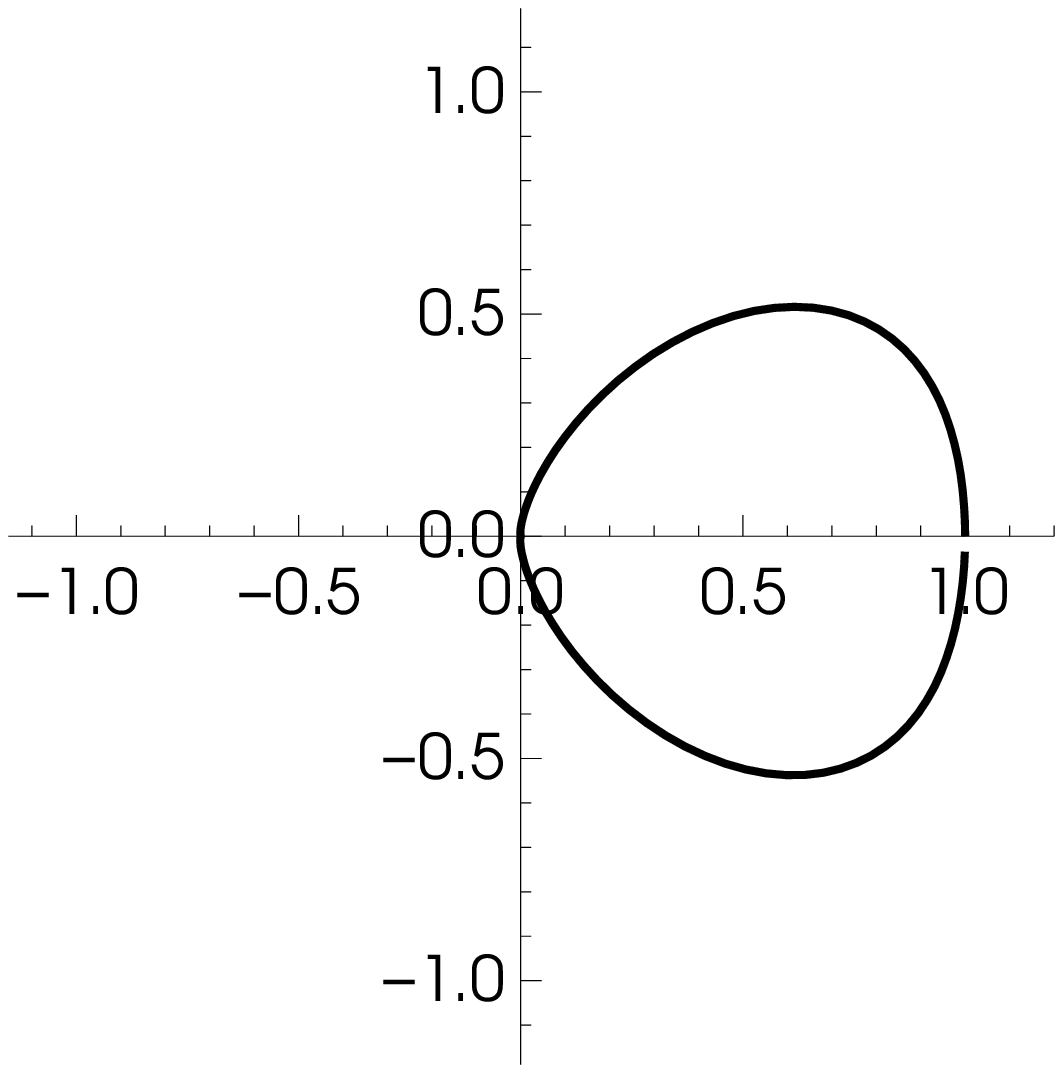}
\includegraphics[width=39mm,height=39mm]{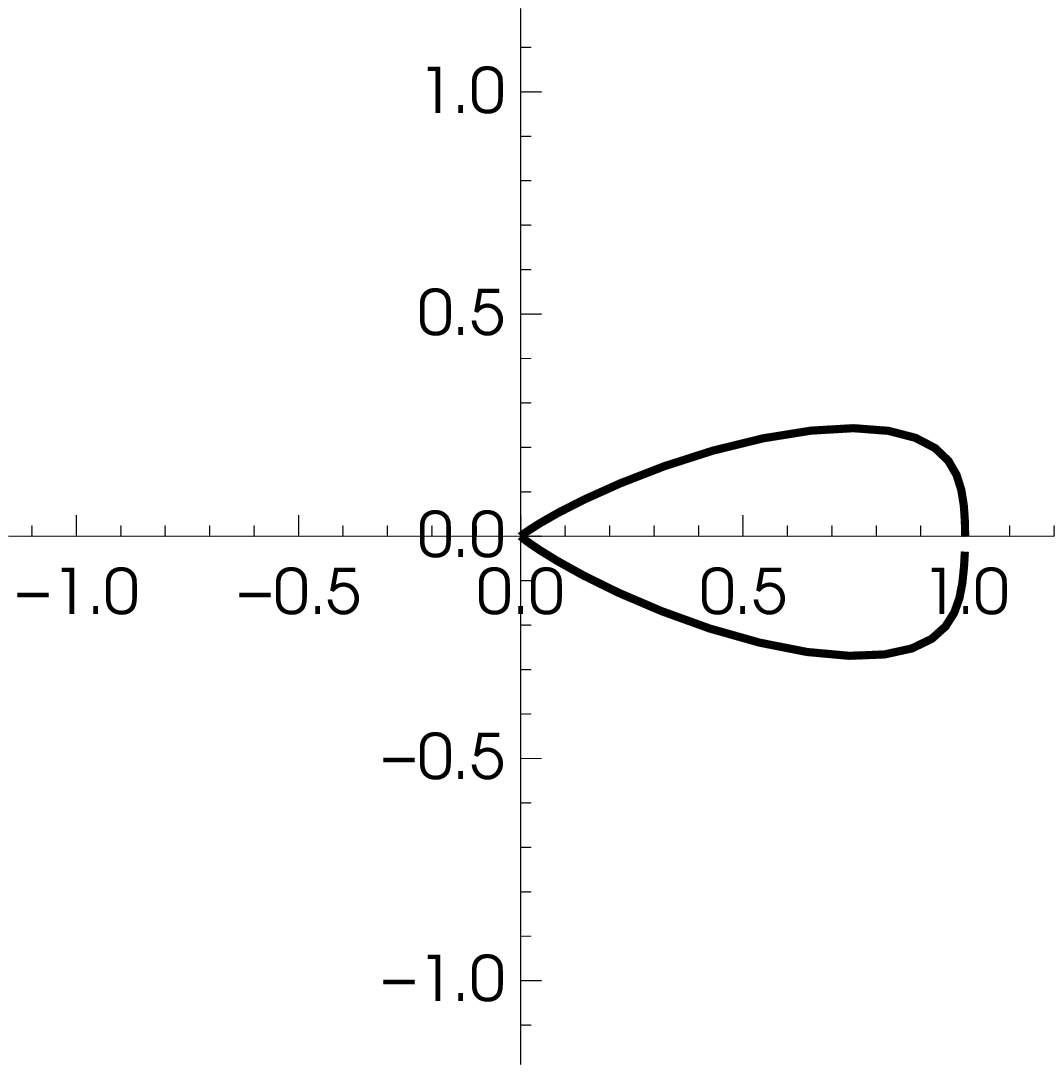}
\caption{Polar diagrams of the
assumed electron pitch-angle distribution $F(E,\beta)$. The angle made
with the
x-axis corresponds to the pitch angle $\beta$ of the electron and the
radial extent corresponds to the magnitude of the distribution. The
electron flux with $\mu =1$ is normalised to 1. Left: isotropic case
$\Delta\mu = 10$.
Middle: intermediate anisotropic case $\Delta\mu = 0.4$. Right: Highly
beamed case $\Delta\mu = 0.1$. }
\label{fig:pole}
\end{figure}
\begin{figure}[htbp]
\centering
\includegraphics[width=39mm,height=39mm]{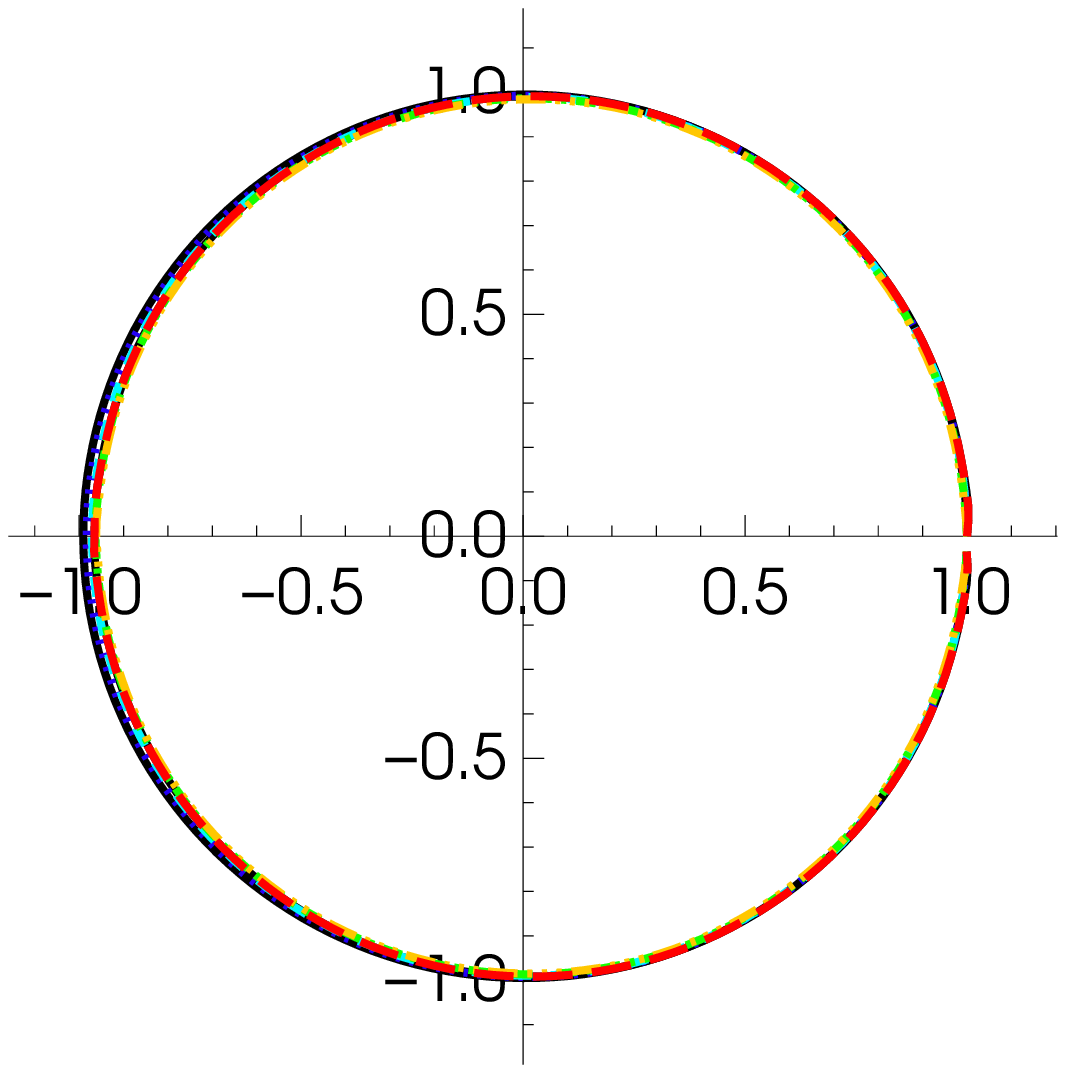}
\includegraphics[width=39mm,height=39mm]{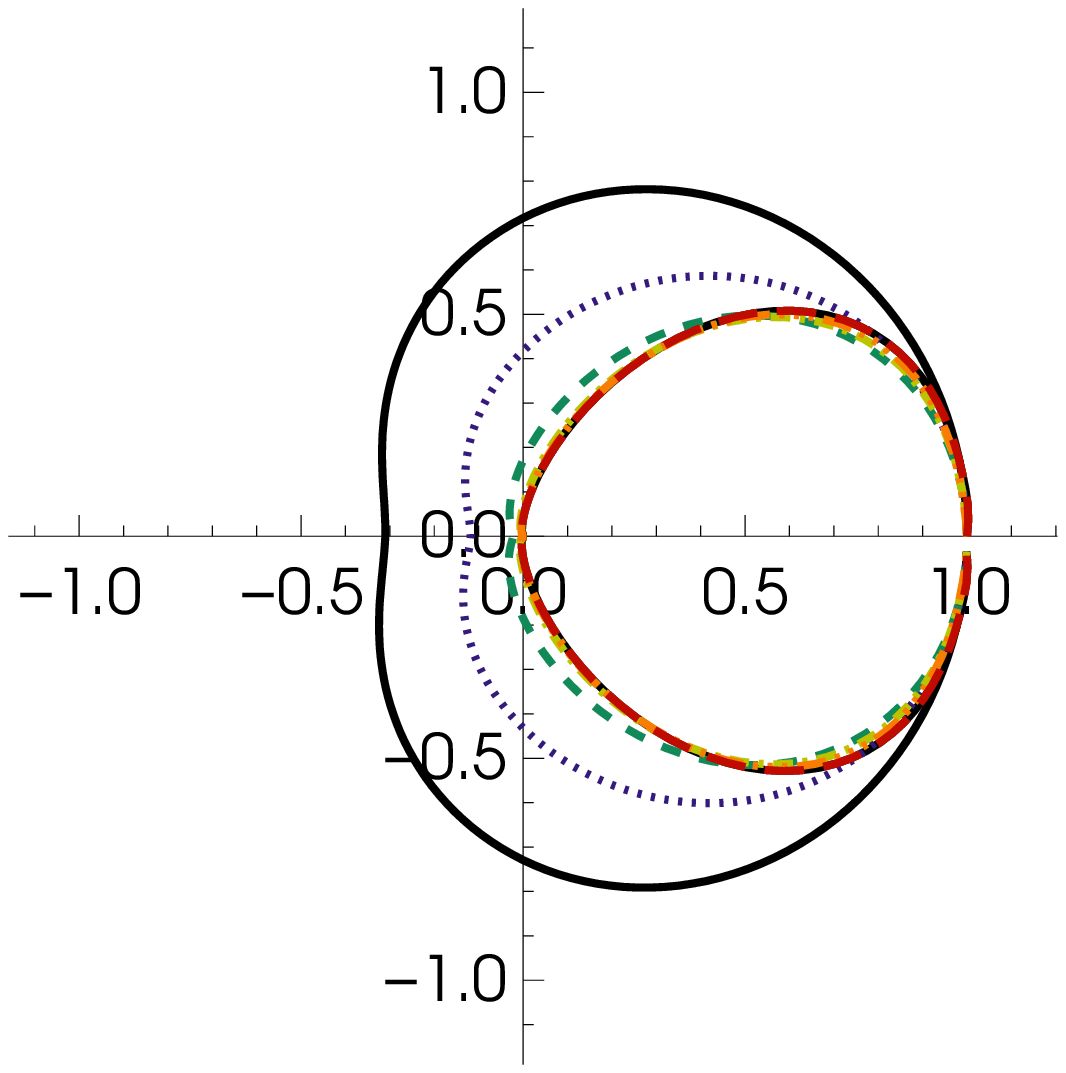}
\includegraphics[width=39mm,height=39mm]{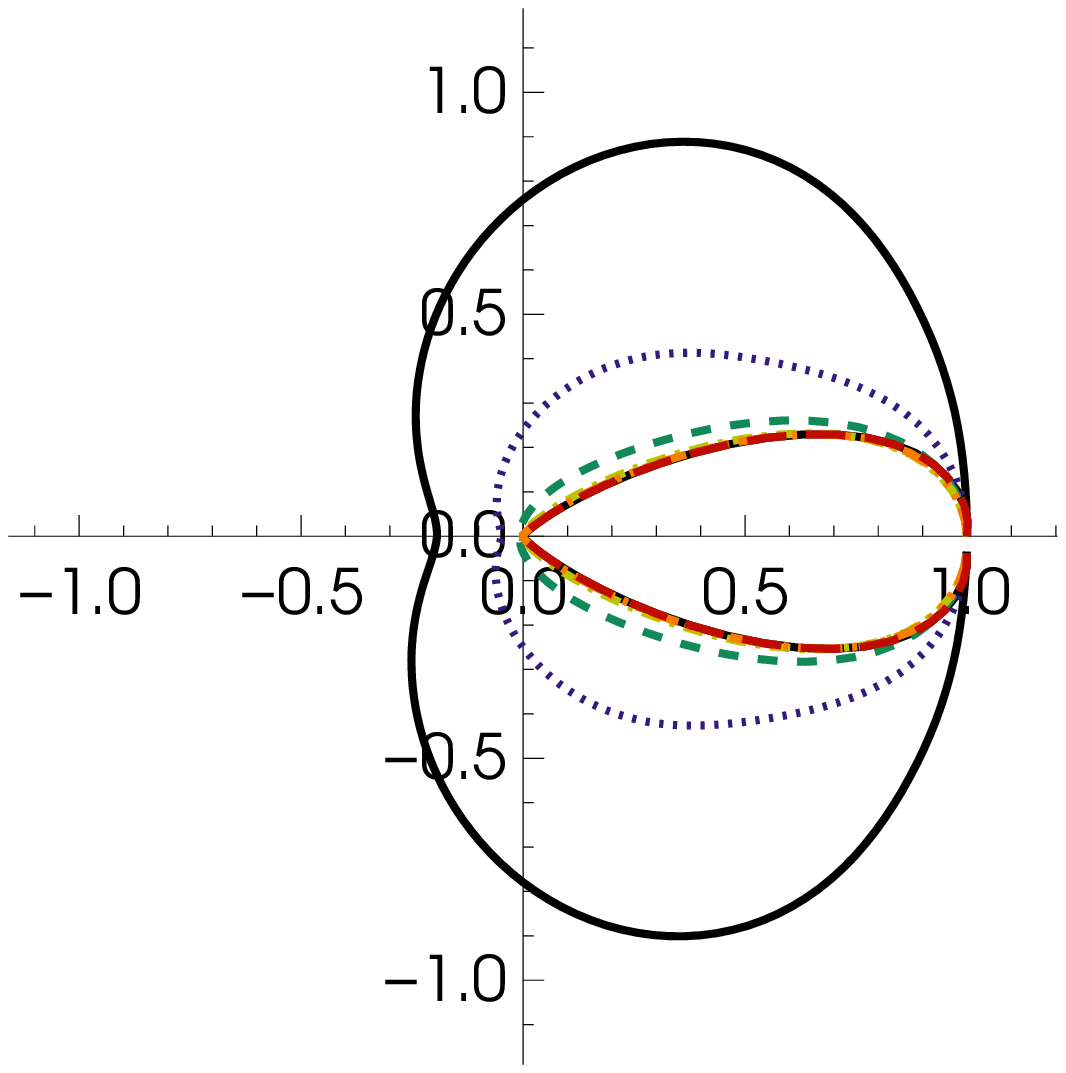}
\caption{Polar diagrams of the emitted photon distribution
$I(\epsilon,\theta_0)$. The angle
made with the x-axis corresponds to the angle $\theta_0$ and the radial
extent corresponds to the magnitude of the photon distribution. The
energy distribution is plotted for several energies: 10~keV (solid black),
40~keV (dotted purple), 150~keV (dashed green), 600~keV
(dashed yellow) and 5 MeV (solid red). The photon flux with $\cos
\theta_0 = 1$ is normalised to 1. Left:
isotropic case Middle: intermediate anisotropic case. Right: Highly
beamed case.}
\label{fig:polp}
\end{figure}
\begin{figure}[htbp]
\centering
\includegraphics[width=39mm,height=39mm]{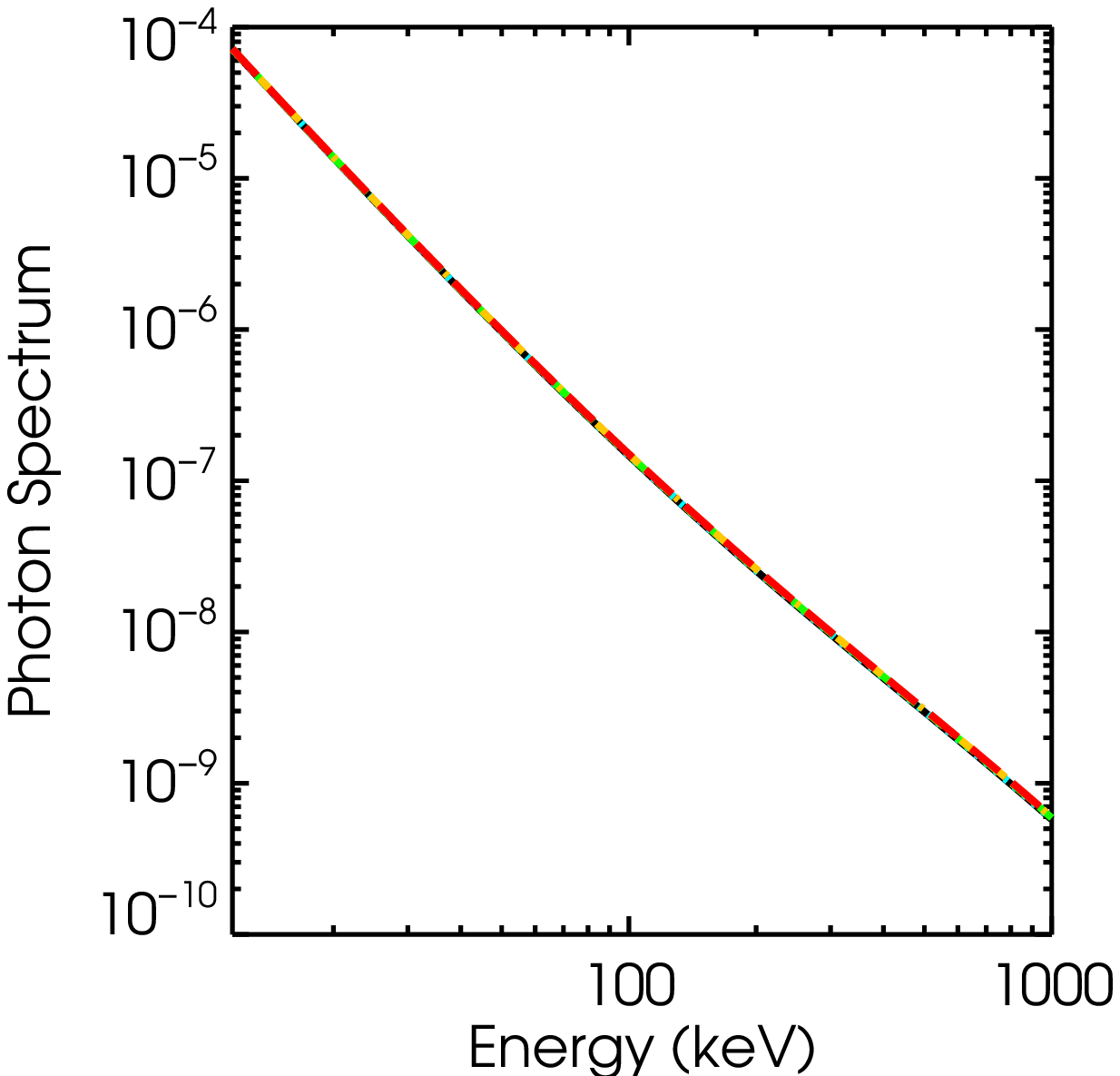}
\includegraphics[width=39mm,height=39mm]{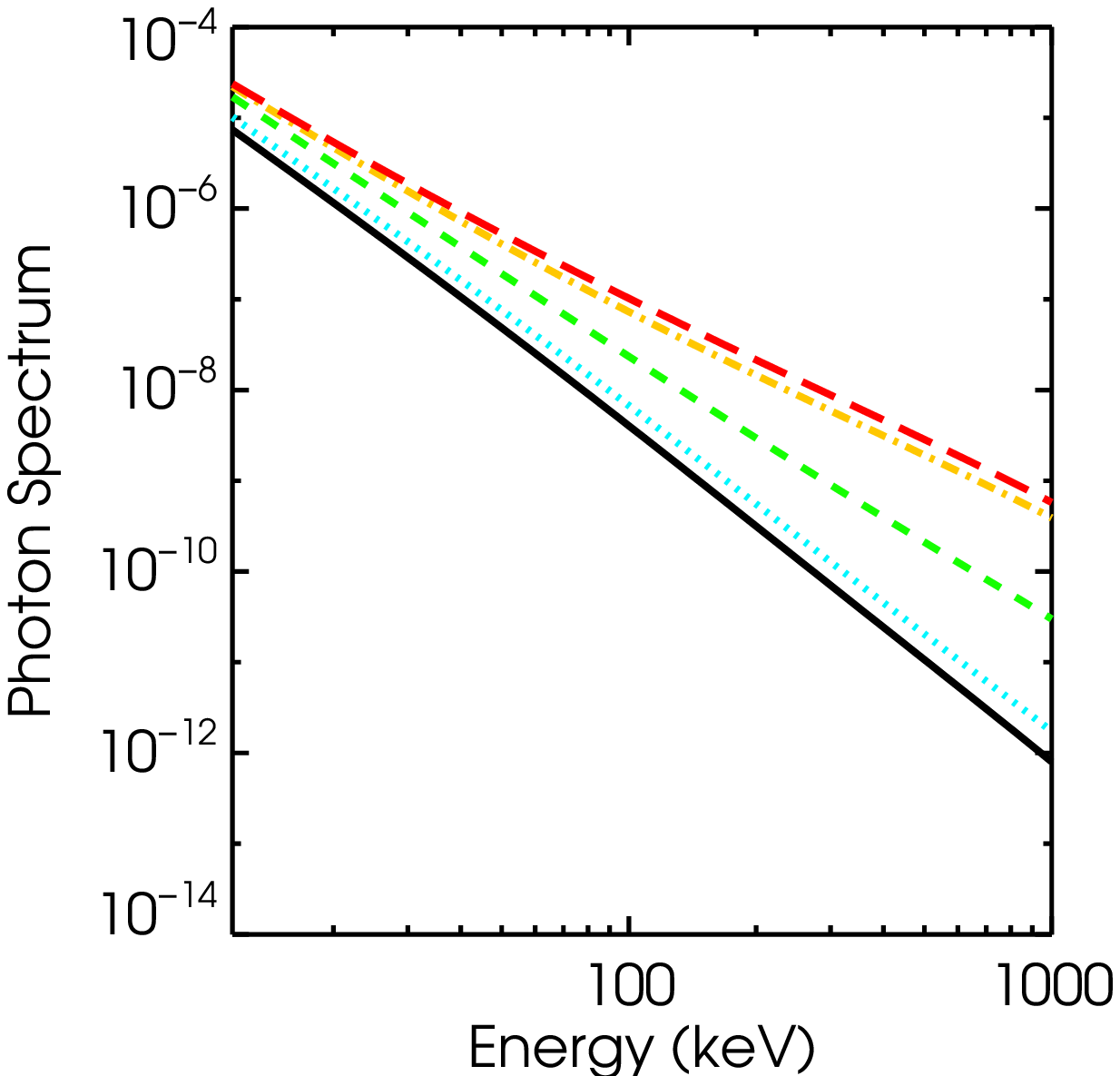}
\includegraphics[width=39mm,height=39mm]{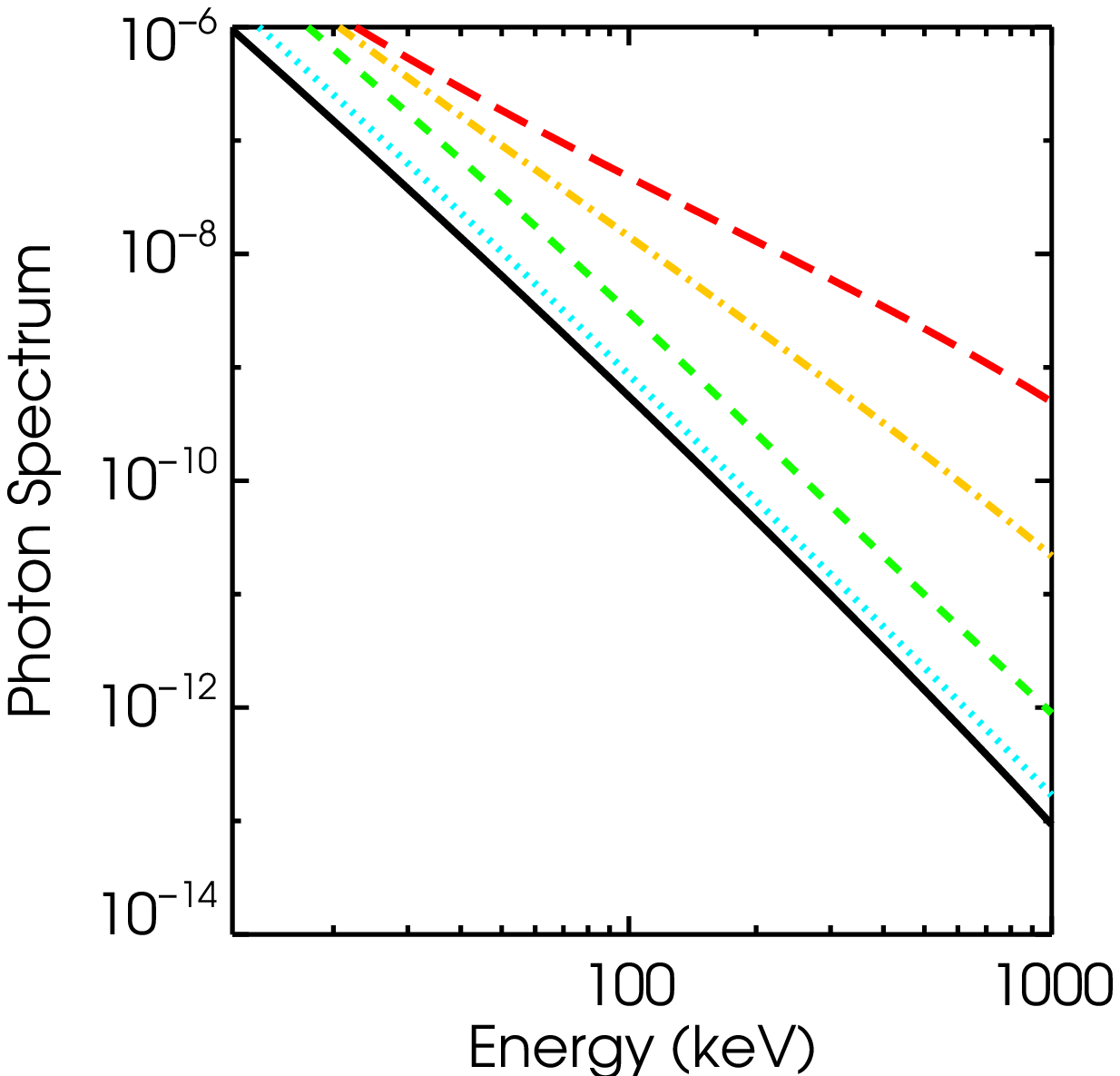}
\caption{Photon flux for several selected
values of $\theta_0$ - $0^\circ$ (red, long dashed),
$45^\circ$ (yellow, dot-dashed), $90^\circ$ (green, dashed), $135^\circ$ (blue, dotted) and $180^\circ$ (black, solid). Left: isotropic case Middle: intermediate anisotropic case. Right: Highly beamed case.}
\label{fig:modflux}
\end{figure}
\begin{figure}[htbp]
\centering
\includegraphics[width=39mm,height=39mm]{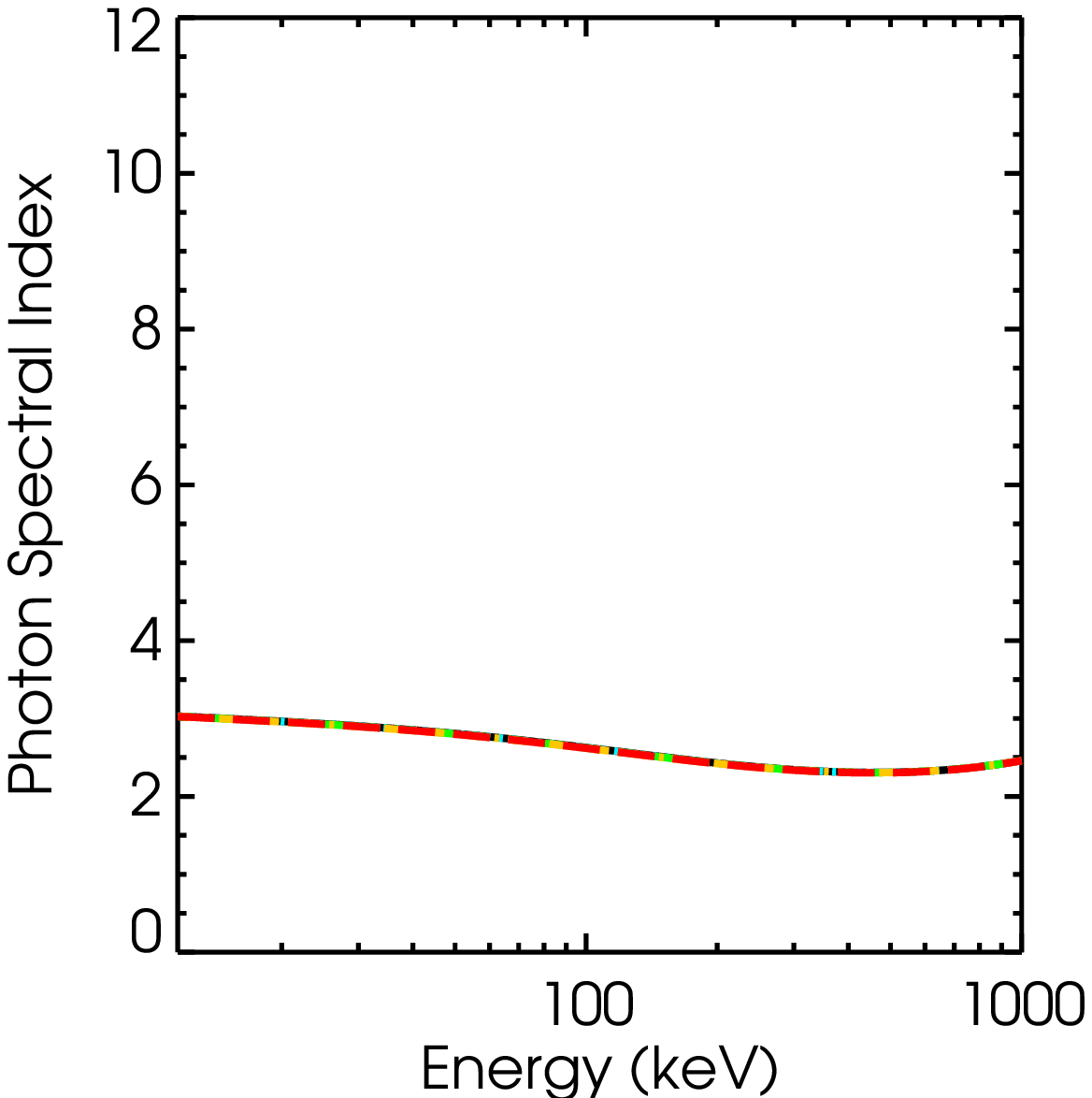}
\includegraphics[width=39mm,height=39mm]{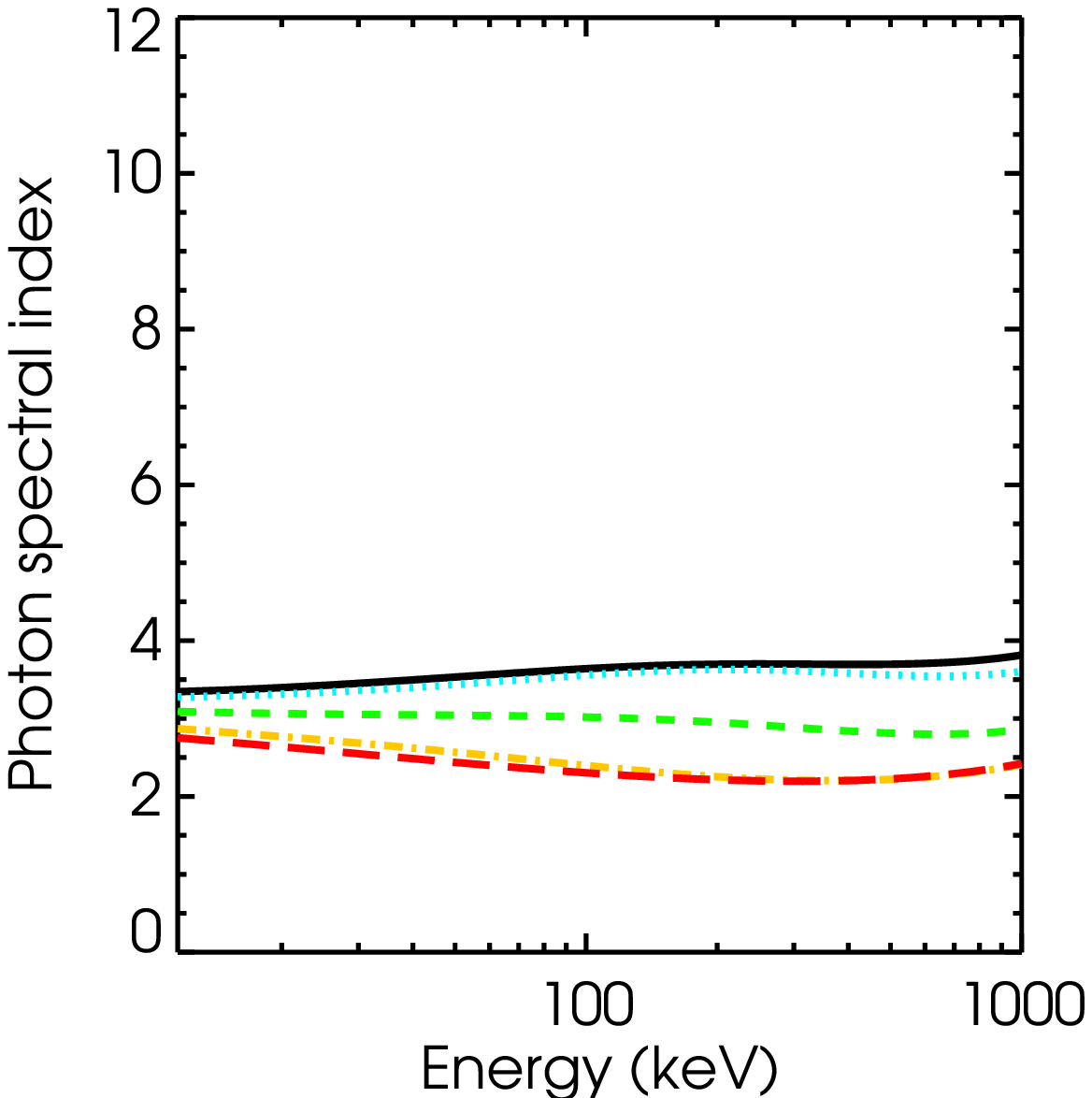}
\includegraphics[width=39mm,height=39mm]{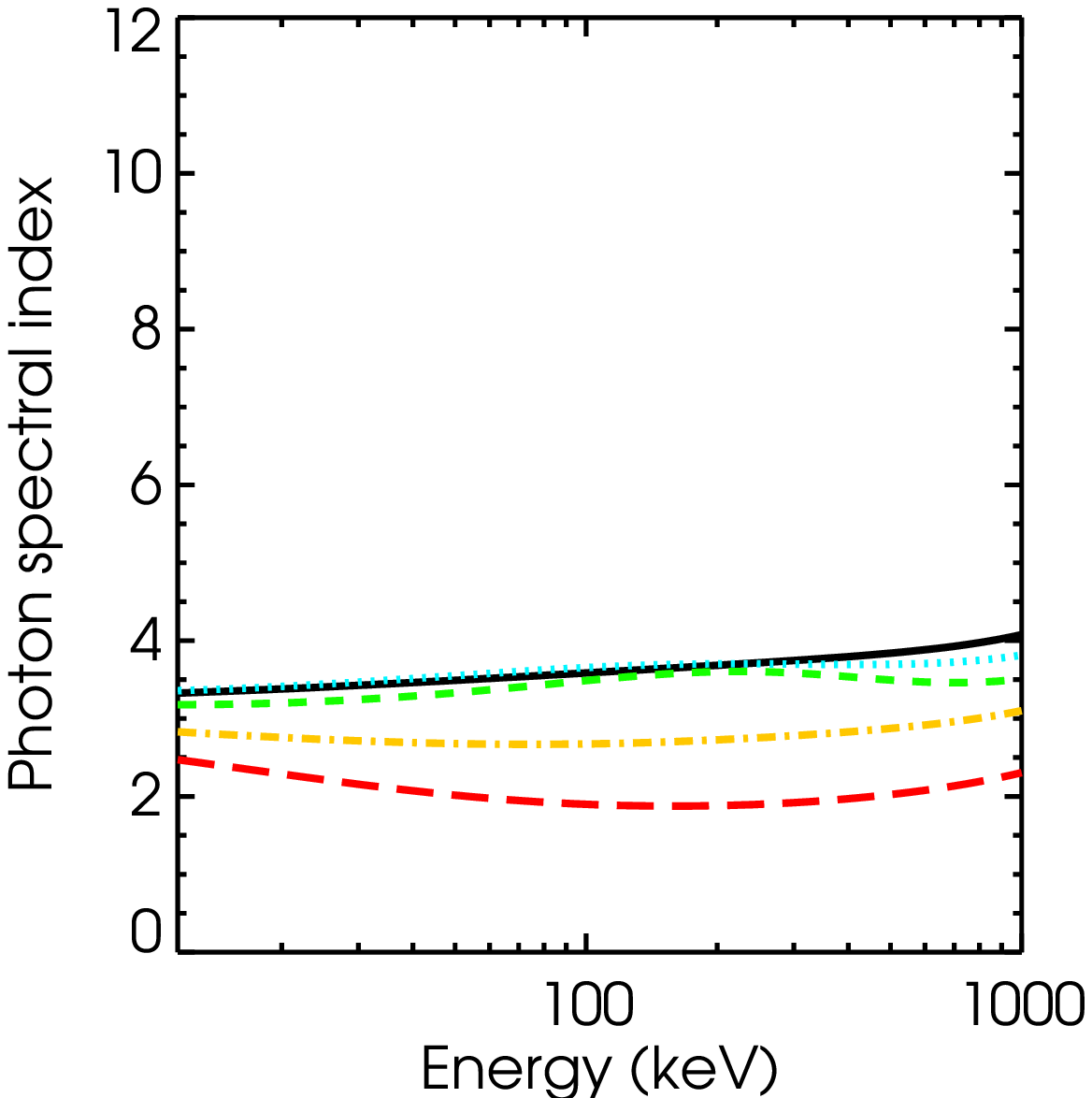}
\caption{Photon spectral index for
several selected values of $\theta_0$ - $0^\circ$ (red, long dashed),
$45^\circ$ (yellow, dot-dashed), $90^\circ$ (green, dashed), $135^\circ$ (blue, dotted) and $180^\circ$ (black, solid).
Left: isotropic case Middle: intermediate anisotropic case. Right:
Highly beamed case.}
\label{fig:modgam}
\end{figure}

This method was performed for the cases of strong ($\Delta\mu = 0.1$)
and intermediate ($\Delta\mu = 0.4$) beaming over an energy range 10 keV
to 5 MeV (Figure 2). After applying this assumed electron spectrum to
the bremsstrahlung cross-section the angular dependant X-ray emission is
found (Figure 3), for lower energies this tends towards being closer to
isotropic than the electron distribution but for high energies it is
reasonably similar to the input electron spectrum. The
emitted flux density (Figure 4) and spectral index (Figure 5) for a
range of angles of observation are then calculated. A Green's matrix
corresponding to the albedo reflection at a range of 
heliocentric angles \cite{2006A&A...446.1157K} was then applied (Figure
6). Emission close to the solar limb shows very little influence from
albedo as expected and the power law in photon energy is
recovered, however emission closer to the disk centre shows a
distinctive hump over the entire energy range due to the albedo
reflection. The influence of albedo can be seen more clearly when
observed photon spectral index is considered (Figure 7), flares close to the disk
centre show a large increase in $\gamma$ above 200 keV, and this is more
pronounced in the strong beaming case. 

\begin{figure}[htb!]
\centering
\includegraphics[width=39mm,height=39mm]{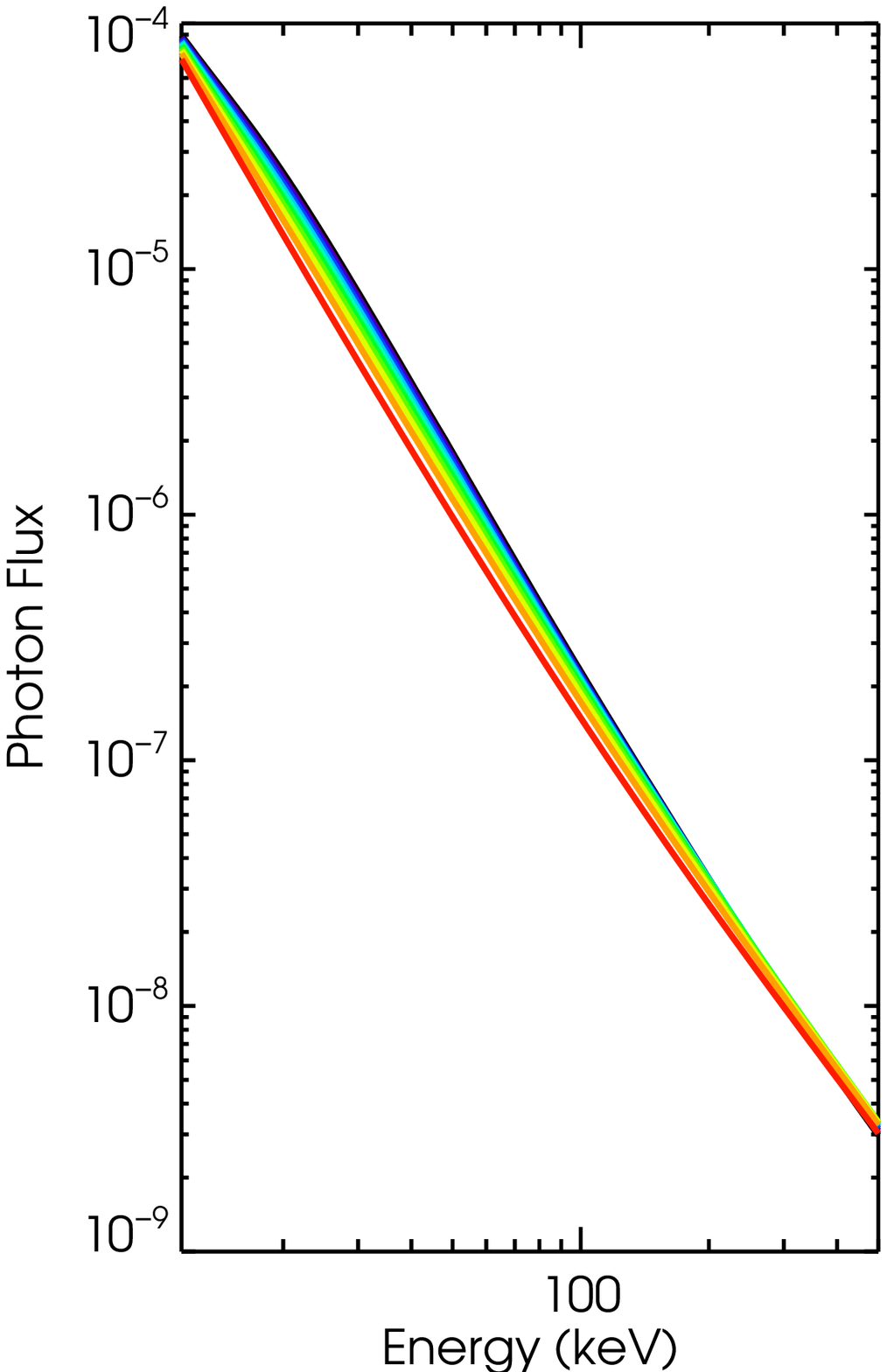}
\includegraphics[width=39mm,height=39mm]{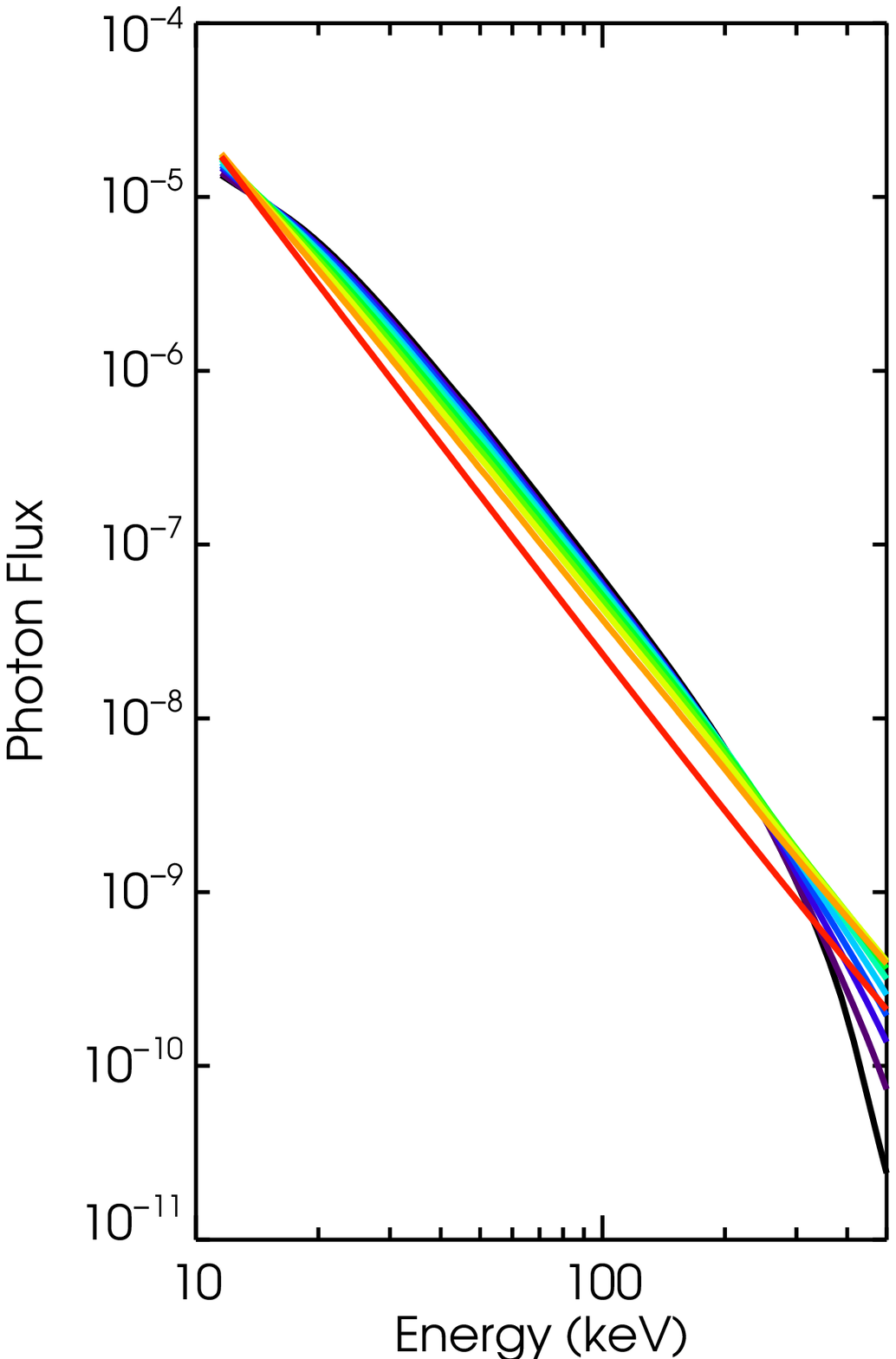}
\includegraphics[width=39mm,height=39mm]{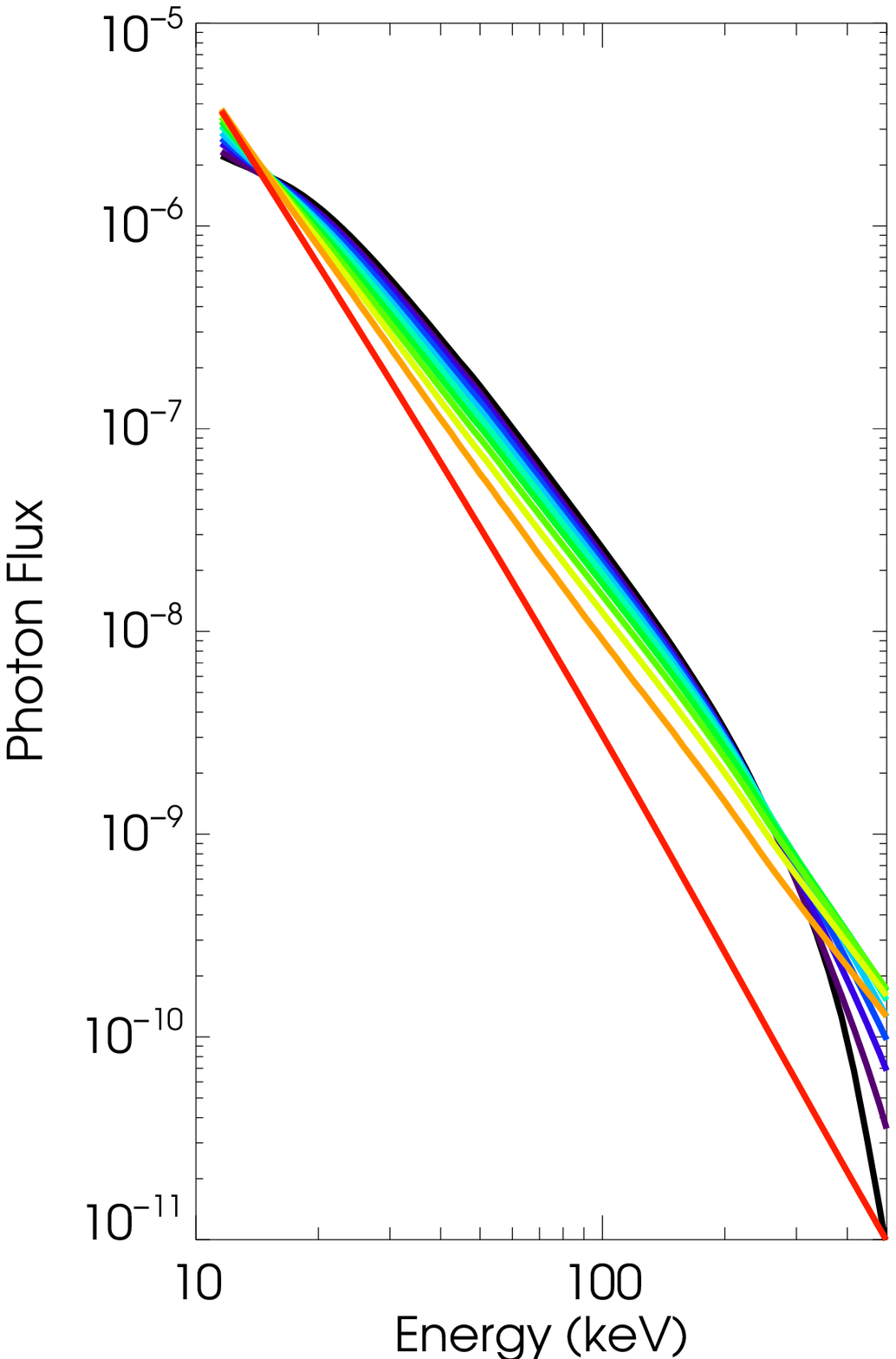}
\caption{Total observed photon flux including reflected albedo
component for flares located at different places
on the disk ranging from disk centre $\cos \theta^{\prime}=1$ (black), $\cos \theta^{\prime}=0.9$ (purple),  $\cos \theta^{\prime}=0.8$ (indigo)  $\cos \theta^{\prime}=1$ to
limb  $\cos \theta^{\prime}=0.2$ (yellow) $\cos \theta^{\prime}=0.1$ (orange) $\cos \theta^{\prime} =0.01$ (red). Left: isotropic case Middle:
intermediate anisotropic case. Right: Highly beamed case.}
\label{fig:modtotf}
\end{figure}

\begin{figure}[htb!]
\centering
\includegraphics[width=39mm,height=39mm]{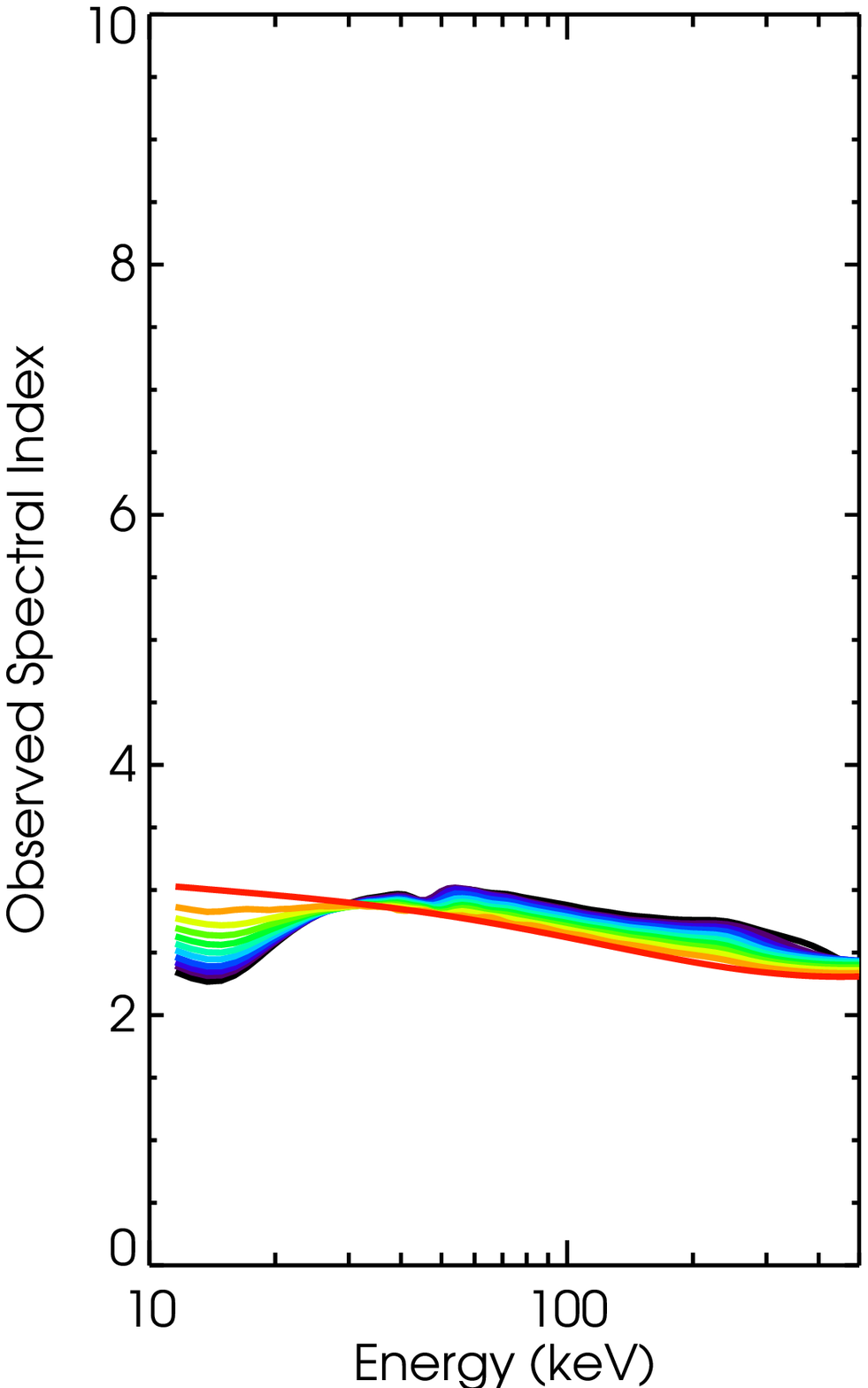}
\includegraphics[width=39mm,height=39mm]{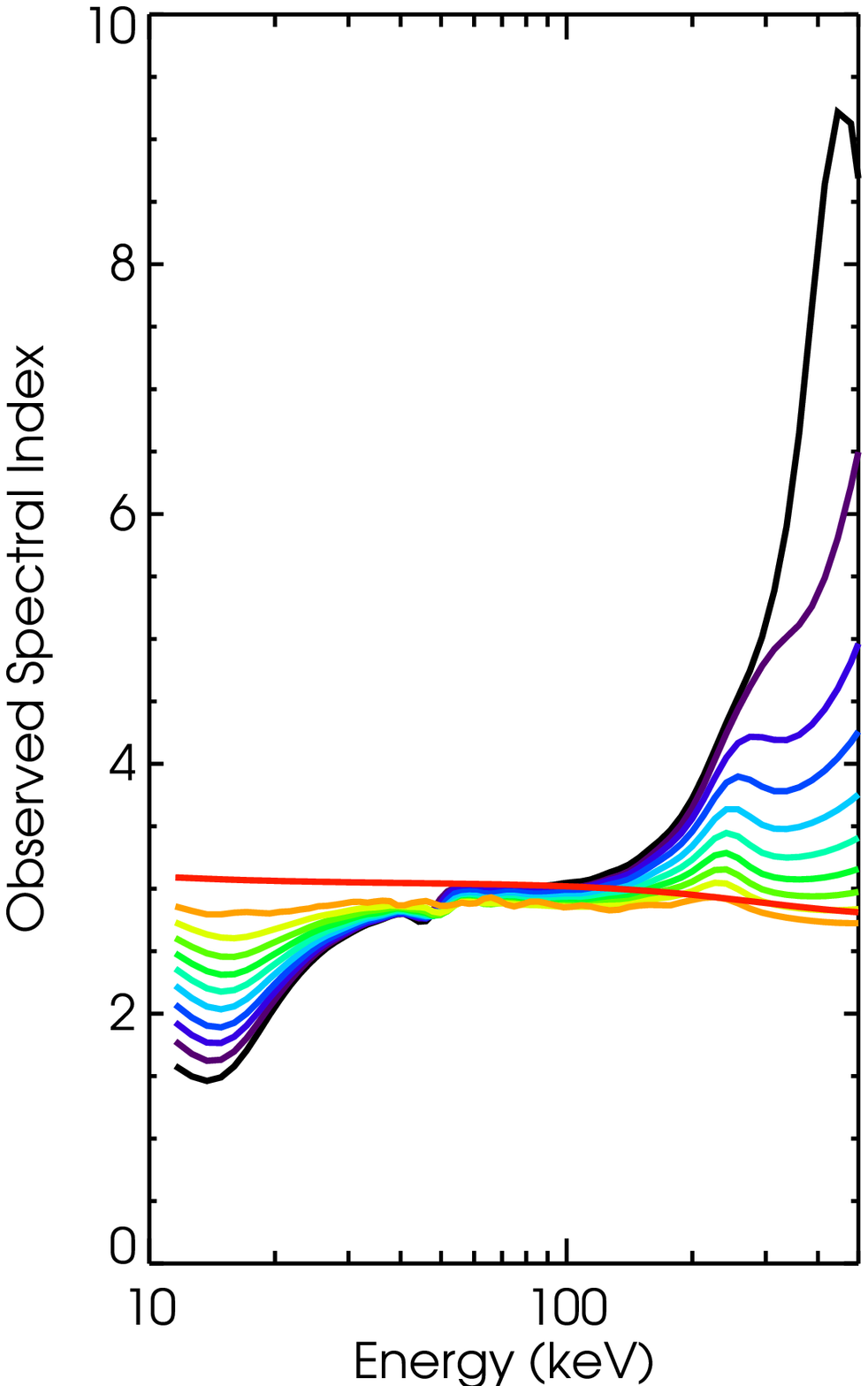}
\includegraphics[width=39mm,height=39mm]{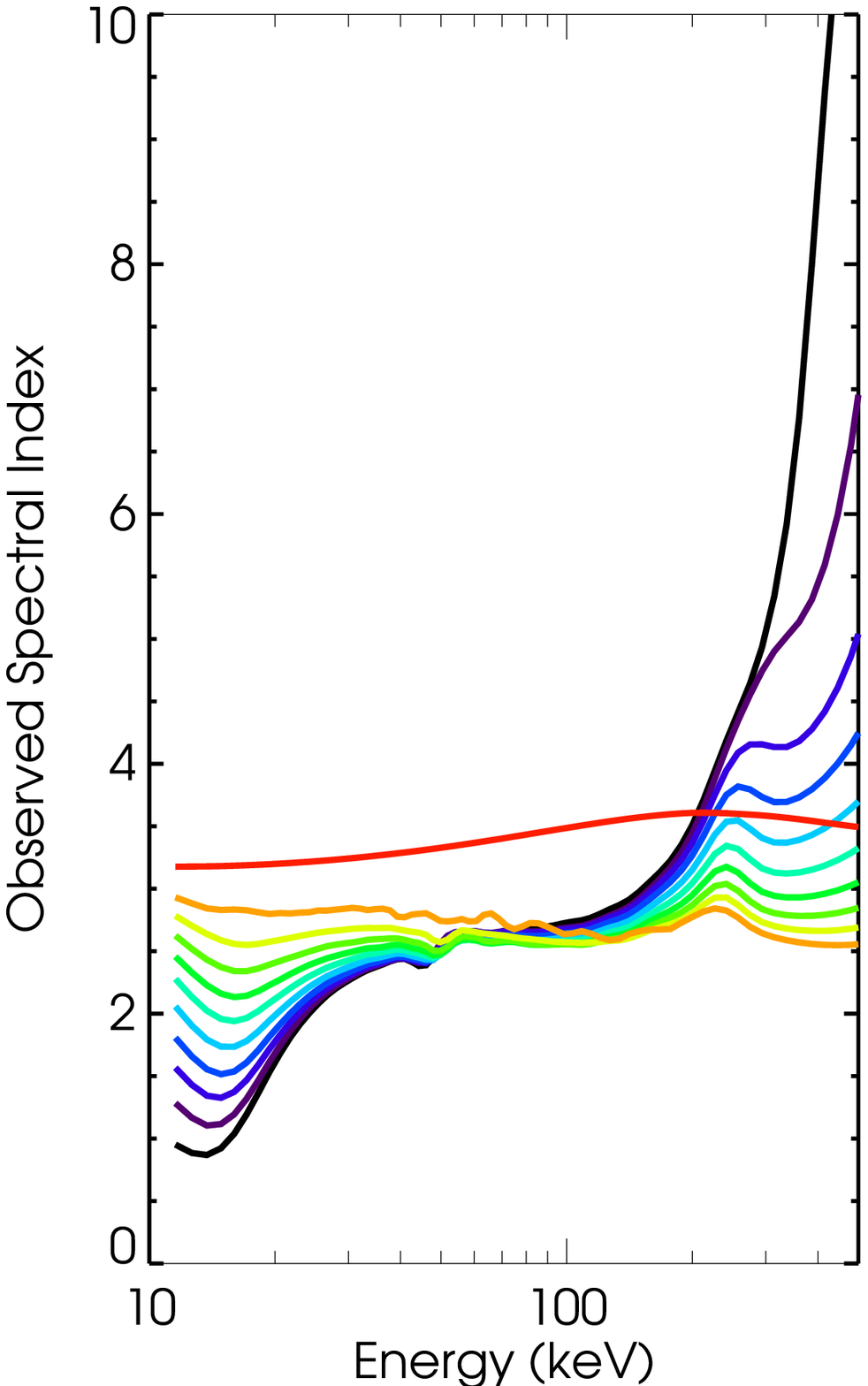}
\caption{Photon spectral index for the total observed
spectrum against photon energy for
flares located at different places on the disk ranging from disk centre
$\cos \theta^{\prime} =1$ (black), $\cos \theta^{\prime}=0.9$ (purple),  $\cos \theta^{\prime}=0.8$ (indigo)  $\cos \theta^{\prime}=1$ to
limb  $\cos \theta^{\prime}=0.2$ (yellow) $\cos \theta^{\prime}=0.1$ (orange) $\cos \theta^{\prime} =0.01$ (red).
Left: isotropic case Middle: intermediate anisotropic case. Right:
Highly beamed case. }
\label{fig:modindex}
\end{figure}

\section{Application to RHESSI data}

The RHESSI data archive was examined for flares with emission 
above $300$~keV with particular attention
paid to flares close to the solar disk centre. These flares are selected
because the forward modelling
suggests that the variation due to beaming is strongest at high energies
and flares closest to the disk centre should have the strongest albedo
reflection and should therefore also show the greatest change due to
beaming. In total eight suitable flares were found (Table 1) which were within $60^{\circ}$ 
of the solar centre (Figure 8) and showed significant \( > 3 \sigma\) counts 
above background  (Figure 9); a number of other flares matching these criteria were
found but they were discounted due to high levels of pulse pileup and
particle contamination.

\begin{table}[htbp!]
\label{tab:flares}
\begin{center}

\caption{Flares suitable for analysis. flare positions are given in arcsecond from the Sun centre.}

\begin{tabular}{ |c| c | c | c | c|c|c|}
\hline
& Flare Date & Start Time (UT) & GOES Class & x position & y
position & $ \mu $ \\
\hline a &20-Aug-2002 & 08:25:21 & M3.4 & 562 & -270 &0.72 \\
\hline b &10-Sep-2002 & 14:52:47 & M2.9 & -622 & -244 &0.72 \\
\hline c &17-Jun-2003 & 22:52:42 & M6.8 & -783 & -148 & 0.52 \\
\hline d &2-Nov-2003 & 17:16:00 & X8.3 & 770 & -343 & 0.51 \\
\hline e &10-Nov-2004 & 02:09:40 & X2.5 & 738 & 116 & 0.69 \\
\hline f &15-Jan-2005 & 22:49:08 & X2.6 & 117 & 325 & 0.93 \\
\hline g &17-Jan-2005 & 09:43:44 & X3.8 & 441 & 301 &0.86 \\
\hline h & 10-Sep-2005 & 21:34:26 & X2.1 & -667 & -255 &0.69 \\
\hline
\end{tabular}
\end{center}
\end{table}

\begin{figure}[htbp!]
\centering
\includegraphics[width=70mm,height=70mm]{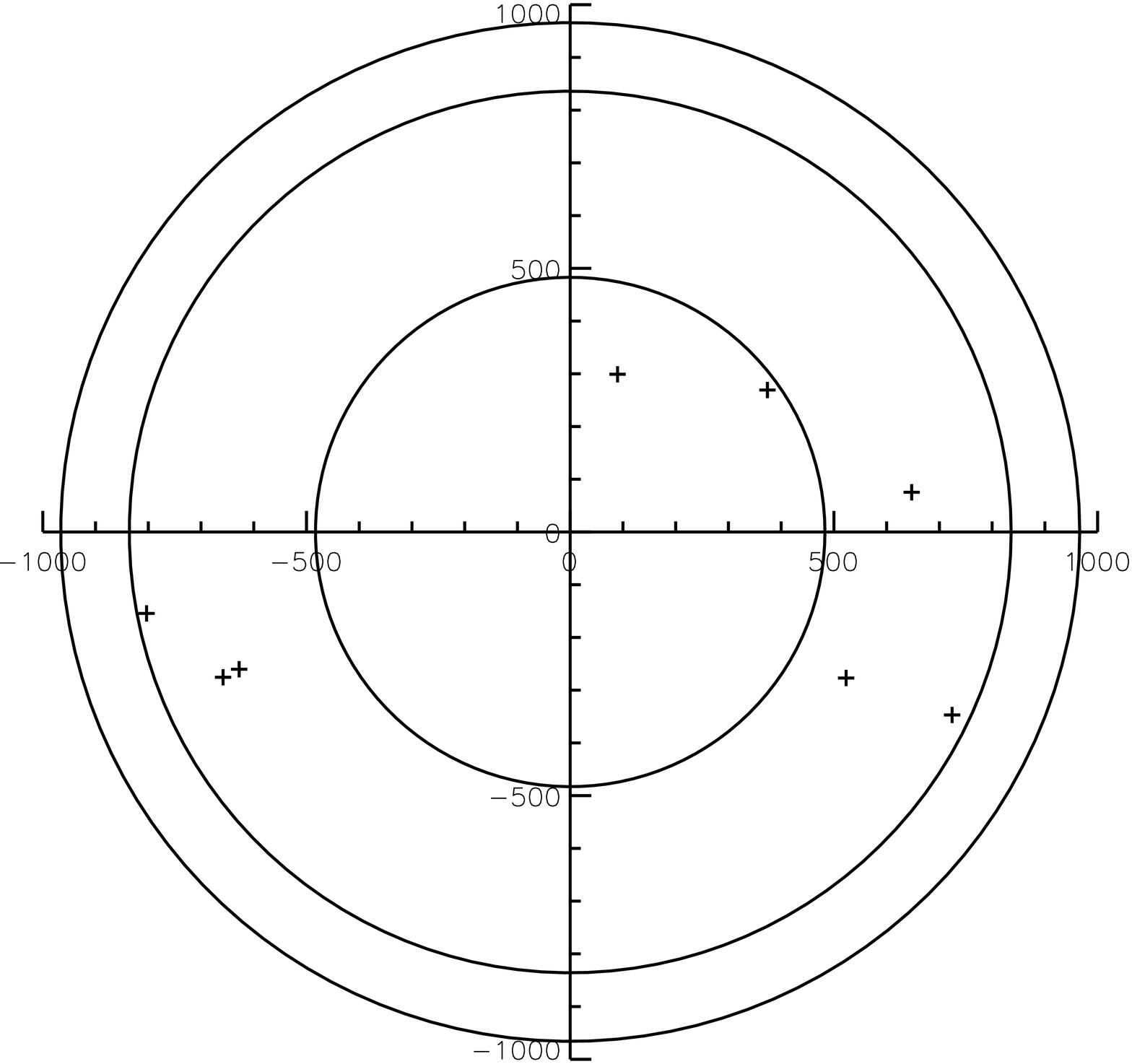}
\caption{Positions of all 8 flares studied on the solar disk. The inner
rings indicate heliocentric angles of $30^\circ$ and $60^\circ$.}
\end{figure}

For each of the flares found the background was removed in the standard
manner \cite{2002SoPh..210..165S}. Counts were accumulated over the
impulsive phase, as the differences in the spectra due to anisotropy in
the electron spectra are greater at higher energies. The time intervals
studied were selected ensuring a high number of high energy counts (Figure 10). A
pseudo-logarithmic binning scheme between 10 keV and 500 keV was used to
initially accumulate the spectra avoiding detectors 2 and 7 due to their
poor resolution \cite{2002SoPh..210...33S}. After background subtraction
had been performed the energy range was further reduced by discarding
the energy bins with counts less than $3\sigma$ above the background.

\section{Inversion method}

The cross-section is similarly divided into two components $Q^{F} $
represents the bremsstrahlung cross-section in the forward direction,
that is the radiation beam\-ed in the same direction as the electron was
travelling and $Q^{B} $ represents the cross-section for the radiation
beamed in the opposite direction to the electron. These matrices are
determined by taking the full angular dependant cross-section and
averaging over a range of angles similar to the method applied to
determine the upward and downward components of the photon flux

\begin{equation}
\displaystyle
Q(\epsilon,E,\theta_0)=\frac{1}{2\pi(1-\cos(\alpha))}
\int^{\alpha}_{\beta=0} Q^{\prime}(\epsilon,E,\beta,\theta_0) \sin\beta
d\beta \;~,
\end{equation}

\noindent with $\alpha= 90^{\circ}$
$Q^B(\epsilon,E)= Q(\epsilon,E,\theta_0=180^{\circ}-\theta')$ and
$Q^F(\epsilon,E)=Q(\epsilon,E,\theta_0=0^{\circ})$

This results in a directly observed count spectrum given by
\begin{equation}
\label{Iobs} I_o(\epsilon) = \frac{{\bar n} V } {4\pi
R^2}\int_\epsilon^\infty(Q^F(\epsilon,E){\overline F}_u(E)\;
+\;Q^B(\epsilon,E){\overline F}_d(E))dE
\end{equation}
and a downward directed photon flux given by
\begin{equation}
I_d(\epsilon) = \frac{{\bar n} V } {4\pi
R^2}\int_\epsilon^\infty(Q^B(\epsilon,E){\overline F}_u(E)\;
+\;Q^F(\epsilon,E){\overline F} _d(E))dE \, .
\end{equation}
A Green's function approach can be used to determine the fraction of
this downward directed photon flux reflected back towards the observer
by albedo \cite{2006A&A...446.1157K}.

The matrix relation between the observed photon spectrum and the
bi-\-direc\-tional electron spectra is now given by

\begin{equation}\label{eq:full}
\mathbf{C = S \left(
Q^{F} + AQ^{B} \qquad Q^{B} + AQ^{F} \right)\left(
{\bar{F_u} \atop \bar{F_d}} \right)} \;~,
\end{equation}
where $\mathbf{A}$ is a discretized matrix representing the Green's
function and $ \mathbf{S}$ represents the photon to count spectral
response of RHESSI.

Equation~(\ref{eq:full}) must be simplified so that the array
containing the cross-section and Green's functions is converted into a
standard two dimensional matrix and the two component electron spectrum
is represented as a one dimensional vector. A method of performing this
transformation is described in \inlinecite{1995ApJ...448L..61H}. The
equation can then be solved using the standard regularised inversion
methods.

A significant problem is determining the electron spectrum which
produced a given photon spectrum. In general there is a linear
relationship between the observed count spectrum represented by vector $
\mathbf{C}$ and the electron distribution which produced it,
$\mathbf{\bar{F}}$, which can be written the form
\begin{equation}\label{eq:mat}
\mathbf{C = MF}
\end{equation}
here $\mathbf{M} $ is a matrix defined by $\mathbf{M = S Q}$ where $
\mathbf{Q}$ is a matrix representing the bremsstrahlung cross-section in
this case $\mathbf{Q = \left(
Q^{F} + AQ^{B} \qquad Q^{B} + AQ^{F} \right)}$
To determine the emitting electron spectrum this must be solved for $
\mathbf{\bar{F}}$, here this is equivalent to the matrix $
\mathbf{\left({\bar{F_u} \atop \bar{F_d}} \right)}$. However this
problem is ill-posed as noise in the data is amplified making a direct
inversion impossible \cite{0266-5611-4-3-004}. A technique for solving
this is regularized inversion \cite{1963sovmat...4..1035}.

As Equation~(\ref{eq:mat}) does not have a unique solution additional
constraints must be imposed by considering the physical properties of
the electron distribution.The solution can then found by solving the
Lagrange multipliers problem
\begin{equation}\label{eq:tik}
||\mathbf{M \bar{F}-C}||^2+\lambda||\mathbf{L \bar{F}}||^2=
\mathnormal{minimum},
\end{equation}
this method is known as Tikhonov regularisation. Here $\mathbf L $
represents the additional constraint, in this case an approximation of
the differential operator is used and $\lambda$ is the regularisation
parameter, a variable which determines the degree of smoothness imposed
on the solution. Equation~(\ref{eq:tik}) can be solved by Generalized
Singular Value Decomposition for arbitrary $\lambda$ and $\mathbf L$
\cite{2004SoPh..225..293K}.

The choice of the regularisation parameter here is determined by
examining the normalised residuals which are
defined by
$r_k=(({\bf{M}}{\overline{\bf{F}}})_k - {\bf{C}}_k)/\delta C_k$.
Then the best value of $\lambda$ by minimizing
\begin{equation}\label{chi2}
\|({\bf{M}}{\overline{\bf{F}}}_{\lambda} - {\bf{C}})(\delta {\bf
C})^{-1}\|^2 = \alpha.
\end{equation}
The choice of $\mathbf L$ is related to the constraint applied to the
solution and in this case should be related to the underlying physics,
here a finite difference matrix representing the first derivative is used.
dependence
The robustness of the solution can be improved if the equation is first
preconditioned \cite{2004SoPh..225..293K}. A forward fit performed on
dependence the data using a standard model of a thermal component plus a broken
power law \cite{2003ApJ...595L..97H}. This estimated electron spectrum,
$\mathbf{{\overline F}}_{0}$ , is used as a starting point for the regularised
inversion. The inversion is performed on the difference between the
data, $\mathbf{C} $ and the fit, $\mathbf{M{\overline
F}_0}$, This modified data
vector and the cross-section matrix are also scaled by a factor of
$\mathbf{\sqrt{M {\overline F}_0}}$. These transforms both make the solution much
flatter and so less prone to errors.

This method has been applied to the problem of inverting the angle
averaged electron spectrum numerous times (e.g.
\opencite{2006ApJ...643..523B},\citeyear{2005SoPh..226..317K}) but can be
extended to determine an angular dependant electron flux. This is done
here by extending the cross-section matrix to take account of the
angular dependence of the bremsstrahlung cross-section and then solving
for an electron flux matrix with two components, one directed down
towards the photosphere, $\bar{F}_d$, and one directed towards the
observer $\bar{F}_u$ \cite{2006ApJ...653L.149K}.

The inversion algorithm is applied to the selected time intervals of the
flare and the upward, ${\bar n}V{\bar F_{u}}$, and downward, ${\bar
n}V{\bar F_{d}}$, electron fluxes calculated. As a first estimate a
thermal plus double power-law fit is performed with ${\overline F}_{u} = {\overline F}_{d}$.

The errors on the electron flux components are calculated by combining
the errors on the count flux and the errors on the background. Random
perturbations are applied to the count flux based on the error and the
electron flux is recalculated. The distribution of these realizations
is then used to estimate the error on the electron flux. The
regularised solution also has finite resolution. The resolution matrix is defined as $R = M_{\lambda}^{-1} M_{true}$,
where ${\overline F}_{true}^{-1} =M_{true}C$ is the true solution to the inverse problem
and ${\overline F}_{\lambda} = M_{\lambda}^{-1} C$ is the regularised solution.
The resolution matrix quantifies the horizontal errors of the solution, so the identity matrix (zero horizontal errors)
correspond to the direct inverse  $M_{\lambda}^{-1}$.  For any practical situations, the regularization 
imposes a spread on the strong peaks centred on the main diagonal, this is an unavoidable occurrence
in any inverse problem. The FWHM of each of the rows of this matrix is taken as the energy resolution for that energy
bin and is considered here as the horizontal error in the electron flux (Figure 11).

The anisotropy was defined to be the ratio of $nV {\overline F}_{d}$ to
$nV {\overline F}_{u}$. Confidence strips for the total anisotropy were calculated
using the same method as the errors in the electron flux (Figure 12). Random
perturbations were applied within the "confidence river" defined by both
the horizontal and vertical error bars.

In order to test the method, we have applied it to simulated electron spectra.
The electron spectra have been assumed to have simple functional forms and equivalent photon spectra
and albedo reflection were added using analytical expressions for electron fluxes.
Random noise, at a similar level to the noise estimated from RHESSI observations, 
has been added to the resulting count spectrum. This simulated spectrum was then inverted 
using the same algorithm as the real data.

Spectra from isotropic initial distributions generally gives a result which is consistent with the input spectrum
within errors and shows a reasonable distribution of residuals. For weak anisotropy the results generally show broader
confidence intervals than for the isotropic case with the same level of noise so that the solution is often
consistent with the weakly anisotropic input spectrum and an isotropic input spectrum. For stronger levels of anisotropy (${\overline F}_{d}/{\overline F}_{u}>10$ at 100 keV),
the method tends to give unphysical negative values for the electron flux. This can be avoided
by increasing the regularisation parameter to force the solution to be smoother and ensure the solution is positive everywhere. However this approach leads to under-regularisation and unacceptably large residuals suggesting that the method
cannot converge on a physically meaningful solution which satisfies the data. We emphasise that there is an upper limit
to the size of the maximum anisotropy detectable as it is difficult to constrain an anisotropy
that is greater than the fractional error in the larger component of electron flux.

The tests also confirm that this method works best for flares
with high energy counts close to the disk centre and that the anisotropy cannot be reliably 
inferred for weak or limb events, as was expected from the forward modelling.   All the inversions of RHESSI data 
show physically sound results with reasonable residuals.

\begin{figure} 
\centerline{\hspace*{0.015\textwidth}
\includegraphics[width=0.515\textwidth,clip=]{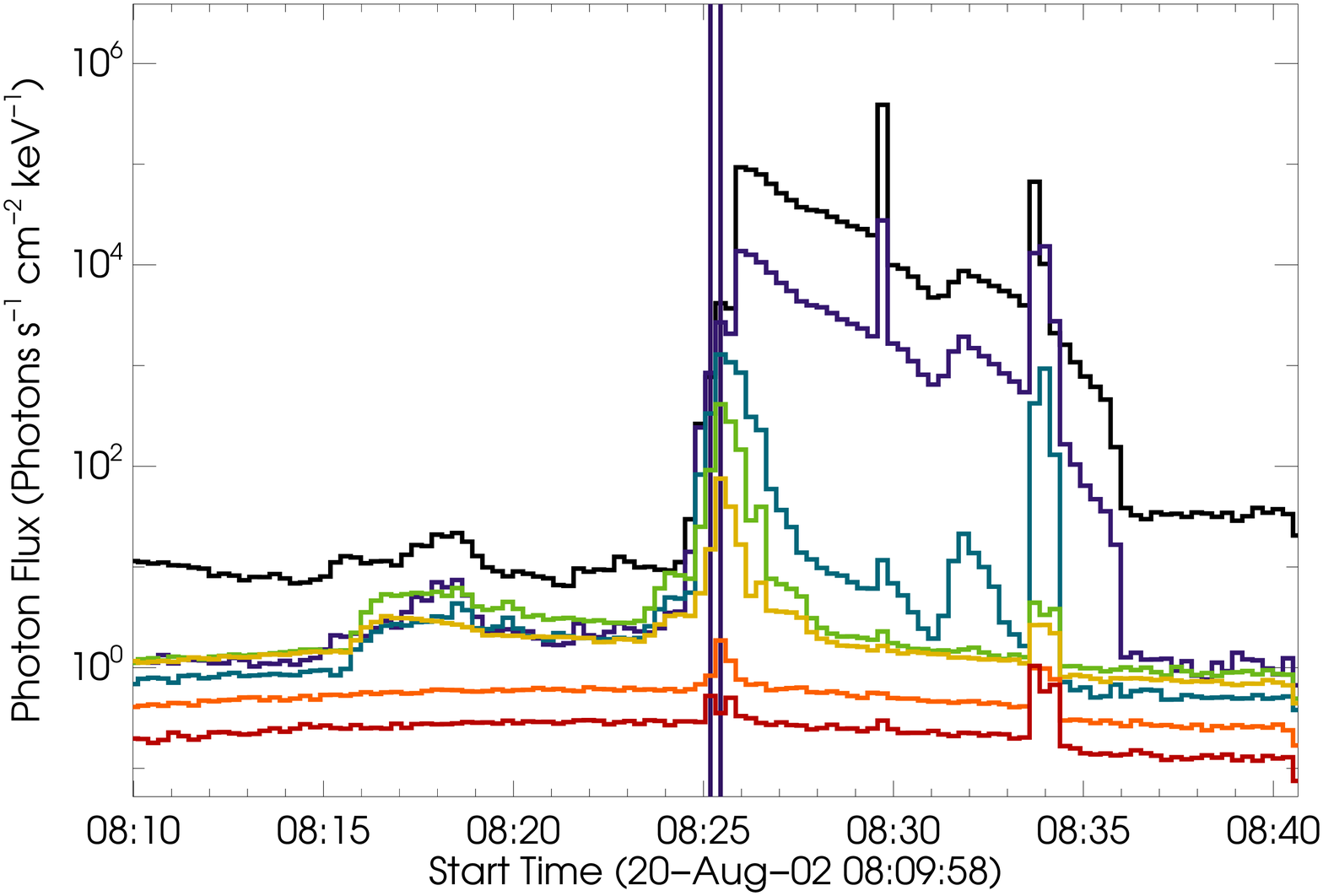}
\hspace*{-0.03\textwidth}
\includegraphics[width=0.515\textwidth,clip=]{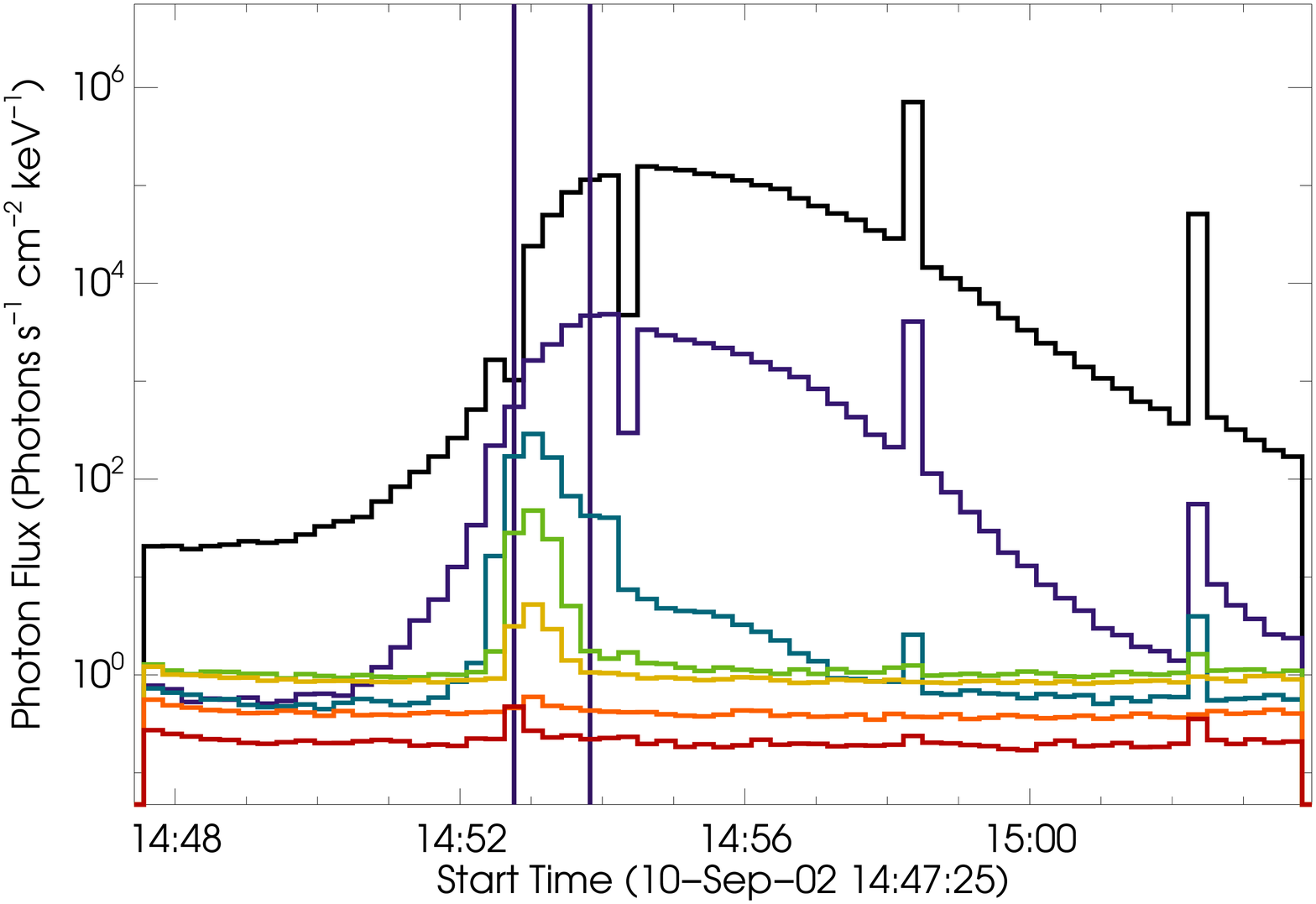}
}
\vspace{-0.32\textwidth} 
\centerline{\Large \bf 
\hspace{0.05 \textwidth} \color{black}{(a)}
\hspace{0.415\textwidth} \color{black}{(b)}
\hfill}
\vspace{0.32\textwidth} 
\centerline{\hspace*{0.015\textwidth}
\includegraphics[width=0.515\textwidth,clip=]{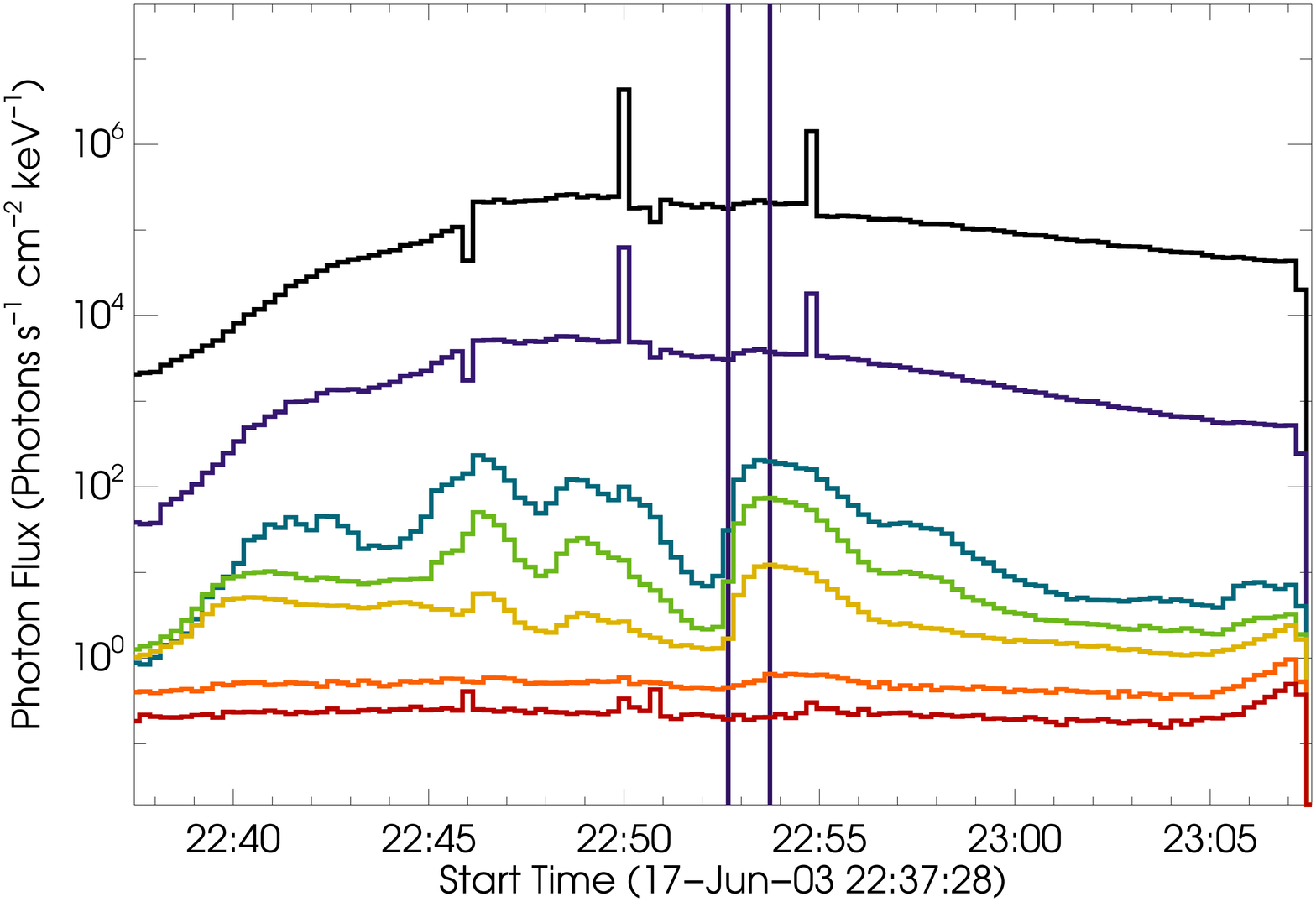}
\hspace*{-0.03\textwidth}
\includegraphics[width=0.515\textwidth,clip=]{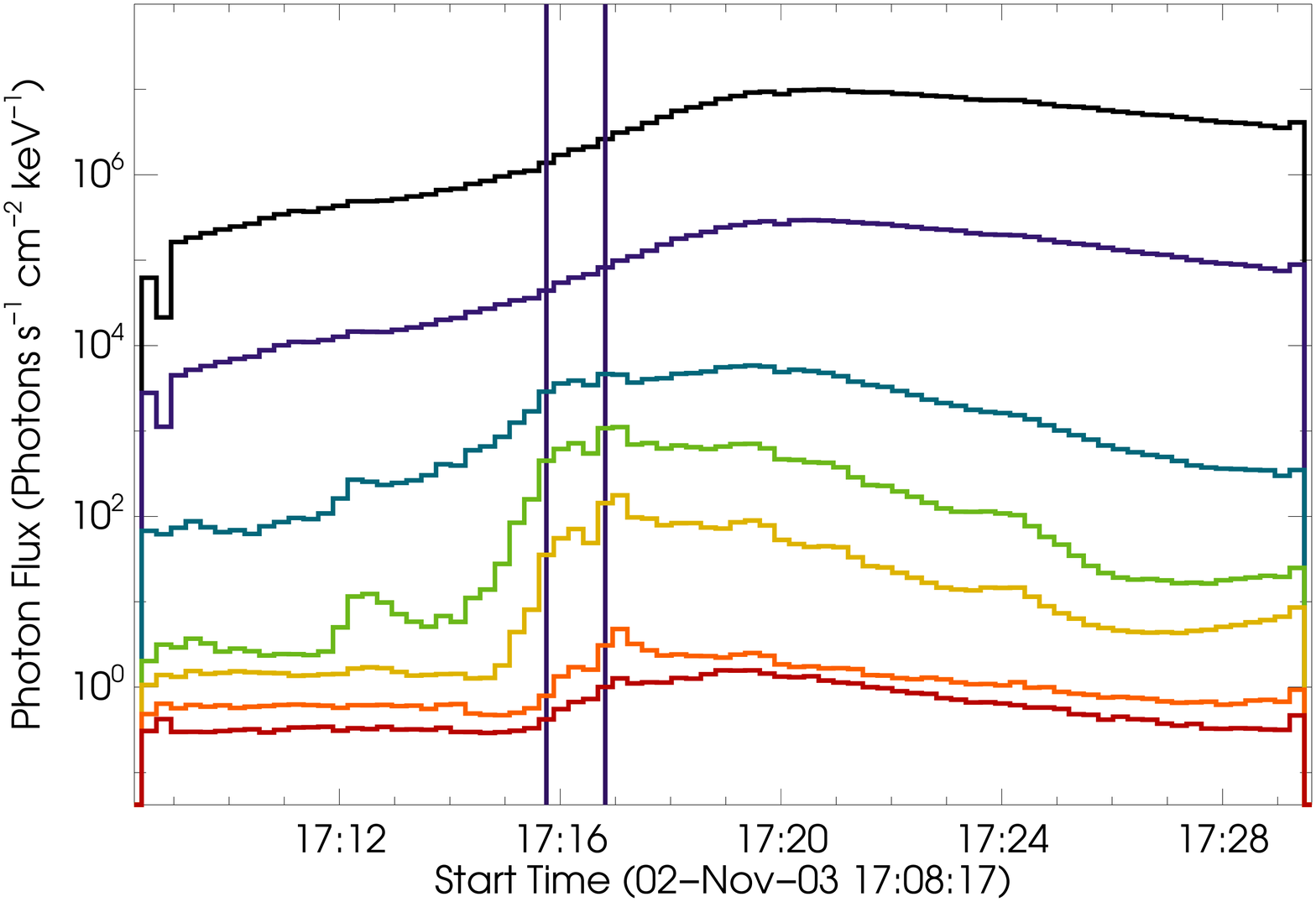}
}
\vspace{-0.32\textwidth} 
\centerline{\Large \bf 
\hspace{0.05 \textwidth} \color{black}{(c)}
\hspace{0.415\textwidth} \color{black}{(d)}
\hfill}
\vspace{0.32\textwidth} 
\centerline{\hspace*{0.015\textwidth}
\includegraphics[width=0.515\textwidth,clip=]{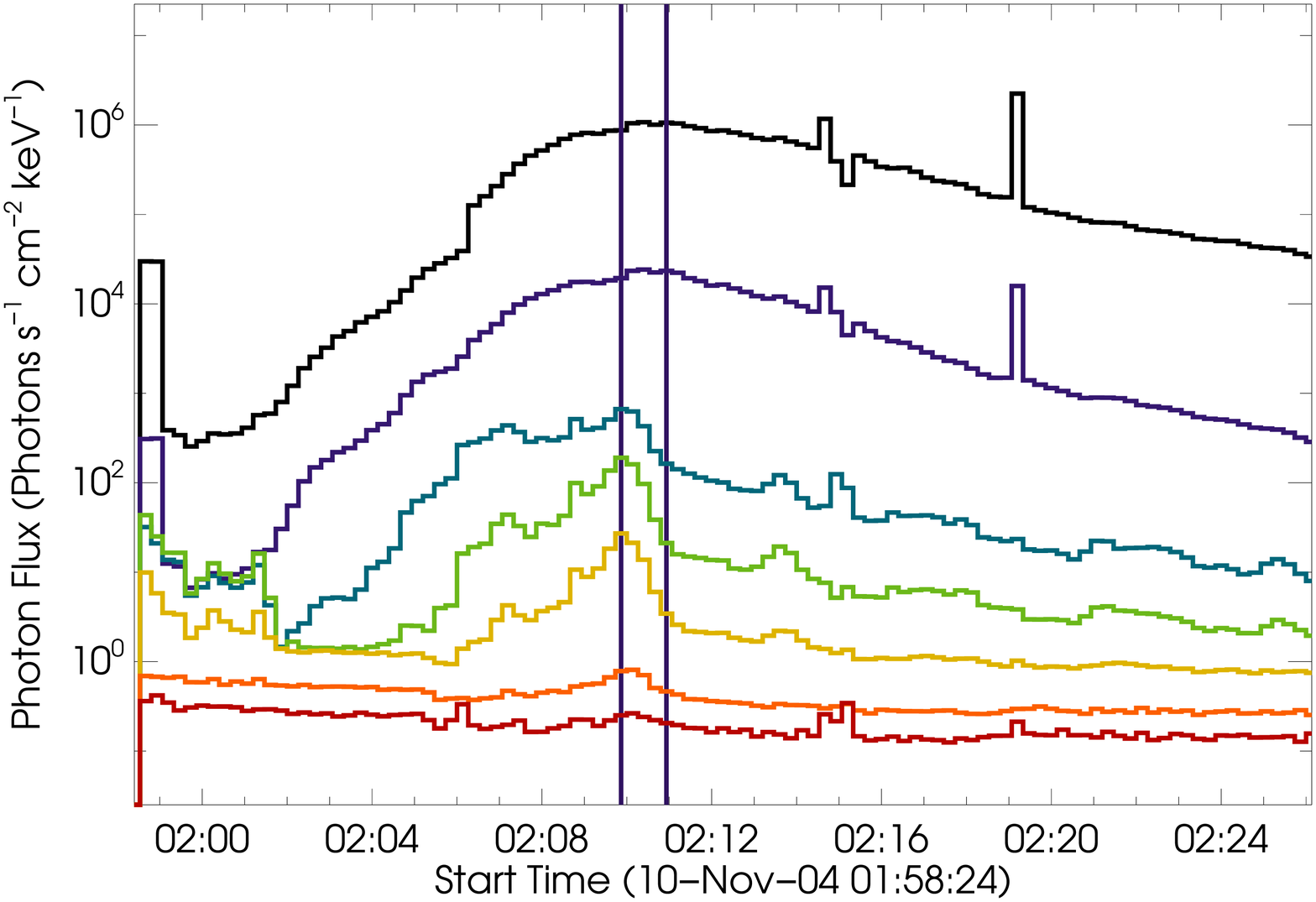}
\hspace*{-0.03\textwidth}
\includegraphics[width=0.515\textwidth,clip=]{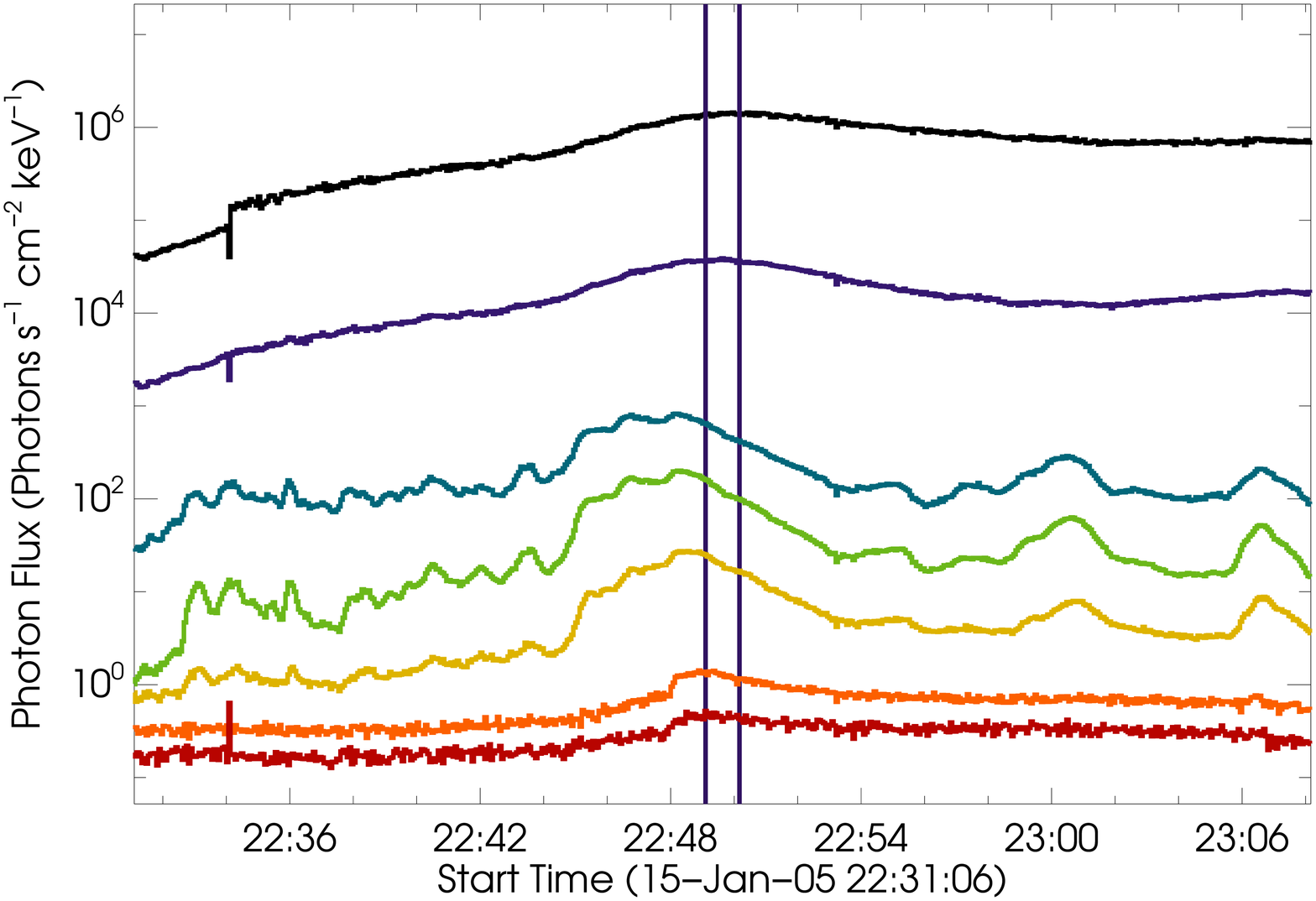}
}
\vspace{-0.32\textwidth} 
\centerline{\Large \bf 
\hspace{0.05 \textwidth} \color{black}{(e)}
\hspace{0.415\textwidth} \color{black}{(f)}
\hfill}
\vspace{0.32\textwidth} 
\centerline{\hspace*{0.015\textwidth}
\includegraphics[width=0.515\textwidth,clip=]{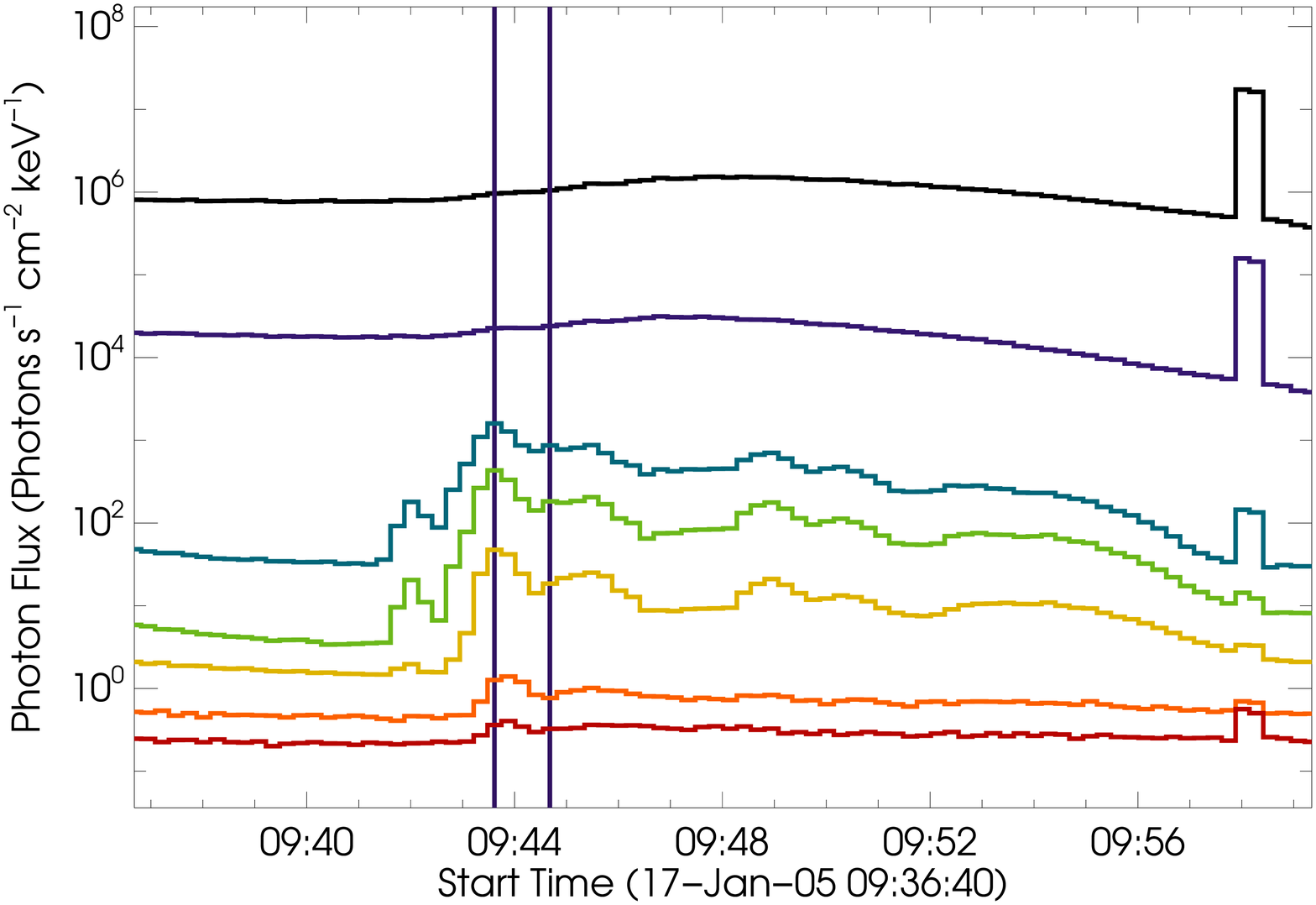}
\hspace*{-0.03\textwidth}
\includegraphics[width=0.515\textwidth,clip=]{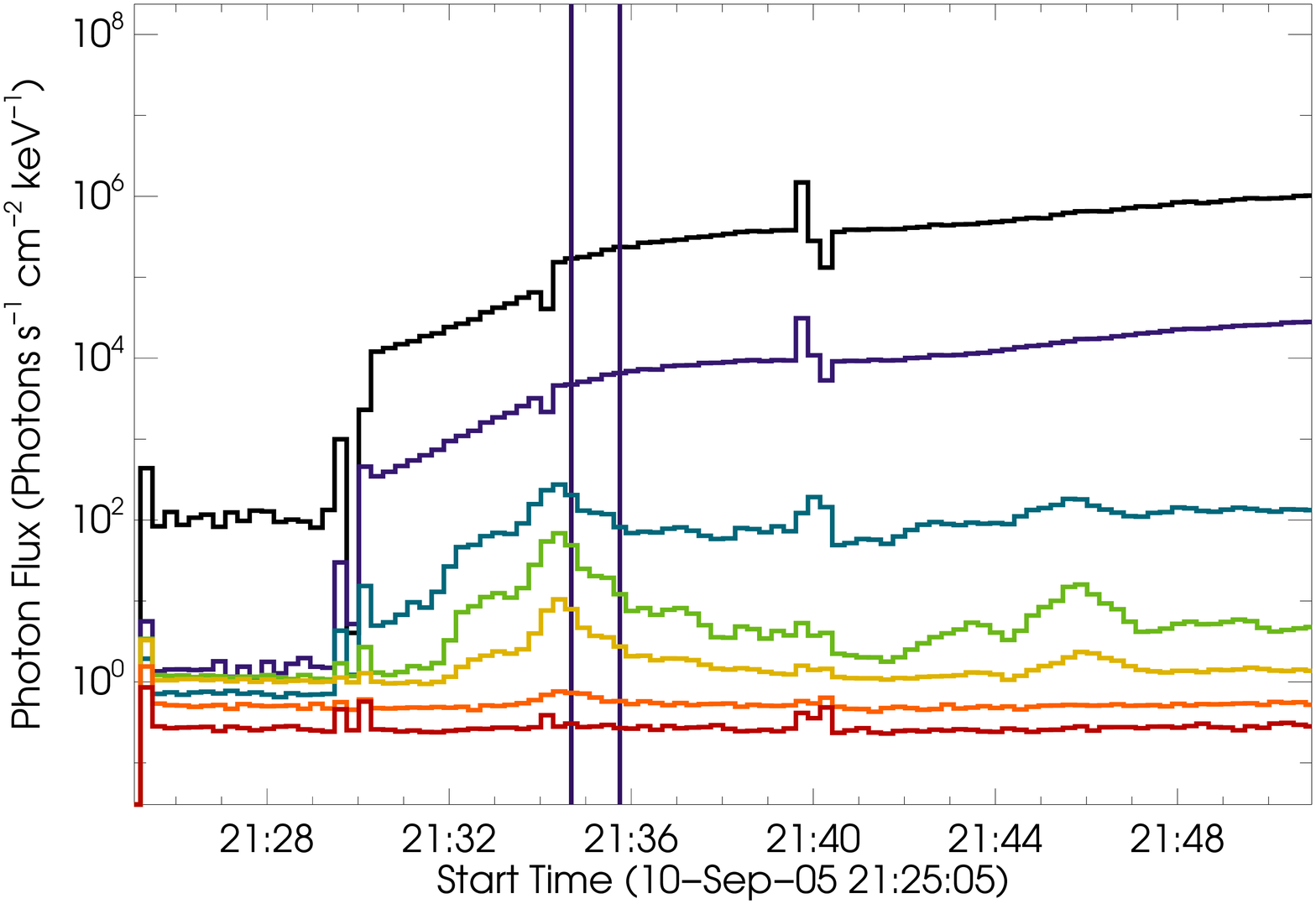}
}
\vspace{-0.32\textwidth} 
\centerline{\Large \bf 
\hspace{0.05 \textwidth} \color{black}{(g)}
\hspace{0.415\textwidth} \color{black}{(h)}
\hfill}
\vspace{0.32\textwidth} 
\caption{RHESSI lightcurves of each of the flares studied in 7 energy
bands- black 7-12 keV, purple 12-25 keV, blue 25-50 keV, green 50-100
keV, yellow 100-300 keV, orange 300-800 keV, red 800-5000 keV. The
vertical lines show the accumulation time interval used. The plots are
semi-calibrated, a diagonal approximation of the RHESSI response is used
to estimate the photon flux from the measured counts. There are still
instrumental artefacts present with the very sharp spikes and dips being
the result of attenuator status changes. All times are in UT.
}
\label{fig:F-panels}
\end{figure}

\begin{figure} 
\centerline{\hspace*{0.015\textwidth}
\includegraphics[width=0.515\textwidth,height=40mm,clip=]{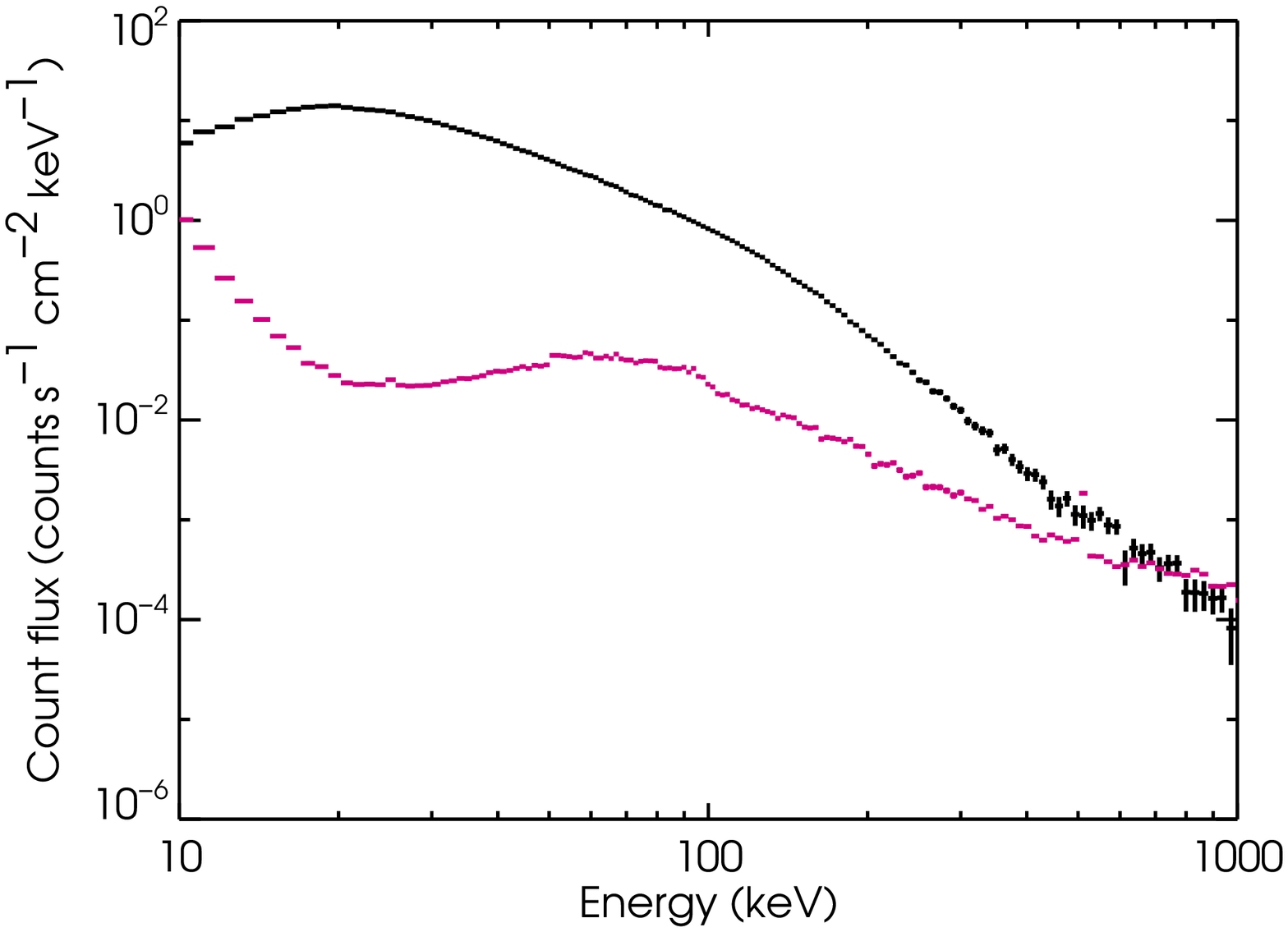}
\hspace*{-0.03\textwidth}
\includegraphics[width=0.515\textwidth,height=40mm,clip=]{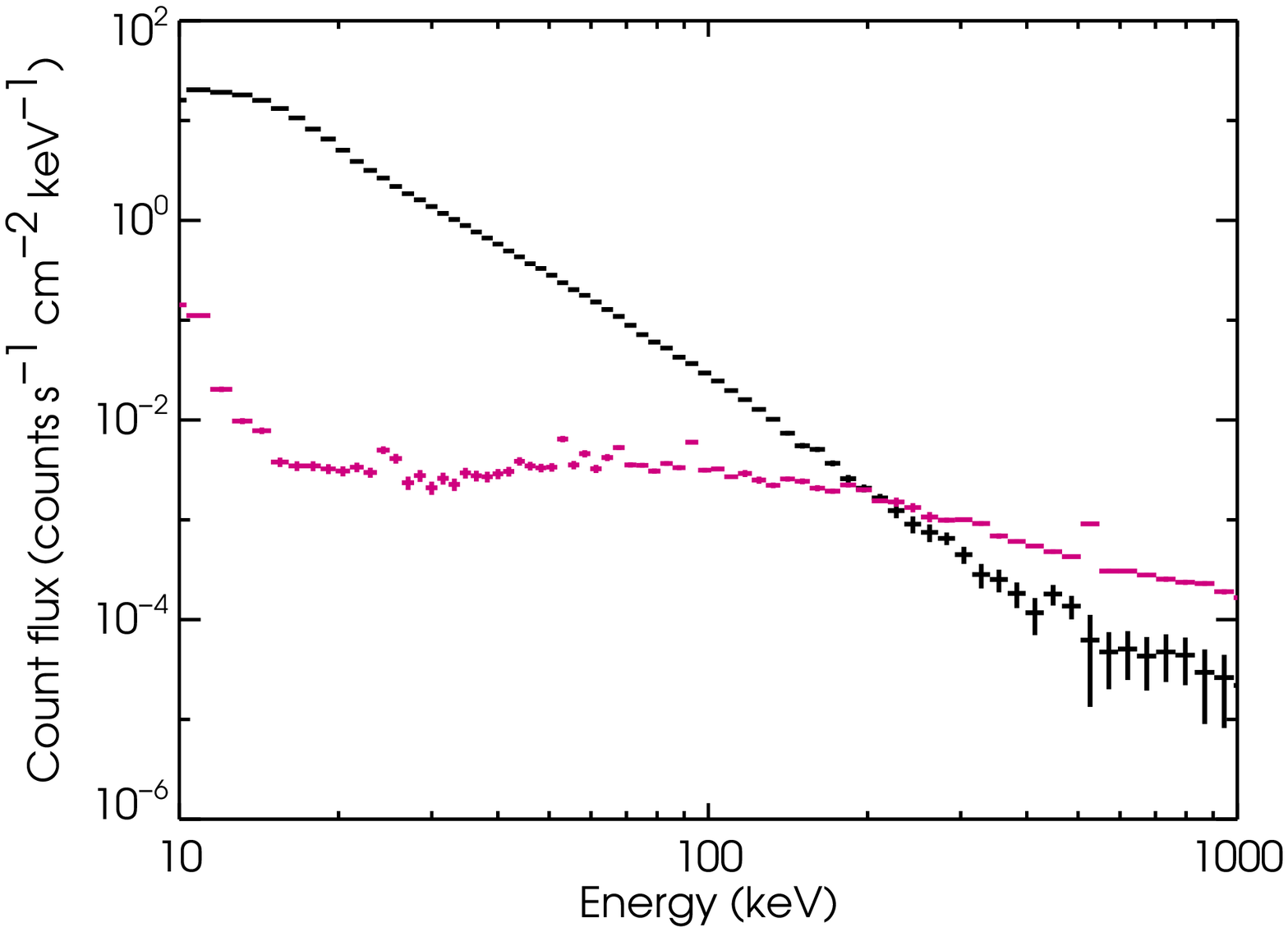}
}
\vspace{-0.30\textwidth} 
\centerline{\Large \bf 
\hspace{0.4 \textwidth} \color{black}{(a)}
\hspace{0.415\textwidth} \color{black}{(b)}
\hfill}
\vspace{0.30\textwidth} 
\centerline{\hspace*{0.015\textwidth}
\includegraphics[width=0.515\textwidth,height=40mm,clip=]{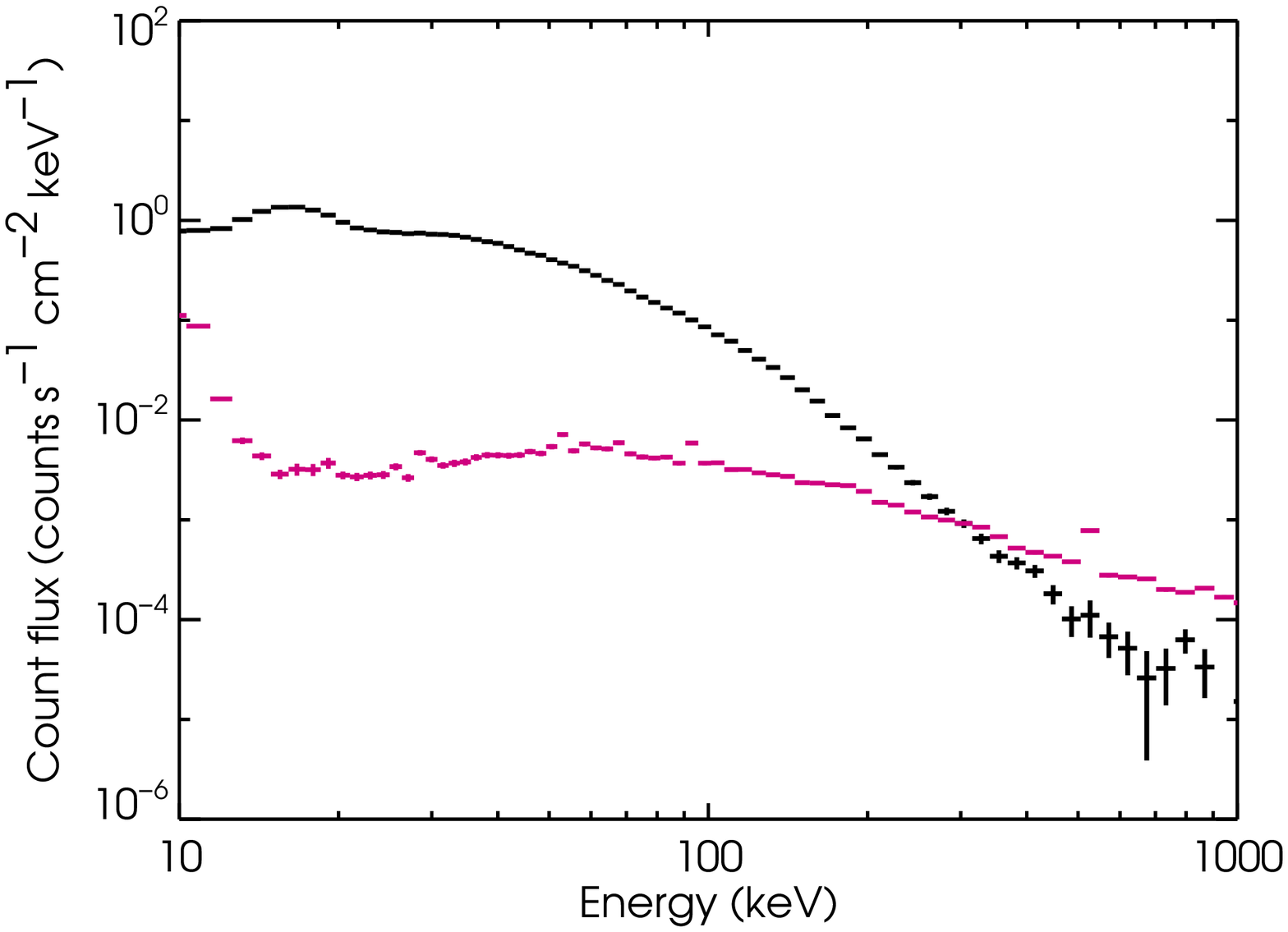}
\hspace*{-0.03\textwidth}
\includegraphics[width=0.515\textwidth,height=40mm,clip=]{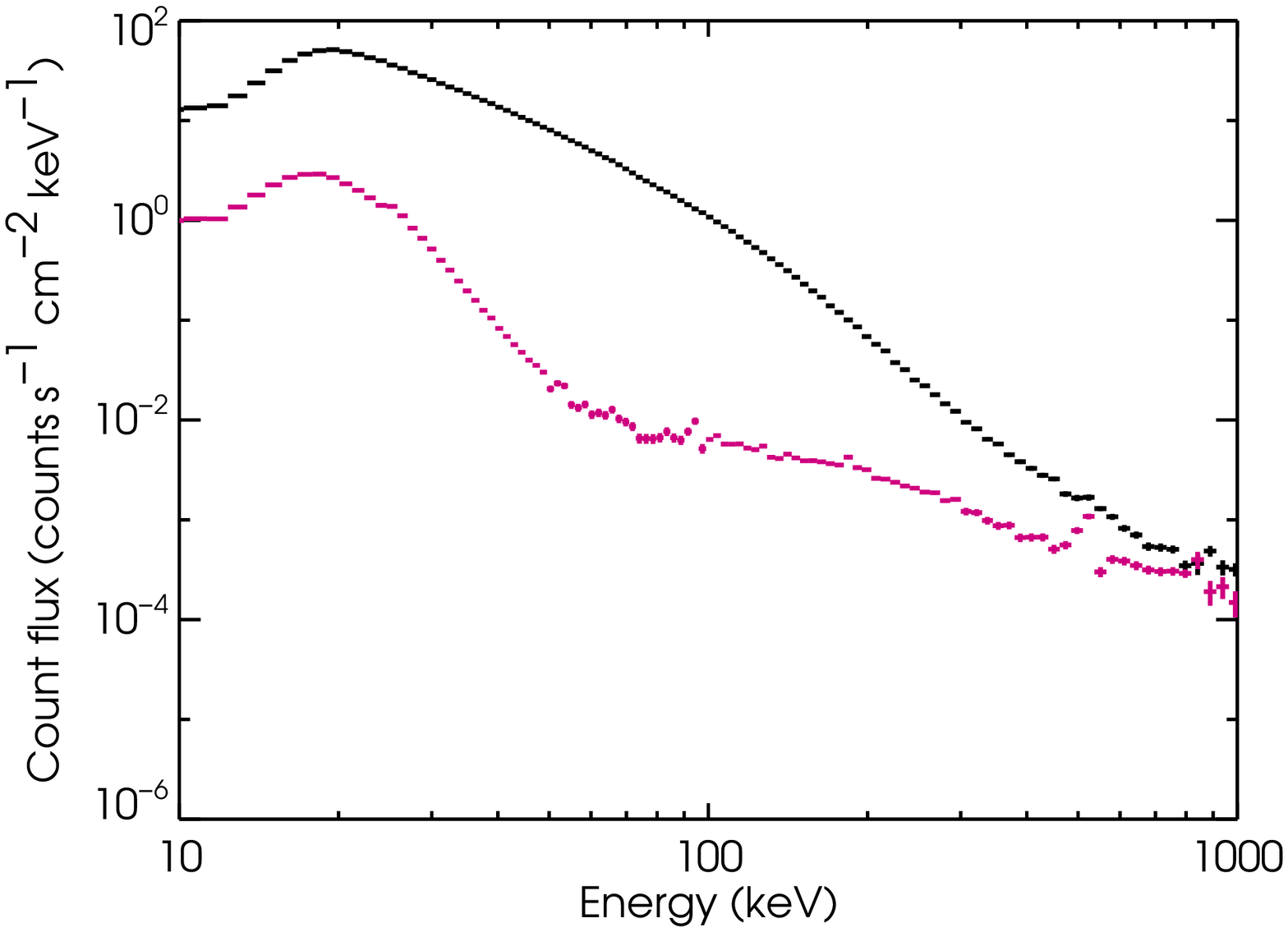}
}
\vspace{-0.30\textwidth} 
\centerline{\Large \bf 
\hspace{0.4 \textwidth} \color{black}{(c)}
\hspace{0.415\textwidth} \color{black}{(d)}
\hfill}
\vspace{0.30\textwidth} 
\centerline{\hspace*{0.015\textwidth}
\includegraphics[width=0.515\textwidth,height=40mm,clip=]{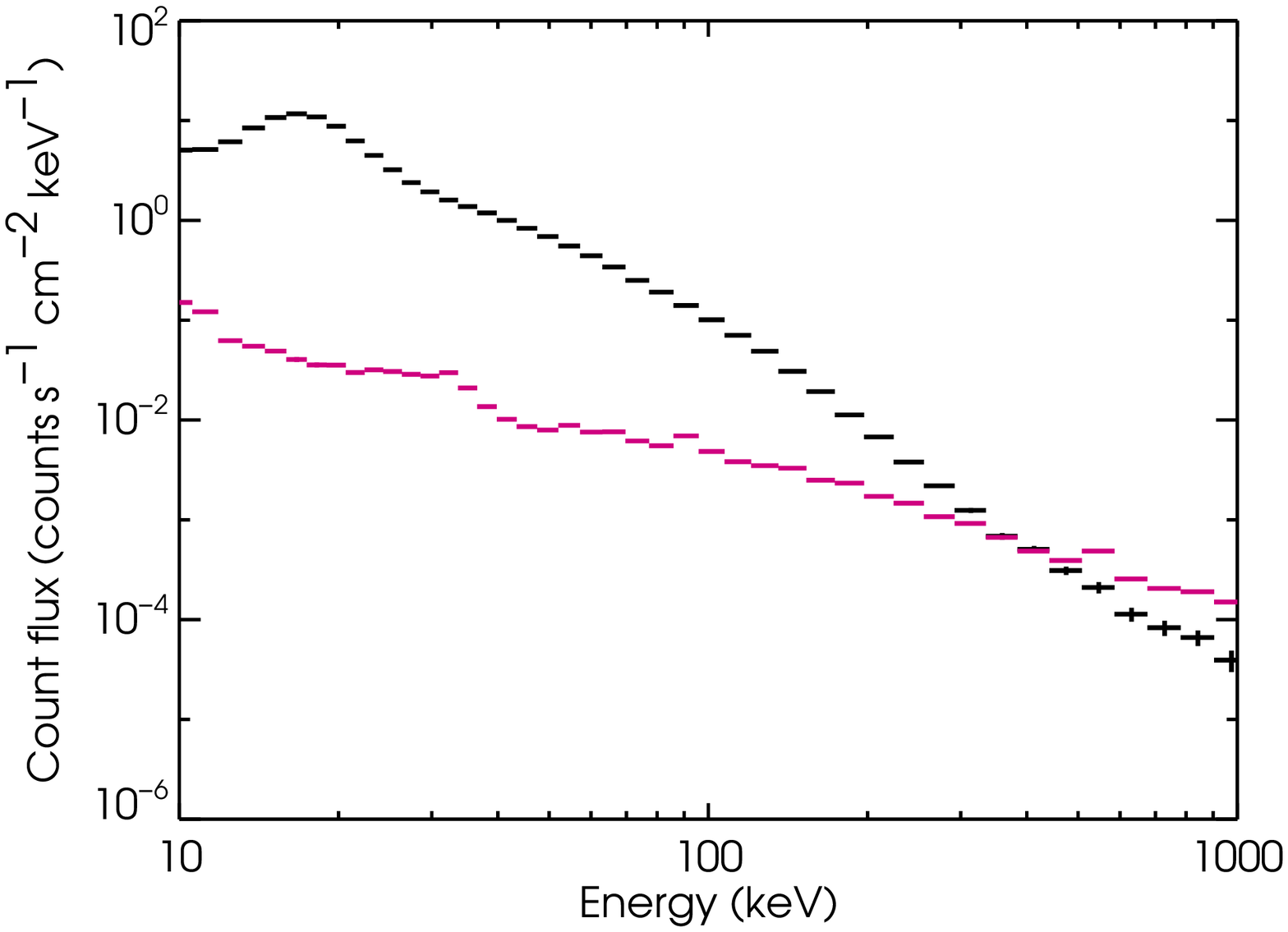}
\hspace*{-0.03\textwidth}
\includegraphics[width=0.515\textwidth,height=40mm,clip=]{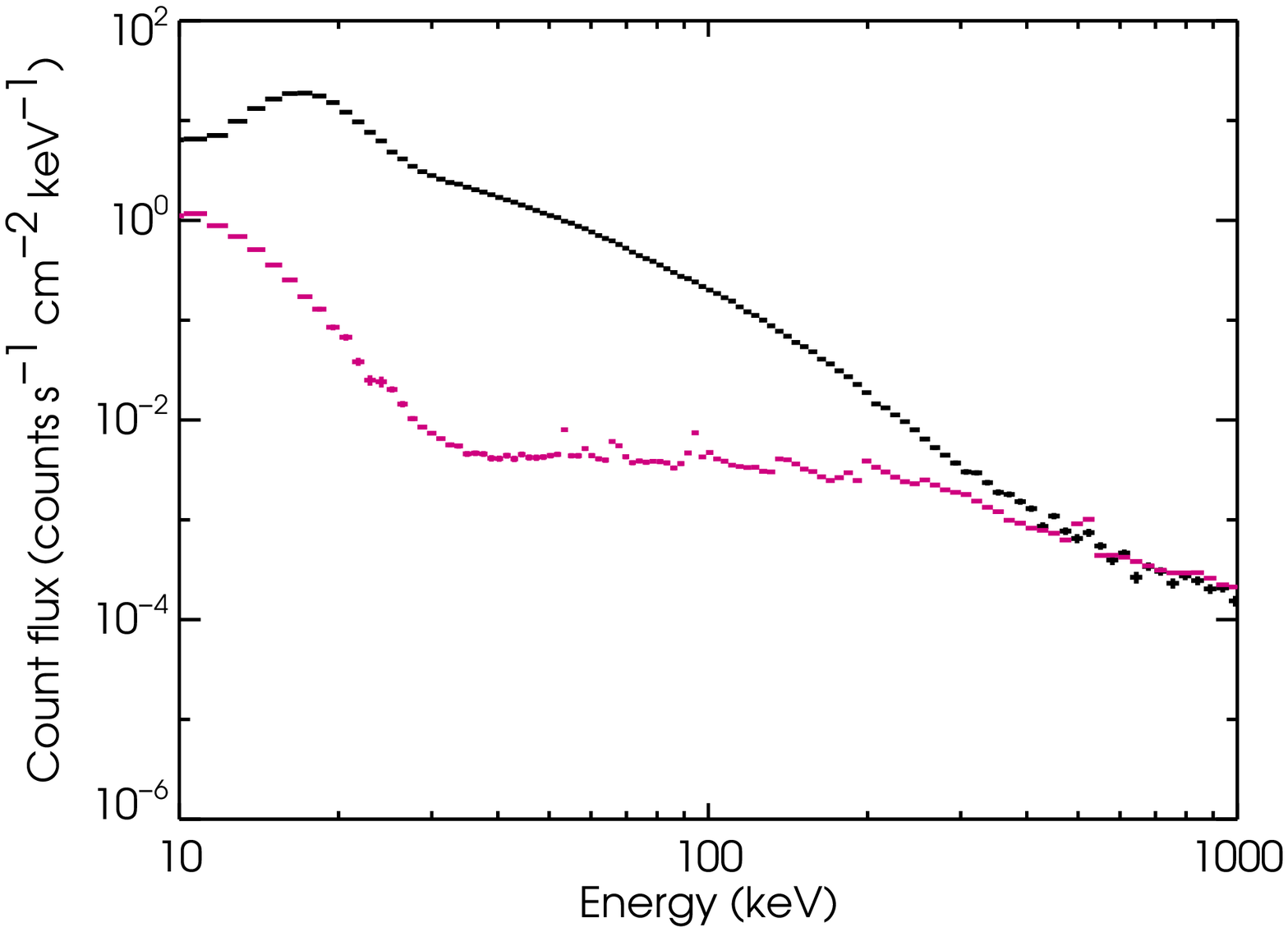}
}
\vspace{-0.30\textwidth} 
\centerline{\Large \bf 
\hspace{0.4 \textwidth} \color{black}{(e)}
\hspace{0.415\textwidth} \color{black}{(f)}
\hfill}
\vspace{0.30\textwidth} 
\centerline{\hspace*{0.015\textwidth}
\includegraphics[width=0.515\textwidth,height=40mm,clip=]{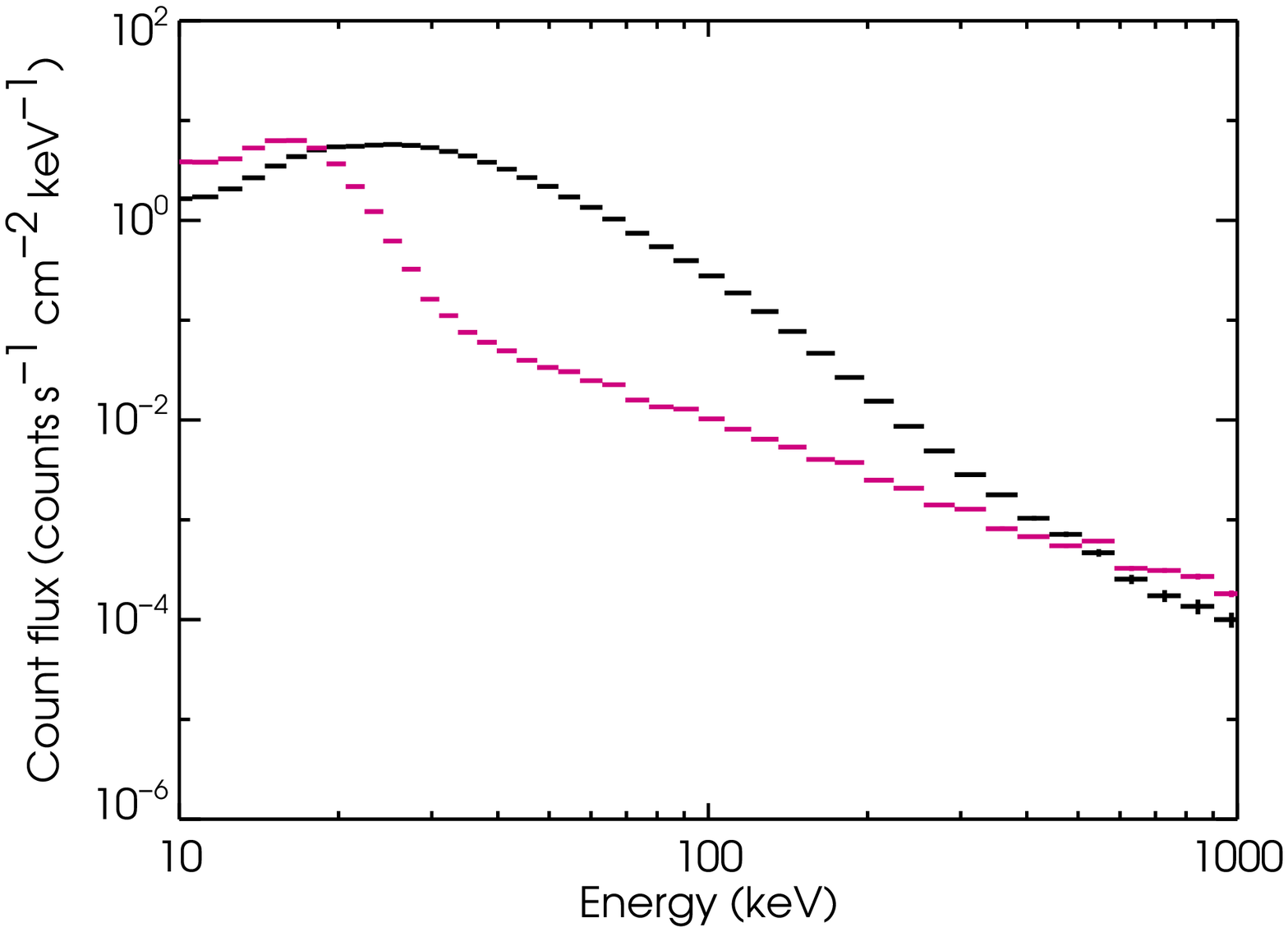}
\hspace*{-0.03\textwidth}
\includegraphics[width=0.515\textwidth,height=40mm,clip=]{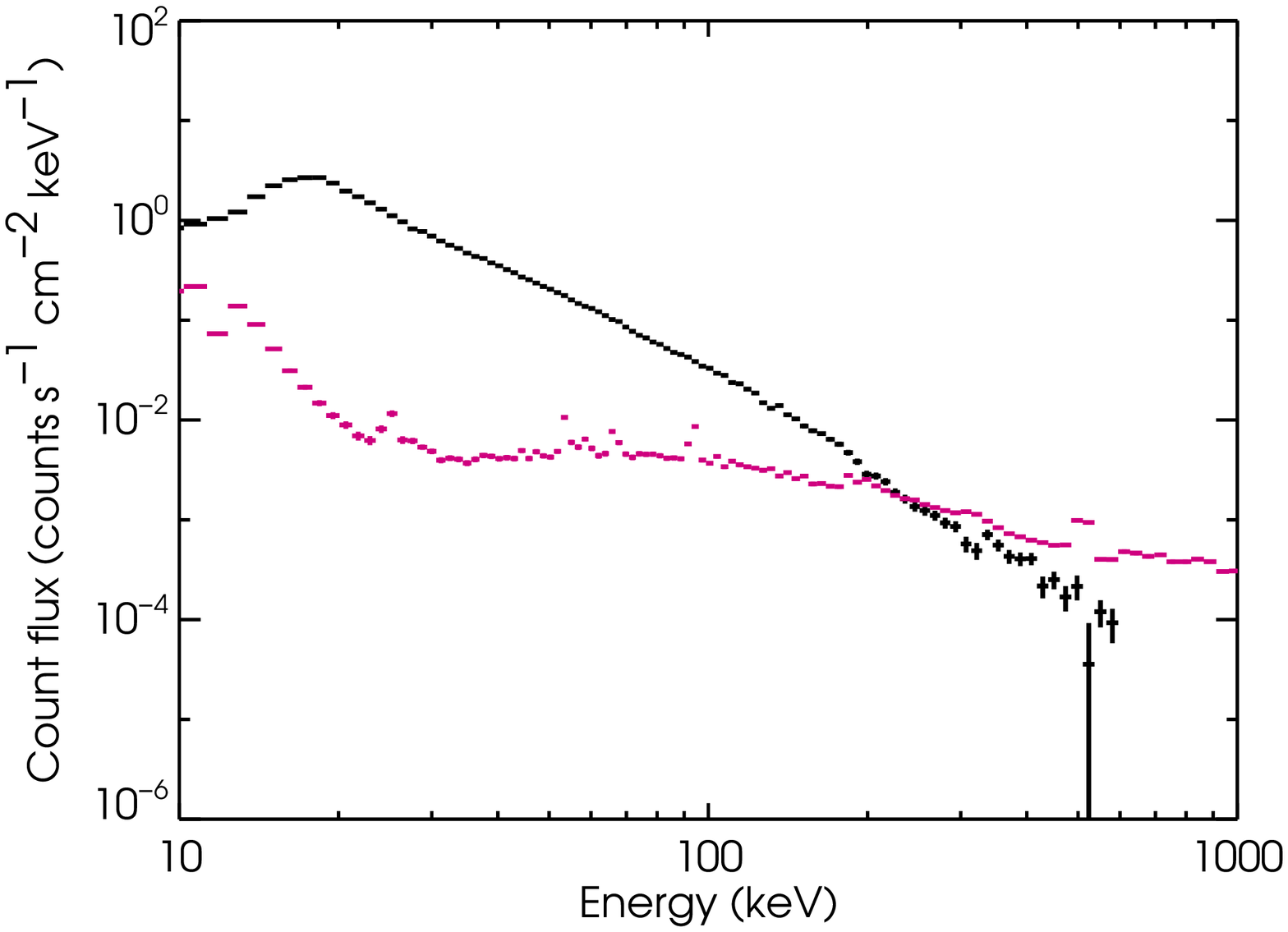}
}
\vspace{-0.30\textwidth} 
\centerline{\Large \bf 
\hspace{0.4 \textwidth} \color{black}{(g)}
\hspace{0.415\textwidth} \color{black}{(h)}
\hfill}
\vspace{0.30\textwidth} 
\caption{ Impulsive phase count spectra accumulated by RHESSI for each
flare studied (see Figure 9). The black line shows the background
subtracted counts and the magenta line the background.
}
\label{fig:C-panels}
\end{figure}

\begin{figure}[htb!]
\centering
\includegraphics[width=60mm,height=130mm]{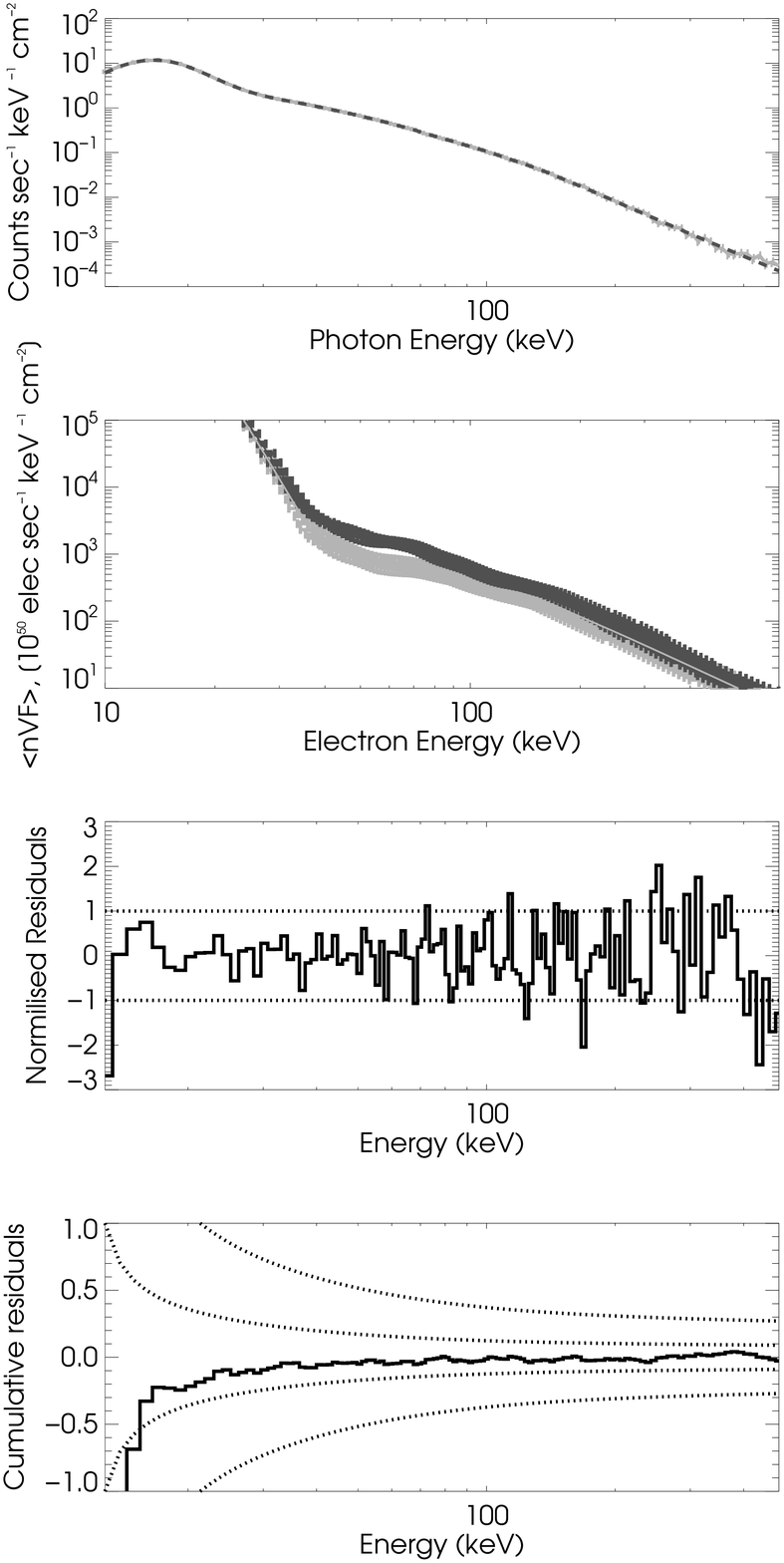}
\includegraphics[width=60mm,height=130mm]{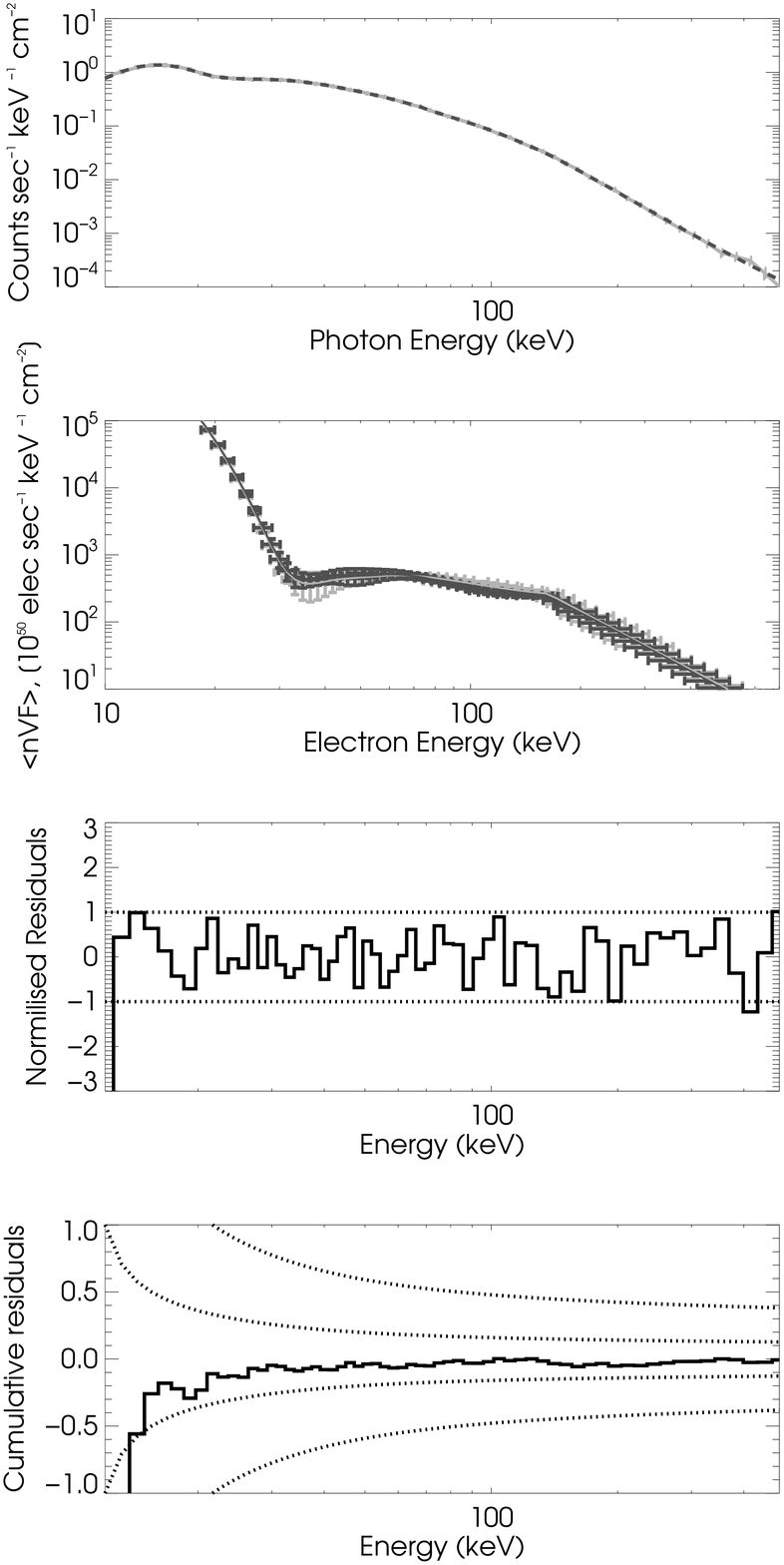}
\caption{Results of the invertion procedure for 64 second intervals from two
selected flares
on 10th November 2004 (left) and 17th June 2003 (right). Top panel
shows the measured count spectrum (full line) overplotted with the count
spectrum corresponding to the calculated regularised electron spectra
(dashed line). The second panel shows regularised electron spectrum with
associated  1-$\sigma$ vertical and horizontal error bars for each point,
the light grey line denotes the upward electron flux and the dark
grey line the downward electron flux. The third panel shows the
normalised residuals for each time interval and the bottom panel shows
the cumulative residuals.}
\label{fig:resid}
\end{figure}

\begin{figure} 
\centerline{\hspace*{0.015\textwidth}
\includegraphics[width=0.515\textwidth,height=40mm,clip=]{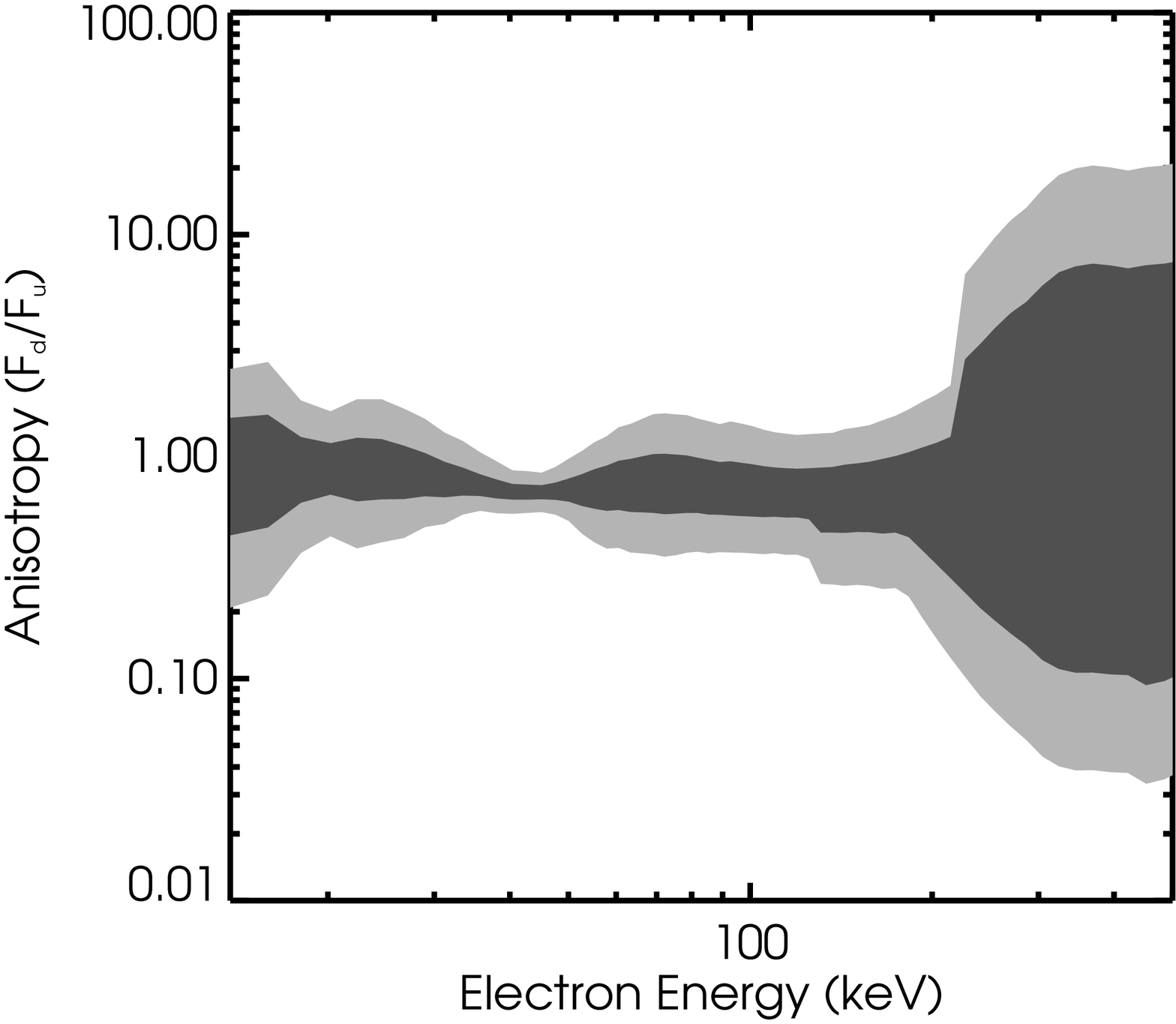}
\hspace*{-0.03\textwidth}
\includegraphics[width=0.515\textwidth,height=40mm,clip=]{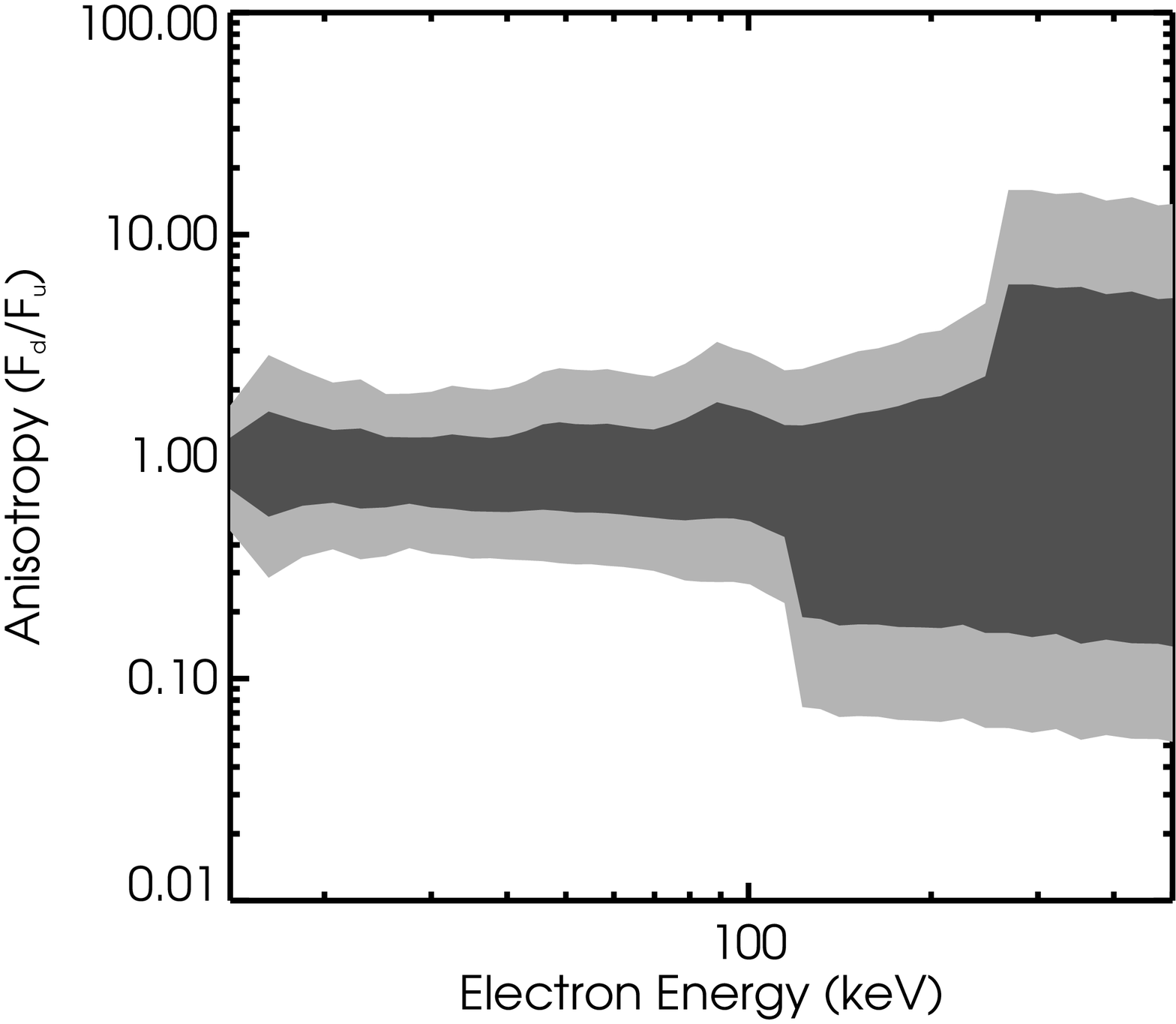}
}
\vspace{-0.3\textwidth} 
\centerline{\Large \bf 
\hspace{0.085 \textwidth} \color{black}{(a)}
\hspace{0.415\textwidth} \color{black}{(b)}
\hfill}
\vspace{0.3\textwidth} 
\centerline{\hspace*{0.015\textwidth}
\includegraphics[width=0.515\textwidth,height=40mm,clip=]{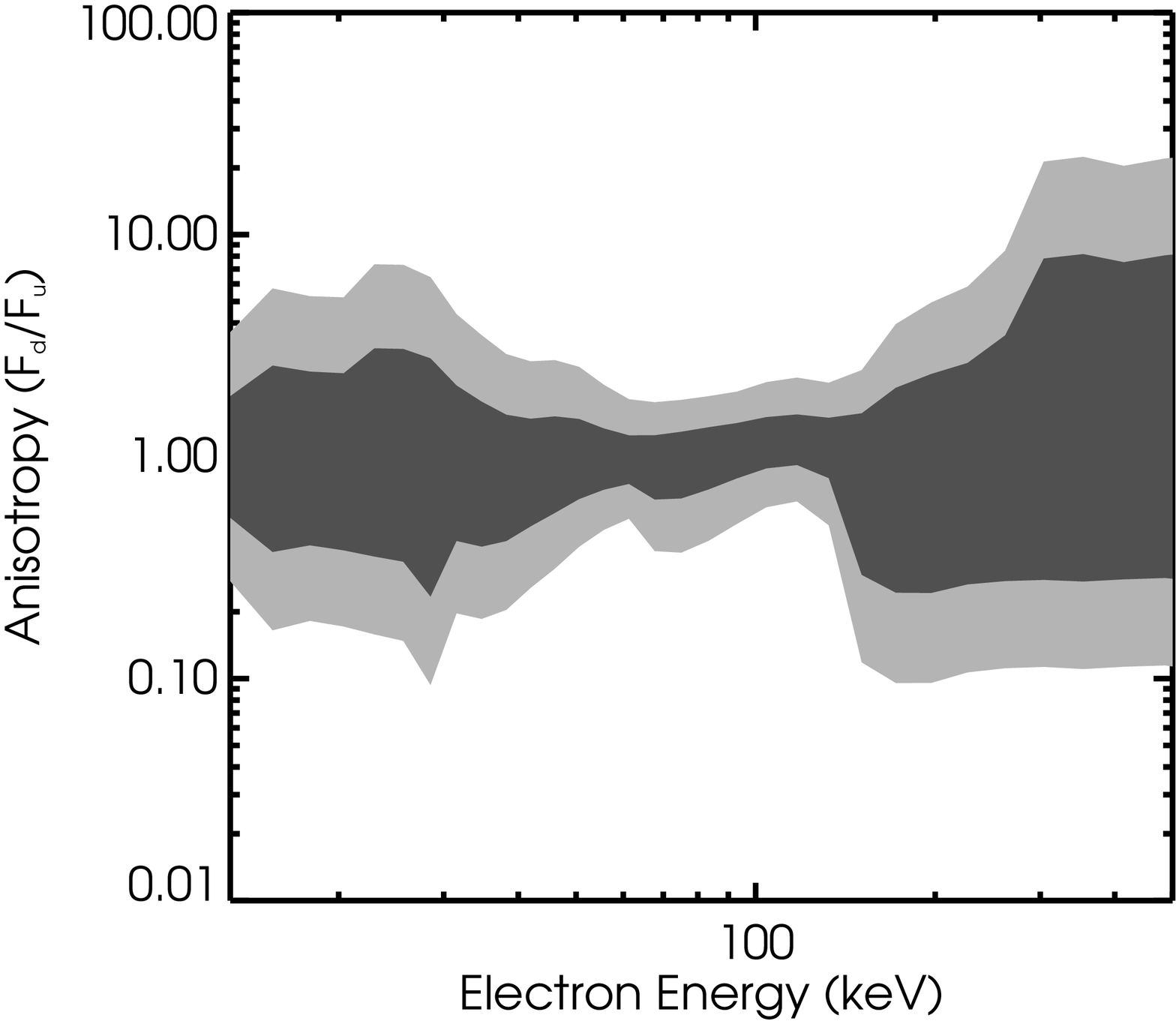}
\hspace*{-0.03\textwidth}
\includegraphics[width=0.515\textwidth,height=40mm,clip=]{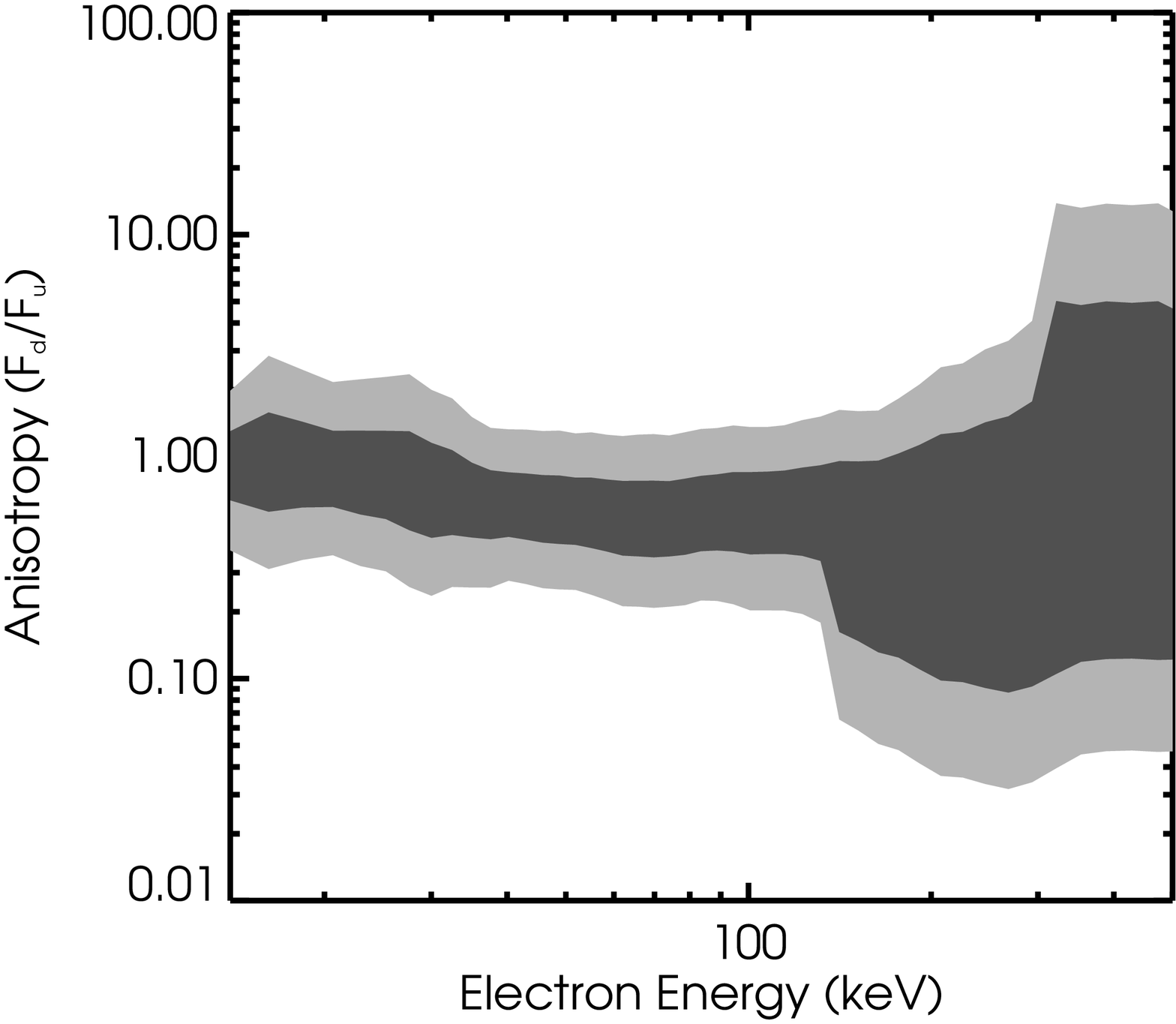}
}
\vspace{-0.3\textwidth} 
\centerline{\Large \bf 
\hspace{0.085 \textwidth} \color{black}{(c)}
\hspace{0.415\textwidth} \color{black}{(d)}
\hfill}
\vspace{0.3\textwidth} 
\centerline{\hspace*{0.015\textwidth}
\includegraphics[width=0.515\textwidth,height=40mm,clip=]{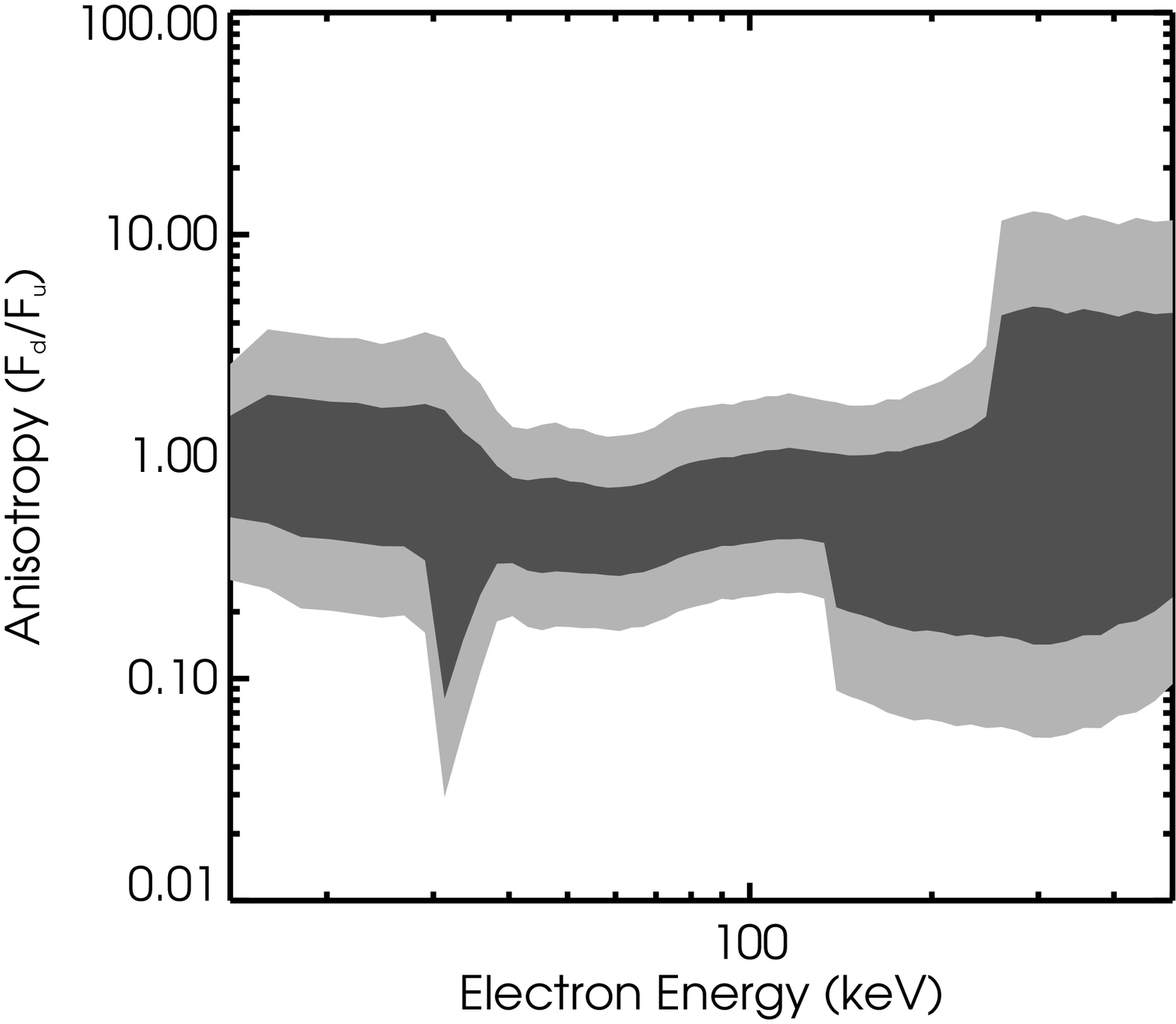}
\hspace*{-0.03\textwidth}
\includegraphics[width=0.515\textwidth,height=40mm,clip=]{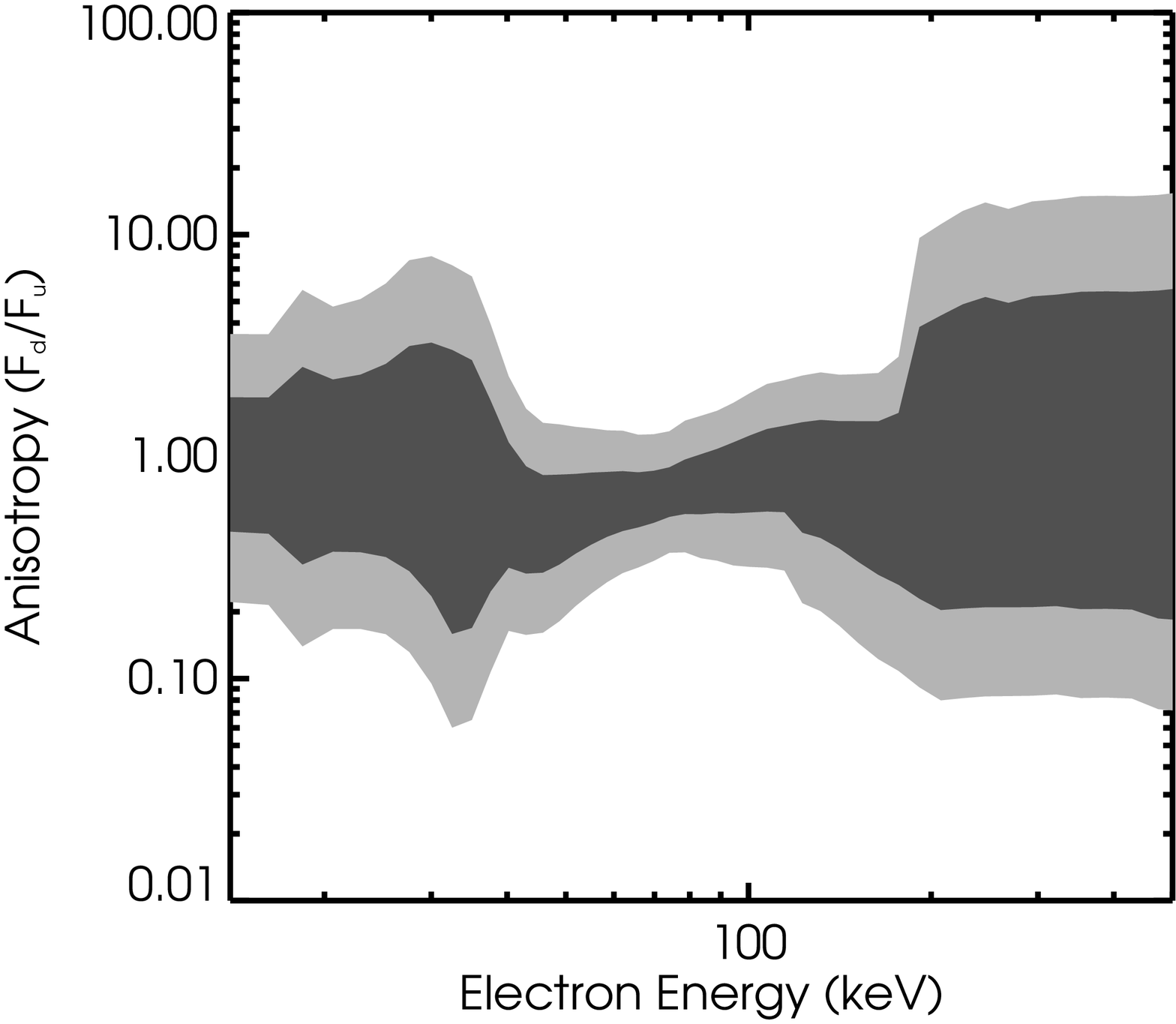}
}
\vspace{-0.3\textwidth} 
\centerline{\Large \bf 
\hspace{0.085 \textwidth} \color{black}{(e)}
\hspace{0.415\textwidth} \color{black}{(f)}
\hfill}
\vspace{0.3\textwidth} 
\centerline{\hspace*{0.015\textwidth}
\includegraphics[width=0.515\textwidth,height=40mm,clip=]{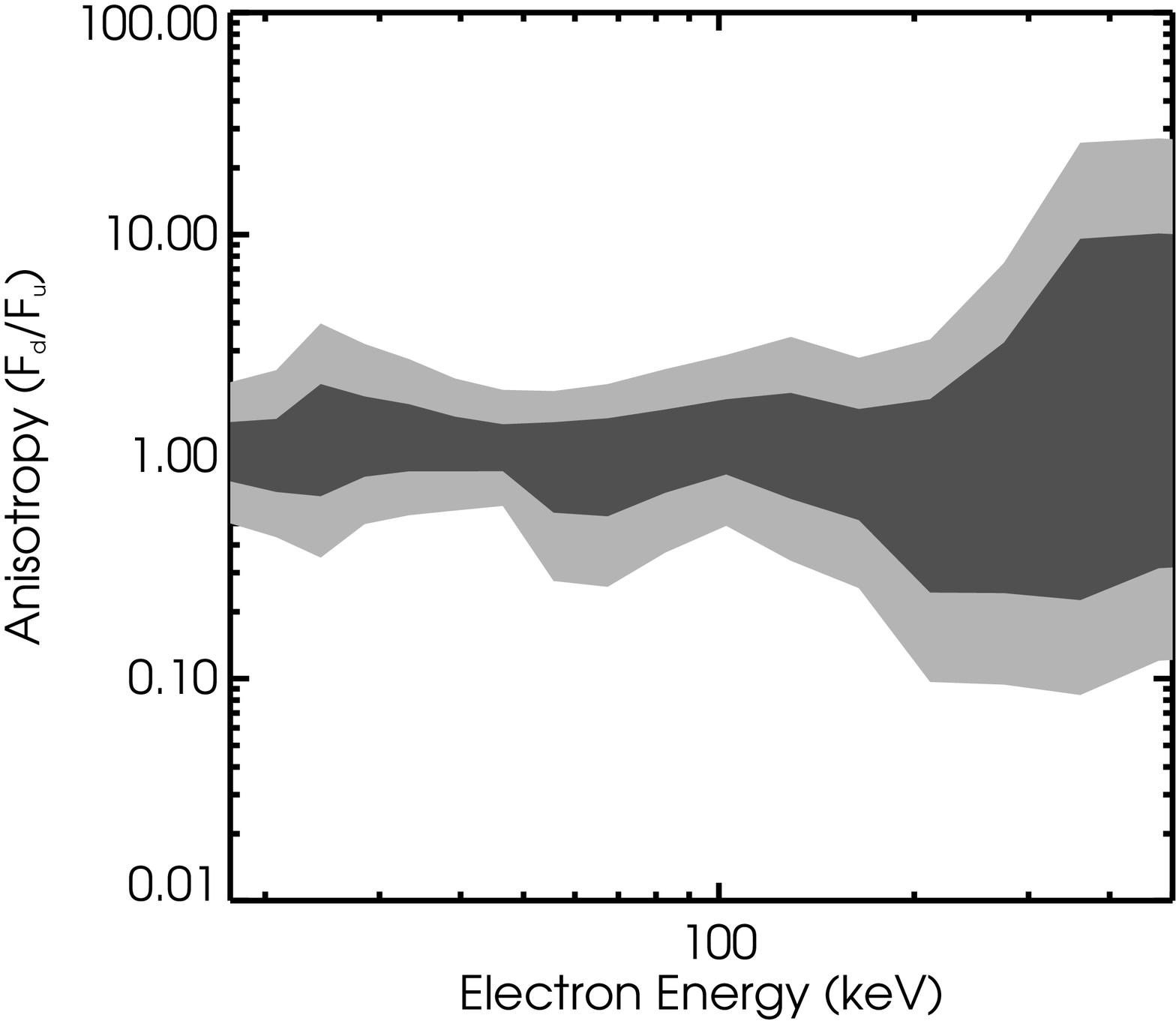}
\hspace*{-0.03\textwidth}
\includegraphics[width=0.515\textwidth,height=40mm,clip=]{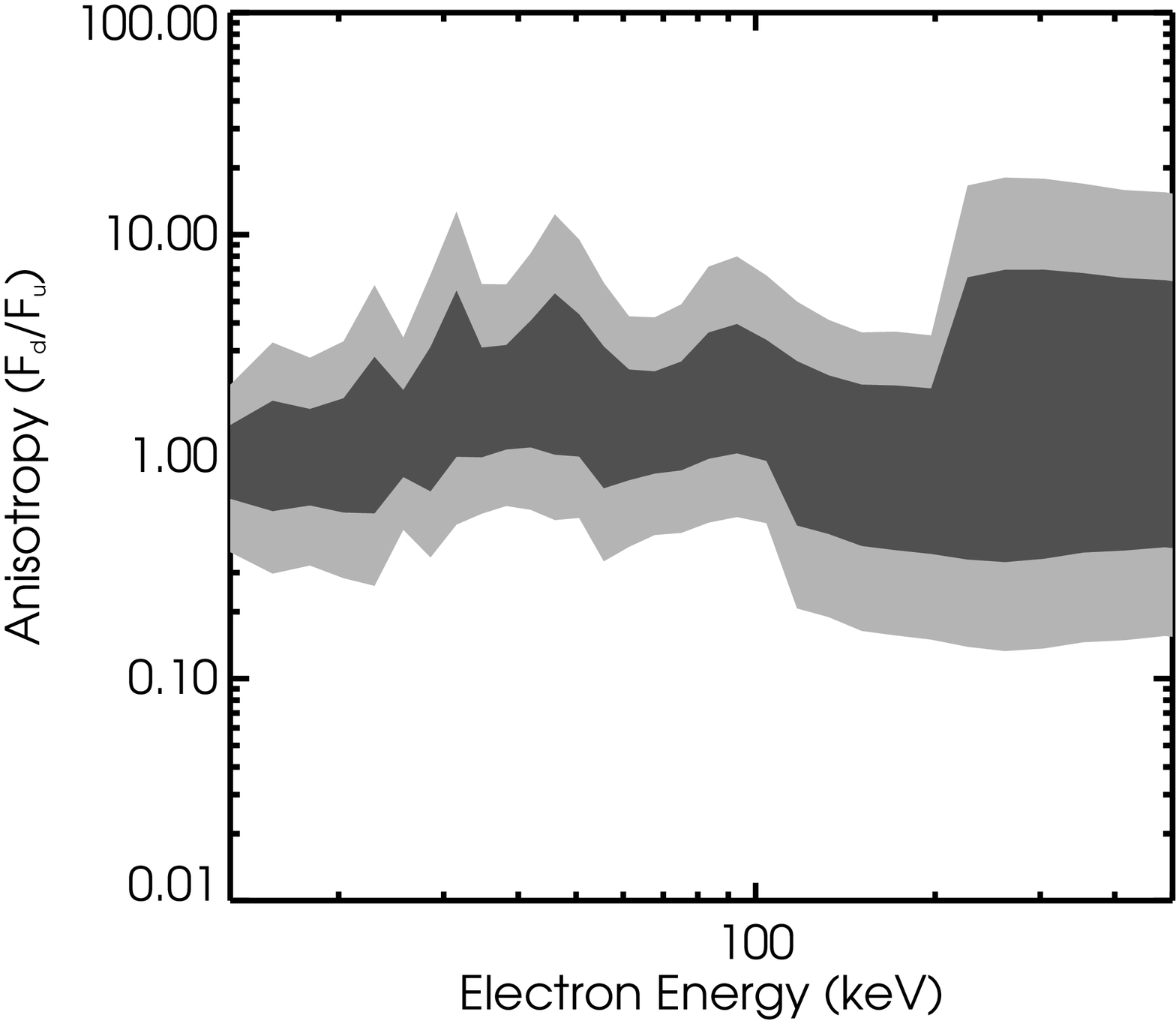}
}
\vspace{-0.3\textwidth} 
\centerline{\Large \bf 
\hspace{0.085 \textwidth} \color{black}{(g)}
\hspace{0.415\textwidth} \color{black}{(h)}
\hfill}
\vspace{0.3\textwidth} 
\caption{The anisotropy of the electron spectrum (defined as $ { \overline F_{d} } /  { \overline F_{u}}$) for all eight flares studied. The dark grey area represents the $
1\sigma $ confidence interval and the light grey the $ 3\sigma $
confidence interval.
}
\label{fig:A-panels}
\end{figure}

\begin{figure}[ht!] 
\centerline{\hspace*{0.015\textwidth}
\includegraphics[width=0.515\textwidth,height=40mm,clip=]{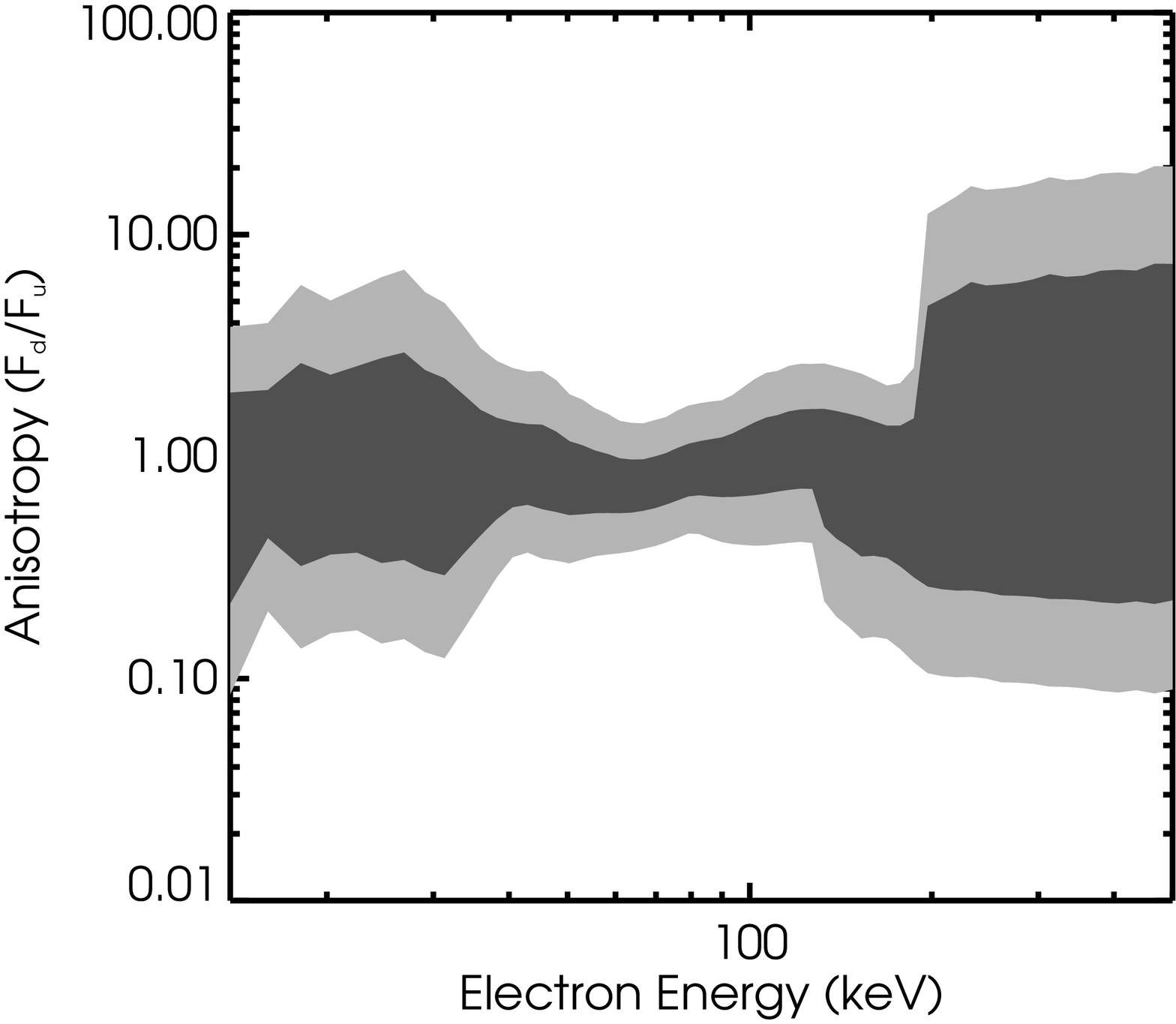}
\hspace*{-0.03\textwidth}
\includegraphics[width=0.515\textwidth,height=40mm,clip=]{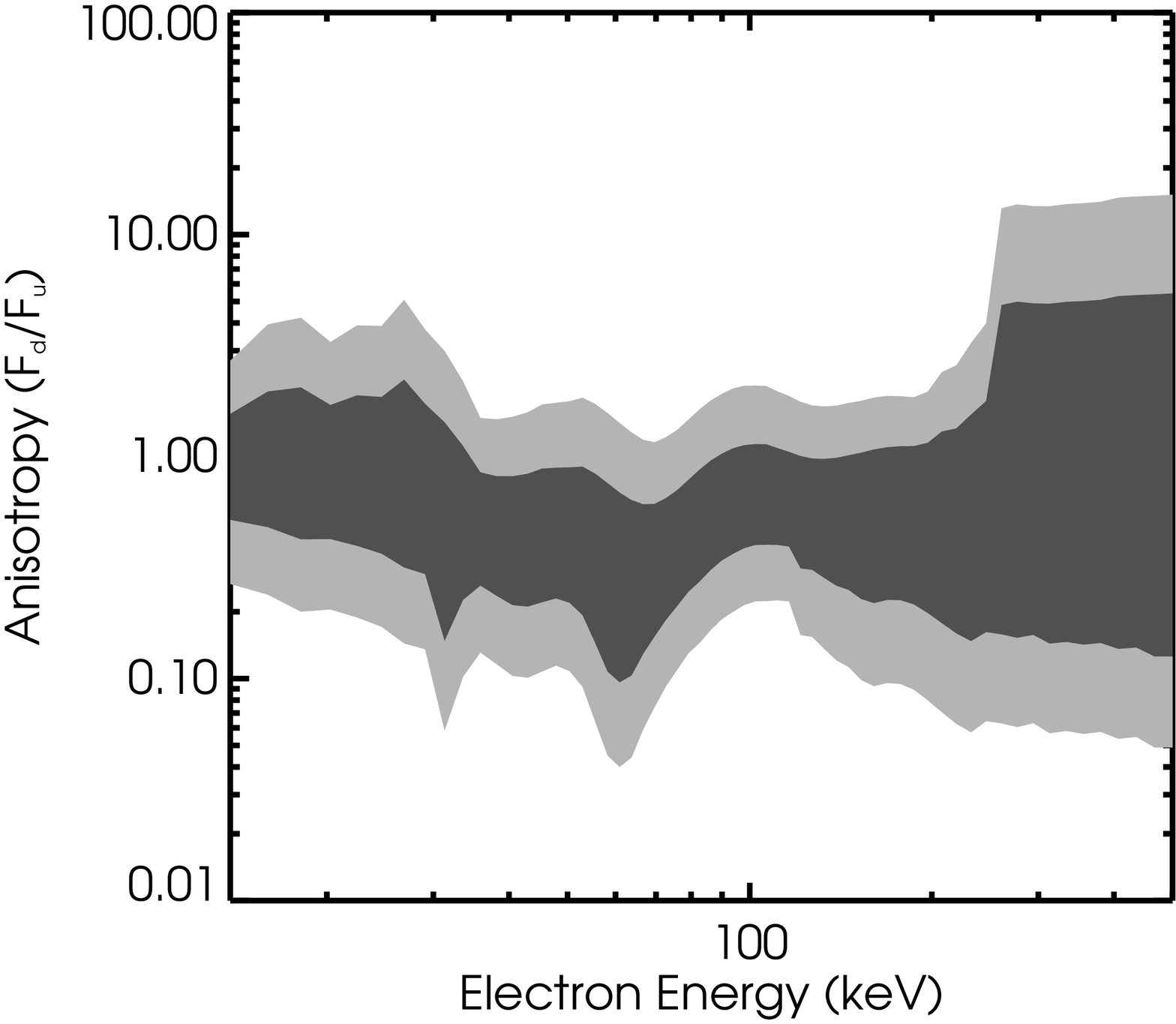}
}
\vspace{-0.3\textwidth} 
\centerline{\Large \bf 
\hspace{0.085 \textwidth} \color{black}{(a)}
\hspace{0.415\textwidth} \color{black}{(b)}
\hfill}
\vspace{0.3\textwidth} 
\centerline{\hspace*{0.015\textwidth}
\includegraphics[width=0.515\textwidth,height=40mm,clip=]{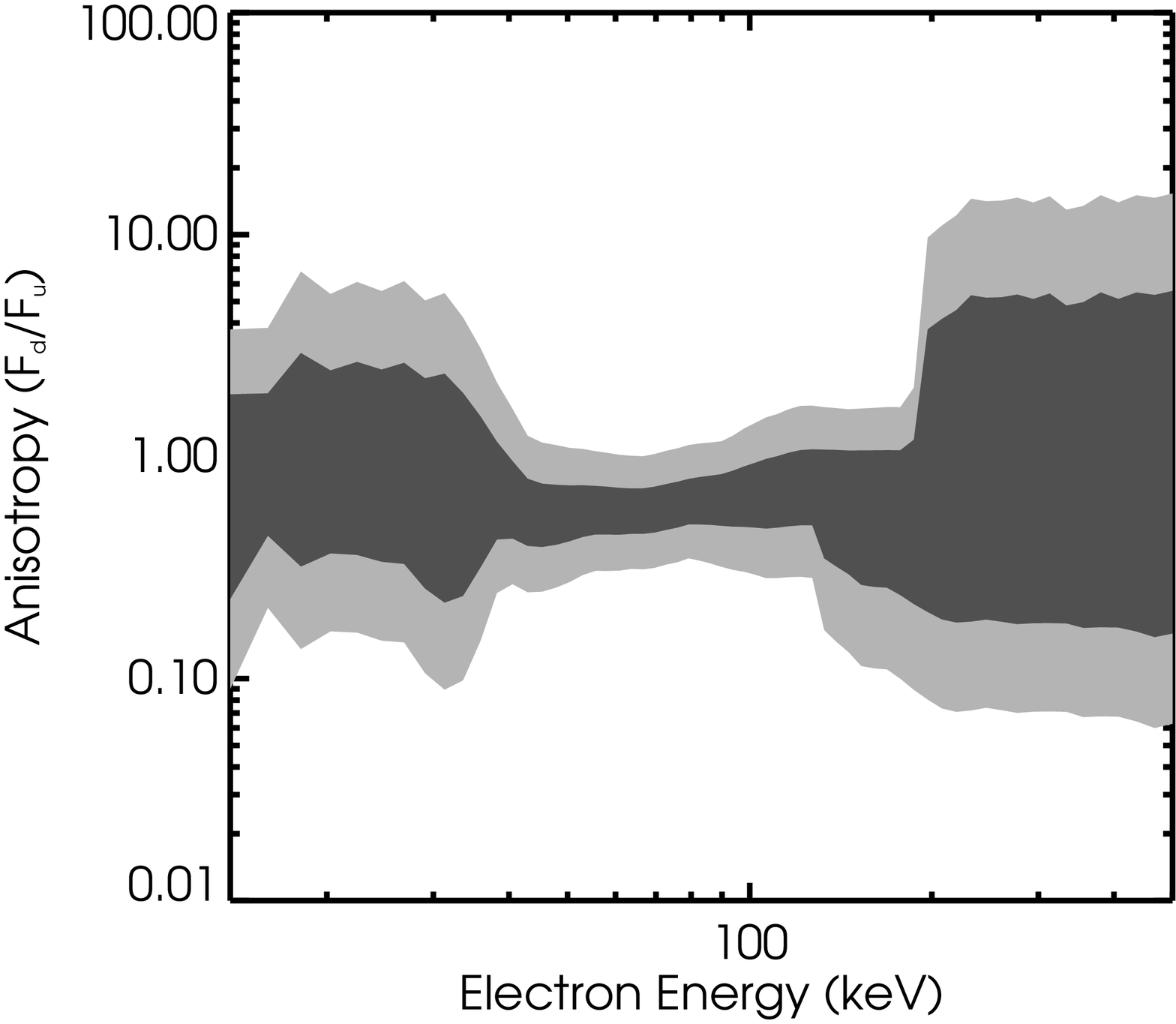}
\hspace*{-0.03\textwidth}
\includegraphics[width=0.515\textwidth,height=40mm,clip=]{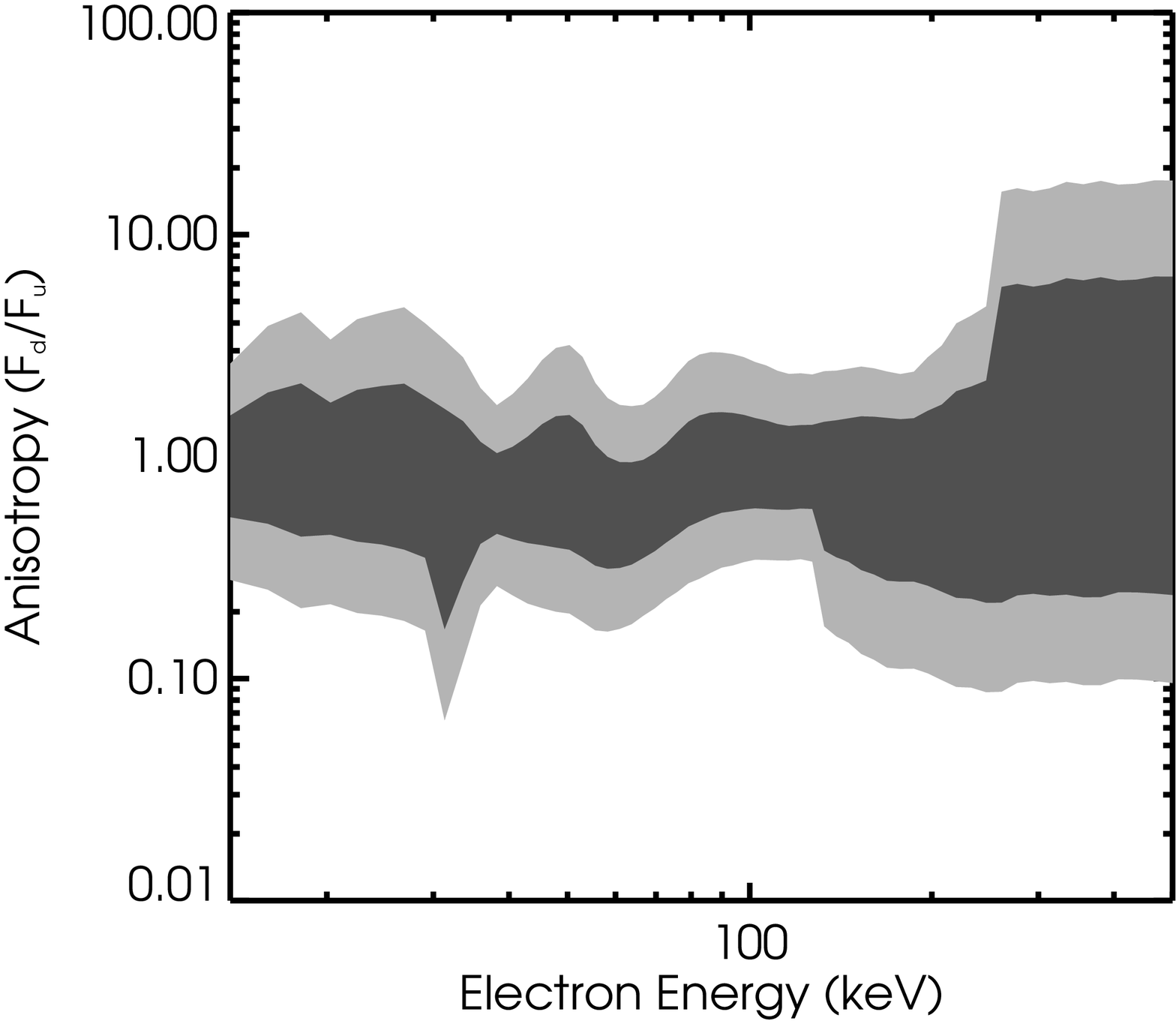}
}
\vspace{-0.3\textwidth} 
\centerline{\Large \bf 
\hspace{0.085 \textwidth} \color{black}{(c)}
\hspace{0.415\textwidth} \color{black}{(d)}
\hfill}
\vspace{0.30\textwidth} 
\centerline{\hspace*{0.015\textwidth}
\includegraphics[width=0.515\textwidth,height=40mm,clip=]{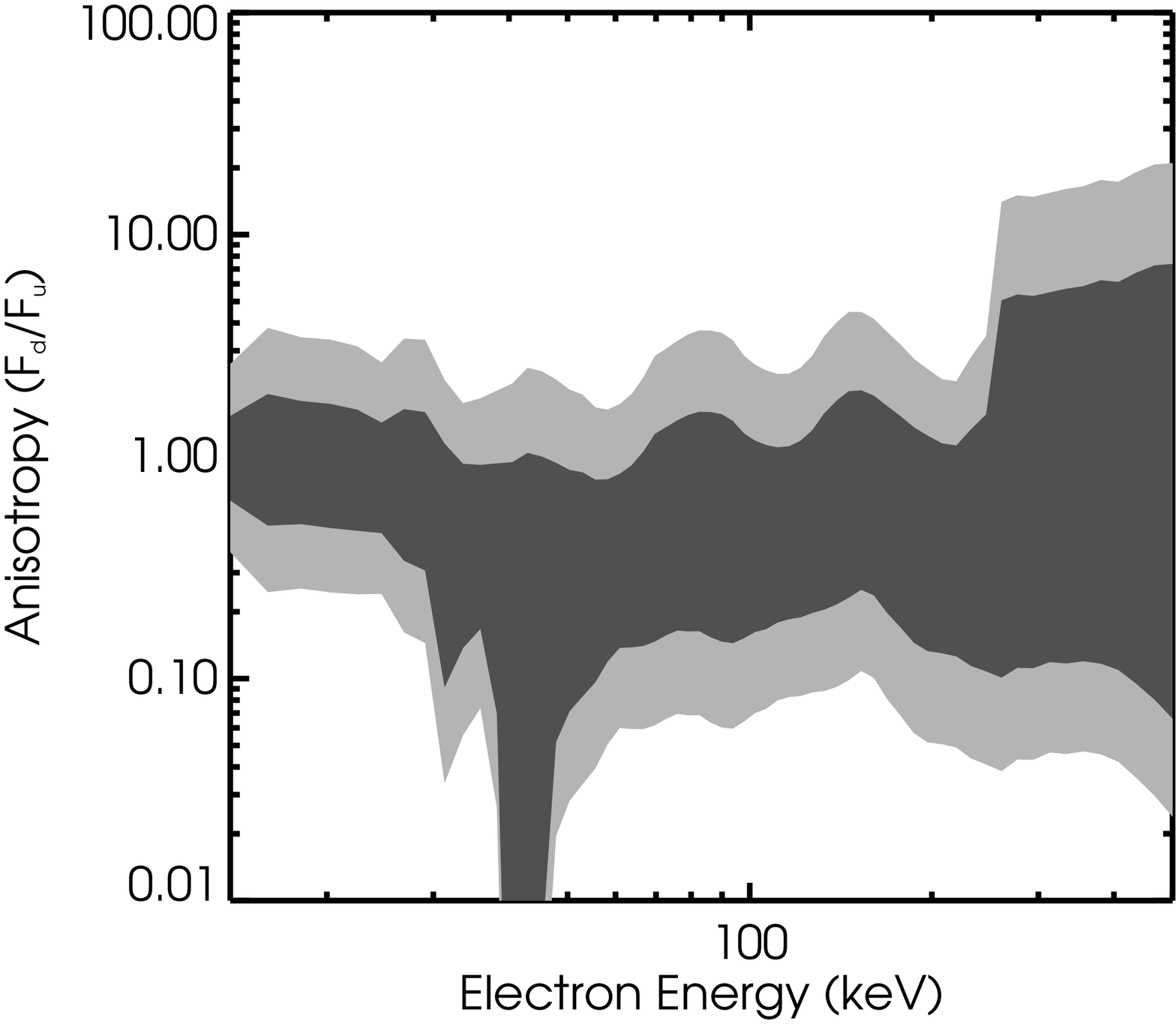}
\hspace*{-0.03\textwidth}
\includegraphics[width=0.515\textwidth,height=40mm,clip=]{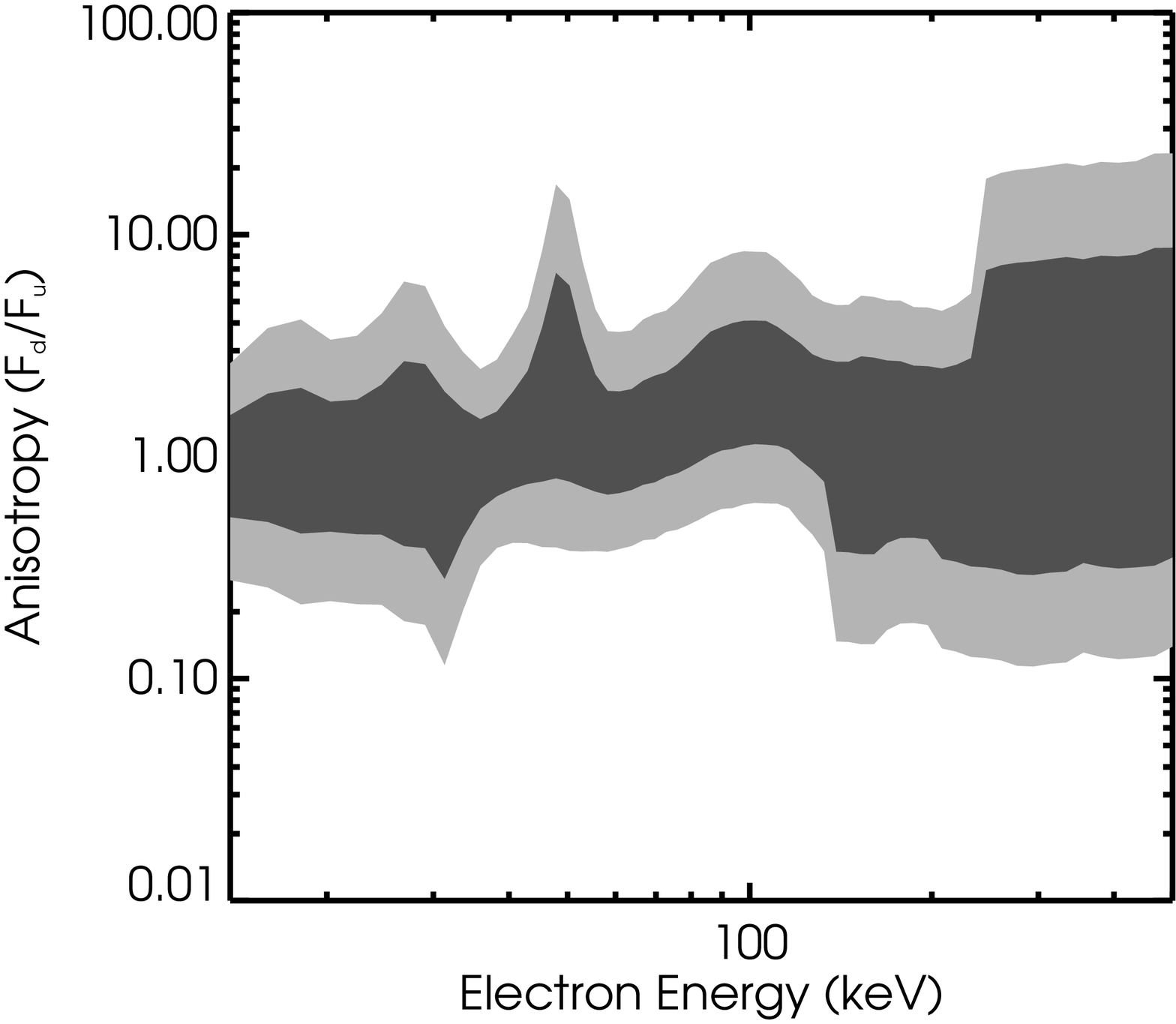}
}
\vspace{-0.3\textwidth} 
\centerline{\Large \bf 
\hspace{0.085 \textwidth} \color{black}{(e)}
\hspace{0.415\textwidth} \color{black}{(f)}
\hfill}
\vspace{0.30\textwidth} 
\centerline{\hspace*{0.015\textwidth}
\includegraphics[width=0.515\textwidth,height=40mm,clip=]{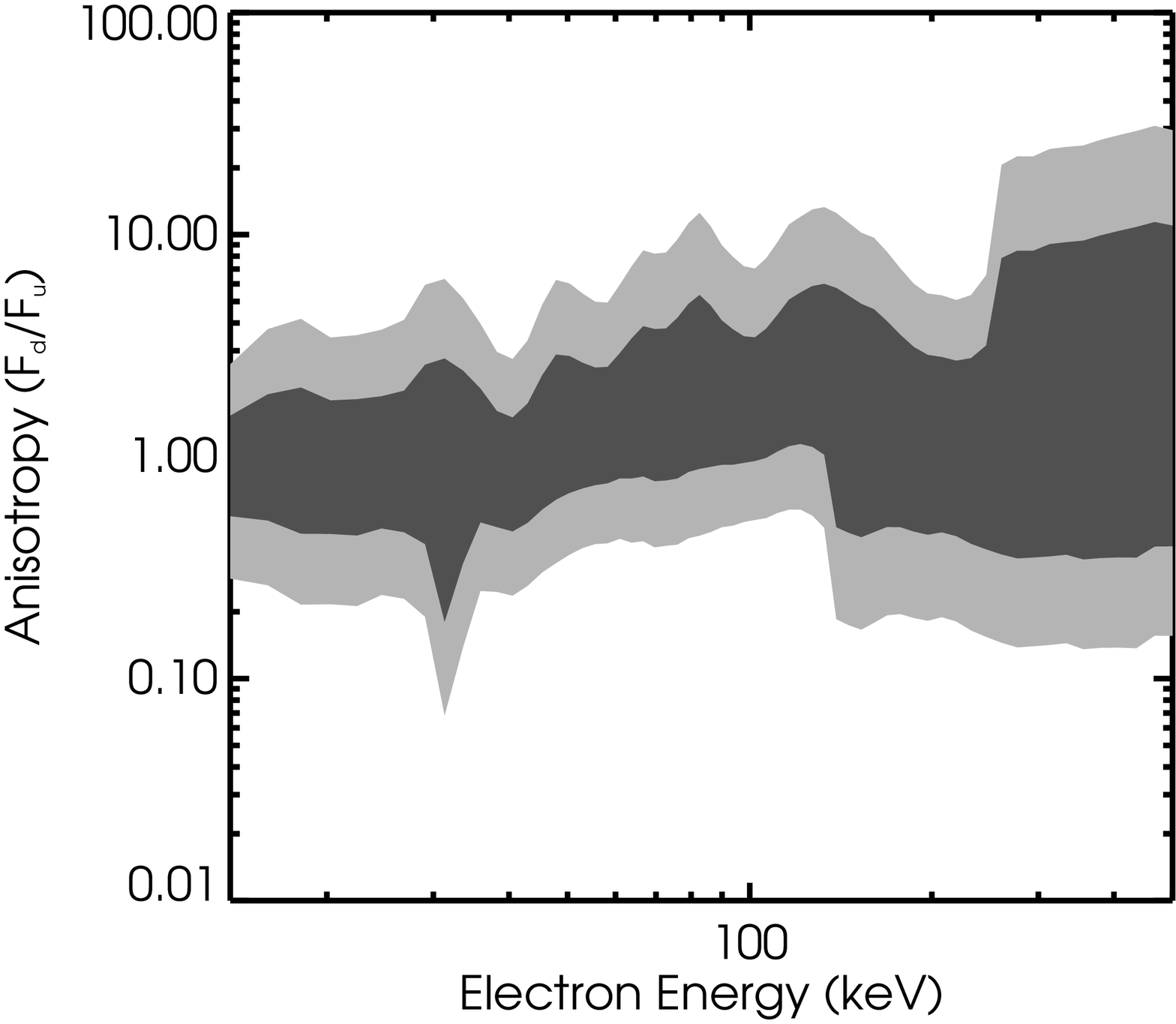}
\hspace*{-0.03\textwidth}
\includegraphics[width=0.515\textwidth,height=40mm,clip=]{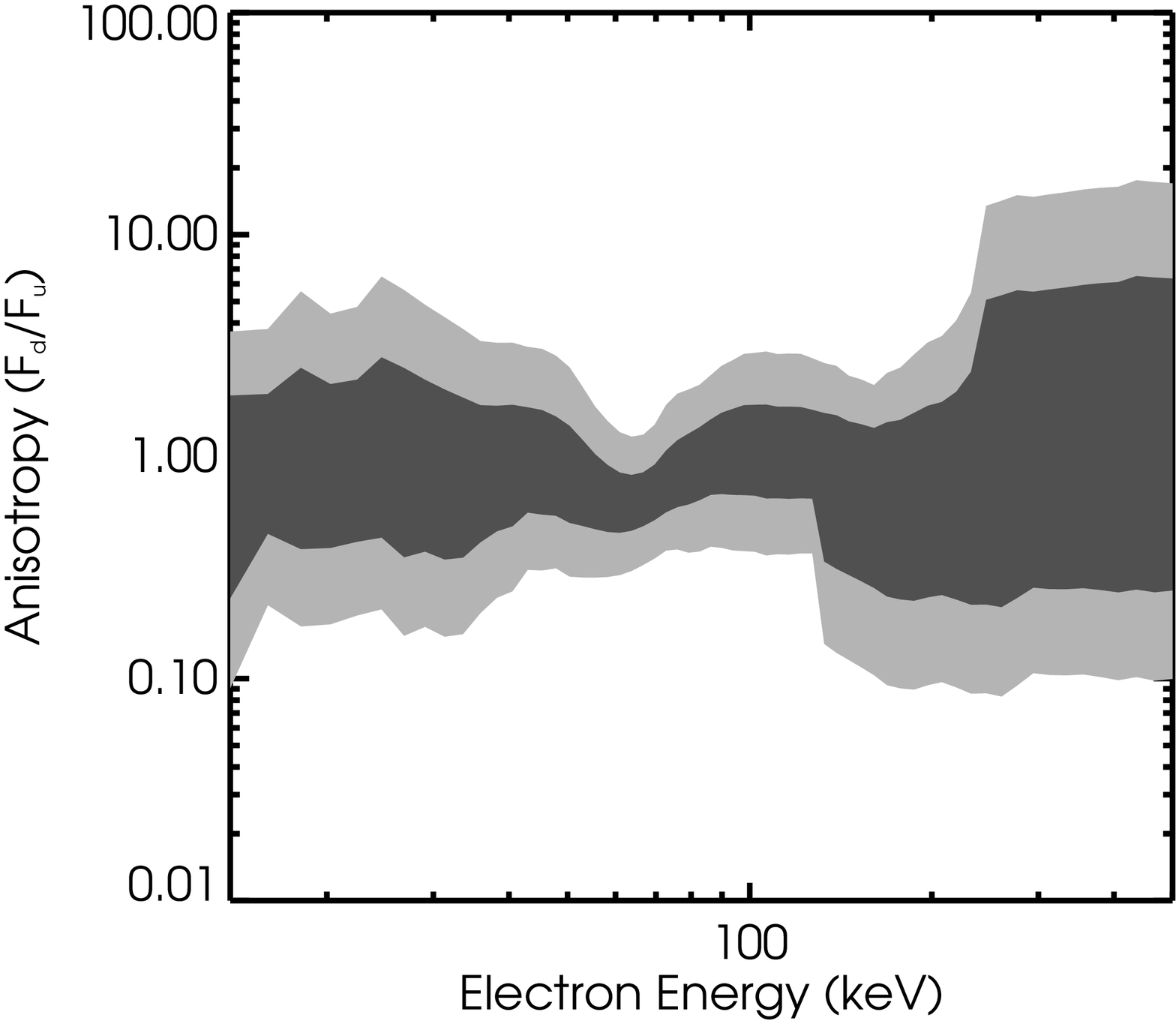}
}
\vspace{-0.3\textwidth} 
\centerline{\Large \bf 
\hspace{0.085 \textwidth} \color{black}{(g)}
\hspace{0.415\textwidth} \color{black}{(h)}
\hfill}
\vspace{0.3\textwidth} 
\caption{The anisotropy of the electron spectrum (defined as
$ { \overline F_{d} } /  { \overline F_{u}}$) for the first eight 4 second time intervals
for the flare occurred on 10 November 2004. The first interval starts at 02:09:40 UT and the
intervals shown here cover the most intense part of the impulsive peak.
The dark grey area represents the $ 1\sigma $ confidence interval and
the light grey the $ 3\sigma $ confidence interval.
}
\label{fig:As-panels}
\end{figure}

\begin{figure}[htb!]
\centering
\includegraphics[width=10cm]{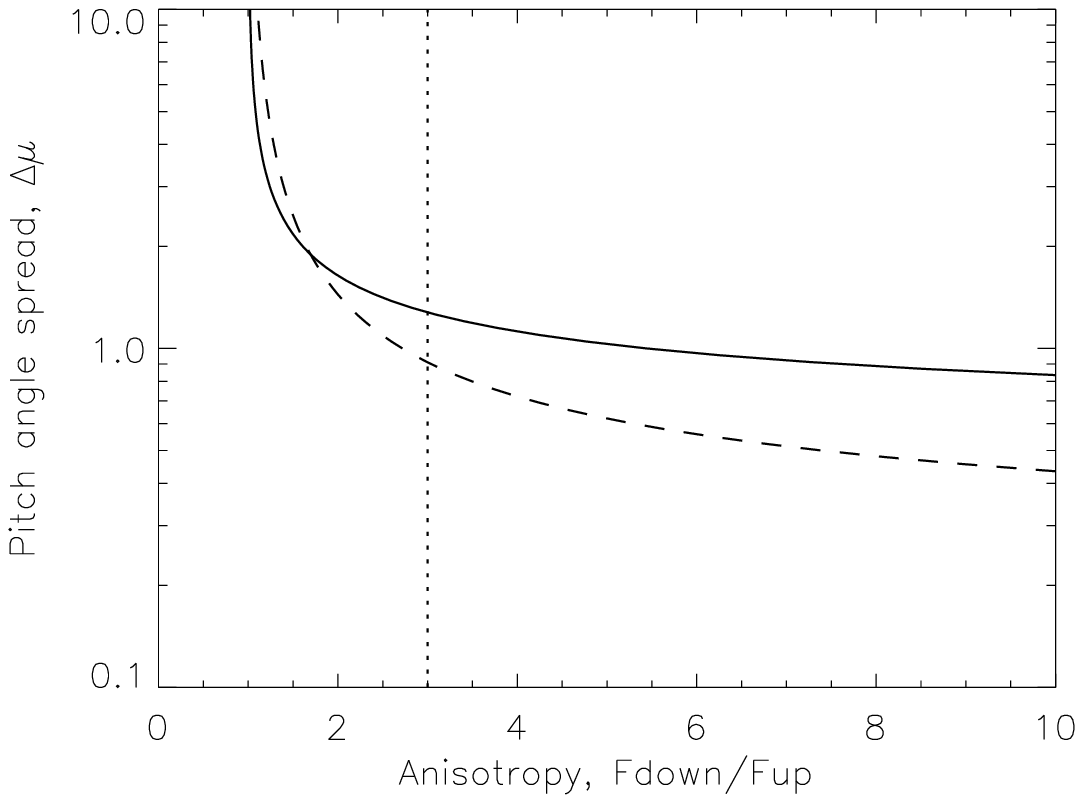}
\caption{Pitch angle spread, $\Delta\mu$, for various anisotropies ${\overline F_{d} }/{\overline F_{u}}$ using ${\overline F(\mu)} \propto  \exp \left( \frac{-(1 - \mu)^2}{\Delta
\mu^2} \right)$ (solid line) and ${\overline F(\mu)} \propto  \exp \left( \frac{-(1 - \mu)}{\Delta
\mu} \right)$  (dashed line). The vertical dotted line shows an anisotropy of 3. }
\label{fig:delmu}
\end{figure}

\subsection{20th August 2002}
This flare was detected on 20th August 2002 around 08:20~UT
with the impulsive peak starting about 08:25 UT. It was detected with a
heliocentric angle of $\sim 43^{\circ}$ equivalent to $\mu = 0.73$. The
flare also shows good count statistics up to 400~keV. As this flare had
attenuator status changes from A0 (open telescope) to A1 (thin shutter
in) at 08:25:16 UT and to A3 (both shutters in) at 08:25:44 UT, the
analysis was only performed over the 16 s period rather than the 64 s
period studied for most flares. There is some particle contamination
over the impulsive phase. This flare was extensively studied by
\inlinecite{2007A&A...466..705K} and the background subtraction used in
this paper is similar to the subtraction described there. This flare was
previously analysed using bi-directional inversion by
\inlinecite{2006A&A...446.1157K}.

\subsection{10th September 2002}
This flare was detected on 10th September 2002 between 14:02 and 15:15 UT
with the impulsive peak starting about 14:52 UT. It was detected with a
heliocentric angle of $\sim 44^{\circ}$ equivalent to $\mu = 0.72$. The
flare also shows good count statistics up to 300~keV. This flare also
has an attenuator status change from A0 to A1 at 14:52:43 UT 
and to A3 at 14:54:16 UT, so the impulsive phase is taken to be
a 32 s time interval between 14:52:47 and 14:53:19 UT.

\subsection{17th June 2003}

This flare was detected on 17th June 2003 starting at approximately
22:30 UT. As RHESSI
shows significant particle contamination during the early stages of this
flare analysis was performed on a later impulsive peak with accumulation
starting at 22:52:42 UT. It was detected with a heliocentric angle of
$59^{\circ}$ equivalent to $\mu = 0.51$. The flare also shows good
count statistics up to 300 keV.

\subsection{2nd November 2003}
This flare was detected on 2nd November 2003. It was detected with a
heliocentric angle of $\sim 59^{\circ}$ equivalent to $\mu = 0.51$. The
flare also shows good count statistics up to 300 keV. RHESSI showed
some elevated particle levels during the impulsive phase of the flare so
analysis was confined to the earlier part of the impulsive phase.

\subsection{10th November 2004}
This flare was detected on 10th November 2004. It was observed with a
heliocentric angle of $46.5^{\circ}$ equivalent to $\mu = 0.69$. The
flare shows no significant particle measurements during the impulsive
phase and low probability of pulse pileup. The flare also shows good
count statistics up to 500 keV.

\subsection{15th January 2005}
This flare was detected on 15th January 2005. It was detected with a
heliocentric angle of $20^{\circ}$ equivalent to $\mu = 0.93$, the
closest of all the flares selected to the disk centre and therefore the
most likely to show evidence of strong downwards directivity. The flare
also shows good count statistics up to 400 keV.

\subsection{17th January 2005}
This flare was detected on 17th January 2005 between 09:30 and 15:15 UT
with the impulsive peak starting about 09:42 UT. It was detected with a
heliocentric angle of $\sim 31^{\circ}$ equivalent to $\mu = 0.86$. The
flare also shows good count statistics up to 300 keV. As this flare
occurs in the tail of a previous flare is has very high background at
low energies. Counts below 18 keV were not accumulated for this flare.
This flare was also previously analysed using bi-directional 
inversion \inlinecite{2006A&A...446.1157K}.

\subsection{10th September 2005}
This flare was detected on 10th September 2005. It was detected with a
heliocentric angle of $\sim 46^{\circ}$ equivalent to $\mu = 0.69$. This
flare showed negligible particle contamination and a low probability of
pulse pileup. The flare also shows good count statistics up to 300 keV.
Due to changes in the attenuator status during the impulsive phase from
A1 to A3 at 21:34:12 UT and the maximum time interval
studied for this flare is 32 s starting at 21:34:26 UT.

\subsection{Short time intervals}
Four seconds is roughly the rotation period of RHESSI and thus the
shortest time interval studied here. Longer time intervals of 8, 16, and
32 seconds were also studied for some flares depending on the length of
the impulsive phase. The results for these time intervals are
very similar to the results for the full impulsive phase. The inversions
for short time intervals
generally show confidence intervals around an anisotropy of 1 at the
$1\sigma$
level extending to around 2 below 100 keV and sharply increasing above
that. As the count statistics are lower for the shorter time intervals
the confidence intervals are wider than for the full impulsive phase.
There was no discernible statistically significant variation in the
level of anisotropy for the duration of the impulsive phase. As an
example the anisotropies for each of the 4 second time intervals for the
flare on 10th November 2004 are shown in Figure~\ref{fig:As-panels}.

\section{Discussion and Conclusions}

This analysis shows consistently that for almost all flares studied by RHESSI
that the recovered ${\overline {F}_d} / {\overline {F}_u}$
is close to unity within the confidence intervals being consistent with
an isotropic pitch angle distribution. For almost every flare downward beaming of a ratio greater
than $\sim 3:1$ is ruled out to $3-\sigma$ confidence below
$\sim 150$ keV. The only clear exception to this is the flare on 20th August 2002 between 30 and 50 keV, 
where the recovered flux appears to be inconsistent with isotropic at the 3-$\sigma$ level
 and suggest a slightly greater ($1.5 - 2$) upward flux. This flare is unusual in several respects.
There is a high level of particle contamination throughout the impulsive phase. 
Several background subtractions were examined  to attempt to account for this. 
This flare is also one of the flattest flares studied, which makes pileup correction
more difficult to estimate. Also, this is one of only two flares studied where the attenuator status
was A1 for the examined time interval. 

These measurements appear to rule out any strong beaming such as would
be expected in the basic collisional thick target model. However as only two
components are recovered and the confidence intervals can be fairly large using this
method, the observations are consistent with a range of possible pitch
angle distributions including fully isotropic distributions, pancake
distributions and weak beaming below the measured confidence level.
The size of the uncertainties could be reduced with better count
statistics and better energy resolution as each of these distributions does show 
different spectral variation. The forward modelling shows very clearly that when there is
significant beaming in the emitting electron population it will result
in a very strong albedo emission. For many cases with substantial
beaming, particularly close to the disk centre, the albedo component can
dominate over the primary component.  In order to directly compare the measured results to the forward model a plot of $\Delta\mu$ against anisotropy  (${ \overline F_{d} } /  { \overline F_{u}}$) is included (Figure~\ref{fig:delmu}) both for the functional form described by Equation 3 and for another commonly used form ${\overline F} \propto  \exp \left( \frac{-(1 - \mu)}{\Delta
\mu} \right)$. It should be stressed that the results in Figures~\ref{fig:A-panels} and \ref{fig:As-panels} are model independent and can be interpreted in terms of a variety of models, including, but not limited to those considered in Figure~\ref{fig:delmu}.

These results are consistent with previous published results which showed little
evidence of directivity below 300 keV.  It should be noted that this study measures anisotropy 
in terms of the electron flux, whereas for other types of study the parametrisation
of anisotropy is  often in terms of the directivity of the X-ray
emission for stereoscopic studies and the centre-to-limb variance for
statistical studies. These are generally related to the electron anisotropy in a model
dependant manner.  As the X-ray emission can be quite broad, particularly for low energies, a large anisotropy in the electron spectrum could result in a low photon spectrum directivity. \inlinecite{2007A&A...466..705K} performed a
centre-to-limb study using RHESSI data and inferred 
a directivity ratio between 0.2 and 5 in the range 15 - 20 keV.  As the emission below $\sim 30 \, keV$ is expected to be predominantly produced by thermal electrons it is expected that the distribution in this energy range should be  isotropic. This is particularly true for the flares on 17 June 2003 and 10 November 2004 which show strong thermal components however this may not be the case for flares which show a weak thermal component  such as the flare on 20 August 2002.

As this study measured X-rays in the energy range 10 - 500 keV the
reliability of the inversion above approximately 250 keV is questionable.
 As can be seen from Figure~\ref{fig:A-panels} the confidence interval
increases significantly at a few hundred keV. Thus it is difficult to make
comparisons with the SMM studies which examined X-ray measurements above
300 keV. However the measurements in this study are for the most part in
agreement with previous studies \cite{1991ApJ...379..381M}.

As electrons propagate through the corona and chromosphere they will be pitch angle scattered 
by Coulomb collisions (e.g. \opencite{1981ApJ...251..781L}; \opencite{1991A&A...251..693M}),
although it seems that collisions will be insufficient to isotropise
an initially beamed distribution \cite{1972SoPh...26..441B,1981ApJ...251..781L}.
These results, therefore, suggest that either the accelerated electron
population is more isotropic, or other transport effects are more important than
anticipated. Specifically, the electron scattering by various
wave-particle interactions could increase the pitch angle spread of the energetic electrons.
Further, if the distribution of energetic electrons is close to
isotropic, the role of return current should be diminished. In addition we note, that
although the return current itself does contribute to the formation of a backward going beam, it
is likely to be more efficient at energies below $\sim 50 $ keV, so that the higher
energy electrons are expected to be weakly affected \cite{2011SSRv..159..107H}.

\begin{acks}

This work is supported by STFC, UK. The European Commission 
is acknowledged for funding from the HESPE 
Network (FP7-SPACE-2010-263086) (EPK).

\end{acks}

%

%


%

%
%

%
%
%
%
%
%

\end{article}
\end{document}